  \documentclass[preprint2]{aastex}

\def\lsim{\mathrel{\rlap{\lower 4pt \hbox{\hskip 1pt $\sim$}}\raise 1pt \hbox
        {$<$}}}
\def\gsim{\mathrel{\rlap{\lower 4pt \hbox{\hskip 1pt $\sim$}}\raise 1pt \hbox
        {$>$}}}

\slugcomment{Accepted to ApJ: October 29, 2008}

\shorttitle{}

\shortauthors{Izutani et al.}

\begin{document}
\title{Explosive Nucleosynthesis of Weak r-Process Elements in
Extremely Metal-Poor Core-Collapse Supernovae}

\author{Natsuko Izutani$^{1}$, Hideyuki Umeda$^{1}$ and Nozomu Tominaga$^{2}$}

\affil{
$^{1}$Department of Astronomy, School of Science, University of Tokyo, Hongo,
Tokyo 113-0033, Japan}
\email{izutani@astron.s.u-tokyo.ac.jp; umeda@astron.s.u-tokyo.ac.jp}
\affil{
$^{2}$National Astronomical Observatory, 2-21-1 Osawa, Mitaka, Tokyo
181-8588, Japan}
\email{nozomu.tominaga@nao.ac.jp}
\affil{Accepted to ApJ: October 29, 2008}

\begin{abstract}
There have been attempts to fit the abundance patterns of extremely 
metal-poor stars with supernova nucleosynthesis models for the
 lighter elements than Zn.
On the other hand, observations have revealed that the presence of EMP stars
with peculiarly high ratio of ``weak r-process elements'' Sr, Y and Zr.
Although several possible processes were suggested for
 the origin of these elements, the complete solution for reproducing those ratios is not found yet.
In order to reproduce the abundance patterns of such stars, 
we investigate a model with neutron rich 
matter ejection from the inner region of the
conventional mass-cut.
We find that explosive nucleosynthesis in a high energy supernova (or ``hypernova'')
 can reproduce 
the high abundances of Sr, Y and Zr but that the enhancements of Sr,
 Y and Zr are not achieved by nucleosynthesis in a normal supernova.
Our results imply that,
if these elements are ejected from a normal supernova,
nucleosynthesis in higher entropy flow than that of the supernova shock
is required.

\end{abstract}

\keywords{supernovae: general  
--- nuclear reactions, nucleosynthesis, abundances}

\section{INTRODUCTION}
The abundance patterns of extremely metal-poor (EMP) stars are useful
in studying nucleosynthesis in massive supernovae (SNe).
Population (Pop) III stars 
are usually considered to be 
massive stars. Some of them might become black holes without 
supernova explosions, but some should have exploded as supernovae 
to initiate the first metal enrichment in the early universe.
The stars born from the gas enriched by the Pop III SNe are 
Pop II stars with low metallicity.
Low-mass Pop II stars have long lifetimes
and might be observed as extremely metal poor (EMP) stars
with [Fe/H] $\lsim -3$.
(Here, [A/B] = $\log_{10}$($N_{\rm A}$/$N_{\rm B}$) - $\log_{10}$($N_{\rm A}$/$N_{\rm B}$)$_\odot$, where
the subscript ``$\odot$'' refers the solar value and
$N_{\rm A}$ and $N_{\rm B}$ are the abundances of elements A and B, respectively.) 
Therefore, a EMP star may reflect the nucleosynthetic result of a 
Pop III SN and constrain properties of the Pop III SN.

There have been attempts to actually fit the abundance patterns of EMP stars
with the supernova nucleosynthesis models. 
For example, using the mixing-fallback model
proposed by Umeda $\&$ Nomoto (2002) (hereafter UN02) and Umeda $\&$ Nomoto (2003) (hereafter UN03)
mimicking aspherical explosion effects (e.g., Tominaga 2009),
they showed that the abundance patterns of the elements from C to Zn
of carbon-normal EMP stars and carbon-rich EMP stars can be successfully
reproduced by energetic core-collapse SN (``hypernova'', hereafter HN), models
and faint SN models, respectively (UN02; UN05; Tominaga et al. 2007),
while those of very metal poor (VMP) stars 
($-3 \lsim$ [Fe/H] $ \lsim -2$)
can be reproduced by normal core-collapse SN models
or the IMF-integration of hypernova and normal core-collapse SN models (Tominaga et al. 2007).
It is important to note that the observed EMP stars are so far all explained
by the pollutions by core-collapse SNe with initial stellar masses
11$ M_\odot \lsim M \lsim 130 M_\odot$ 
and no evidence of pair instability SNe\footnote{Pair instability SNe explode by the explosive oxygen burning, before the
onset of Fe core-collapse.} with initial
stellar masses 140$M_\odot \lsim M \lsim
300 M_\odot$ (UN02, see also Chieffi $\&$ Limongi 2002;
UN03; Umeda $\&$ Nomoto 2005, hereafter UN05; Heger $\&$ Woosley 2002, 2008).

 These previous SN models do not eject elements heavier than Zn in a sizeable
amount, and 
this is consistent with the abundance of some EMP stars.
However, there are also EMP stars showing enhancements of neutron-capture elements.
Some of them show abundance patterns almost identical to
the solar system r-process pattern for Sr and heavier elements (e.g., 
Sneden et al. 2000; Hill et al. 2002).
One example of such a star is CS22892-052 and called as a main ``r-process
star'' (Sneden et al. 2003).
The process producing heavy neutron capture elements ($\sim$Ba-U)
is referred to as ``main'' r-process (e.g., Truran et al. 2002; Wanajo \&
Ishimaru 2006).
On the other hand, there are other EMP stars that require another 
neutron-capture process referred sometimes as 
``LEPP'' (lighter element primary process) or ``weak r-process''
(Travaglio et al. 2004; Wanajo $\&$ Ishimaru 2006).
Travaglio et al. (2004) reported EMP stars with abundances of Sr,
Y, and Zr which
 cannot be explained by the s-process or main r-process.
In the weak r-process stars, the elements 
with intermediate mass (37$\leq$Z$\leq$47, i.e., from Rb to Ag) elements
show moderate enhancements with respect to heavy ones (Z$\geq$56, i.e.,
heavier than Ba).
More recently Franc\c ois et al. (2007) showed several other examples
of the weak r-process stars.
There are evidences of the existence of weak r-process but 
its origin is unknown.  

 Several possible mechanisms to produce the weak r-process elements are
proposed.
Wanajo et al. (2001) presented calculations of r-process nucleosynthesis
in neutrino-driven winds from a proto-neutron stars.
They showed that the abundance pattern of weak r-process is reproduced
when the main r-process nucleosynthesis is failed.
The nucleosynthesis in neutrino-driven winds was also studied
in Hoffman et al. (1996) by taking the electron fraction $Y_{\rm e}$
of the wind matter as a free parameter.
They showed that weak r-process elements may be synthesized
for low $Y_{\rm e}$ ($<$0.47).
Although the production of
weak r-process elements in neutrino-driven winds was suggested, it is difficult to
give a detailed yield because the physical conditions and
ejected mass depend on unknown supernova explosion mechanisms.   

Explosive nucleosynthesis in low $Y_{\rm e}$ matter was also studied
in the context of multi-dimensional explosion models (Janka et al. 2003; 
Pruet et al. 2003). They showed that small amounts of low $Y_{\rm e}$ 
($\lsim $0.46)
matter as well as high $Y_{\rm e}$ ($\lsim$0.56) matter are ejected 
from a hot bubble just outside a proto-neutron star.
The high $Y_{\rm e}$ ($\lsim$0.56) matter is suggested to be 
ejected even in the one-dimensional cases (e.g., Fr\"olich et al. 2006).
On the other hand, the ejection of low $Y_{\rm e}$ matter is driven
by the convection in the hot bubble, and thus essentially the multi-dimensional
phenomenon.
Janka et al. (2003) suggested that the low $Y_{\rm e}$ matter contains
the weak r-process elements (Sr, Y and Zr) to explain the Galactic abundances, but
a detailed nucleosynthesis calculation did not confirm the production of these 
elements (Pruet et al. 2005).

In this paper, we investigate the physical conditions to
produce sufficient amounts of the weak r-process elements (Sr, Y and Zr)
and discuss whether core-collapse SNe 
with a slight modification can be
compatible with the observed abundances of the weak r-process elements
in the EMP stars.
In order to do this, we assume that small amount of low $Y_{\rm e}$ matter
is ejected by the multi-dimensional effects, which 
may be driven by the convection in a hot bubble (Janka et al. 2003) 
or jets in a jet-like explosion (e.g., Maeda $\&$ Nomoto 2003)
or a collapsar model (e.g., Pruet et al. 2003, 2004; Popham, Woosley
$\&$ Fryer 1999). 
The entropy of the low $Y_{\rm e}$ matter flow may depend on the
ejection mechanism.
We assume that the matter flow has the same entropy as the supernova shock wave. Jet-like explosion or collapsar models may describe HNe.
However, they contain many unknown parameters, and the innermost $Y_{\rm e}$ of
the ejecta depends on those parameters.
On the other hand, the simulations of Janka et al. (2003) contain less input parameters, so the obtained $Y_{\rm e}$ profile is more reliable, though 
their simulations are about normal SNe.
We are interested in the EMP stars, and their progenitor may be more massive and explode energetically, i.e., may become HNe.
Therefore, we vary $Y_{\rm e}$ beyond the range given by the simulation in Janka et al. (2003).
Although we have multi-dimensional effects in mind,
we only perform one-dimensional calculations in this paper,
because it is often useful to make a large parametric
search to disclose the essence of physics.

In $\S$ 2, we show observational trends of [Sr,Y,Zr/Fe].
In $\S$ 3, we describe our progenitor and explosion models.
In $\S$ 4, we present weak r-process nucleosynthesis and specify conditions
mention our assumption applying to our models in order to reproduce 
reproducing the observational [Sr,Y,Zr/Fe]. 
We also compare our yields with 4 EMP stars which have
peculiarly high [Sr,Y,Zr/Fe].
In $\S$ 5, summaries and discussions are given.

\section{OBSERVATIONAL DATA}

 Since we are interested in the weak r-process elements in the EMP stars,
we select stars with [Fe/H] $\lsim -2.8$ from Cayrel et al. (2004) and
use their data from carbon to zinc.
Observations of [Sr,Y,Zr/Fe] are taken from Honda et al. (2006) for HD122563
and Franc\c ois et al. (2007) for the other stars.

Taking the previous works on the weak r-process into consideration,
we use two abundance ratios as a diagnostic to distinguish 
``main r-process stars'' and ``weak r-process stars''.
They are relative numbers of Sr and Ba, Sr/Ba, and Y and Eu, Y/Eu. We use
$\log(N/N_{\rm H})_{\odot}$ from Anders $\&$ Grevesse (1989).
A main r-process star CS22892-052 has [Sr/Ba] = -0.57 and 
[Y/Eu] = -1.16.
Therefore, if a EMP star has [Sr/Ba] $> -0.57$ and [Y/Eu] $> -1.16$, 
we consider the star as a weak r-process star, otherwise as a main r-process
star (see also Aoki et al. 2005).
Figure 1 shows [Sr/Fe], [Y/Fe], and [Zr/Fe] vs. [Fe/H]
of the weak r-process stars. Among 21 selected stars, 20 stars are 
with $-1 \lsim$ [Sr,Y,Zr/Fe] $\lsim 1$. 

\section{METHOD $\&$ MODEL}

The calculation method and other assumptions are the same as described in
Umeda et al. 2000 (hereafter UNN00), UN02, UN05 and Tominaga et al. 2007, 
except for the size of the nuclear reaction networks.
In this paper, we adopt the Pop III progenitors as in UN05 and
apply the model with $M$ = 13 $M_\odot$ and $E_{51}$ = 1.5 (hereafter model-1301),
the one with $M$ = 25 $M_\odot$ and $E_{51}$ = 1 (hereafter model-2501), and
the one with $M$ = 25 $M_\odot$ and $E_{51}$ = 20 (hereafter model-2520).
Model-1301 and model-2501 are normal SN models,
and model-2520 is a HN model.
Detailed nucleosynthesis is calculated as a postprocessing
after the hydrodynamical calculation with a simple $\alpha$-network.
The isotopes included in the post process calculations are 809
species up to $^{121}$Pd (see Table 1).
We note that a neutrino process during explosive burning (Yoshida et al. 2008; Woosley $\&$ Weaver 1995) is not taken into account.
The abundance distributions 
after the SN explosion for
model-1301, model-2501 and model-2520 are shown in Figure 2.
We obtain the final yields by setting the inner boundary of the ejected matter,
a mass-cut ($M_{\rm cut}$) as we describe in the next section.
 
\subsection{MASS-CUT}

In this section we summarize the abundance pattern of SN ejecta when
a ``conventional'' mass-cut is adopted.  The ``conventional'' means that the mass-cut is chosen to eject 
a reasonable amount of $^{56}$Ni.
Although previously the mass-cut is often chosen to eject 0.07 $M_\odot$ of $^{56}$Ni
reproducing the brightness of normal SNe as SN1987A
(e.g., Nomoto et al. 2003b), 
we set $M_{\rm cut}$ = 1.59 $M_\odot$ for model-1301,
1.76 $M_\odot$ for model-2501 and 2.31 $M_\odot$ for model-2520 to yield  
[Si/Fe]  $\simeq$ [Si/$^{56}$Ni] $\sim$ 0.4.
As a result, 13$M_\odot$ model eject a similar amount of $^{56}$Ni to SN~1987A
($\sim$ 0.07 $M_\odot$), 
but 25$M_\odot$ models eject large amounts of
$^{56}$Ni ($\sim$ 0.5 $M_\odot$)\footnote{In the models assuming the
mixing-fallback effects, the actual amount of the ejected $^{56}$Ni mass
is smaller than this value. See e.g., Table 3 and section 4.4 below.}.
In Figure 3, the abundance patterns from Si to Ru are compared with those of EMP stars.
This figure shows that the adopted $M_{\rm cut}$s yield a rather good 
agreement between the predicted and the observed abundance ratios for most of
the elements above Si.
On the contrary, [Sr,Y,Zr/Fe] in the models are much lower than those
observed in the ``weak r-process stars''.

\section{WEAK r-PROCESS NUCLEOSYNTHESIS}

As mentioned in $\S$ 3, the models with the ``conventional'' mass-cut do not reproduce 
$ -1 \lsim$ [Sr,Y,Zr/Fe]  $\sim 1$.
In this section, we study the conditions to produce weak r-process elements.
We take into account of the uncertainty of $Y_{\rm e}$ 
and assume some ejection of matter from regions below $M_{\rm cut}$.

\subsection{$Y_{\rm e}$ Uncertainty and Mass Ejection from the Region below 
$M_{\rm cut}$}

Recent theoretical multi-D hydrodynamical simulations of core collapse
supernova have shown that the presupernova value of $Y_{\rm e}$ can be modified during the explosion, even significantly, in the innermost zones of the 
exploding envelope.
In Figure 4 we schematically depict the value of $Y_{\rm e}$
before and after the explosion.
The presupernova value shown is for the model when the central density is
$\sim$ $10^{10}$g cm$^{-3}$.
After that time, the electron capture significantly reduces $Y_{\rm e}$ ($\lsim$0.4)
in the inner part, but the very neutron-rich matter is rarely ejected.

Recent simulations have shown that not only neutron-rich($Y_{\rm e}$ $<$ 0.5)
but also proton-rich($Y_{\rm e}$ $>$ 0.5) regions appear after explosion(e.g.,
Fr$\ddot{\rm o}$lich et al. 2006).
The $Y_{\rm e}$ distributions based on an actual 2D-simulations is, for
example, given in Figure 4 of Pruet et al. (2006).
Figure 4 shows the assumed $Y_{\rm e}$ profile mimicking the results of such
simulations. 
Those simulations have shown that a density just above a proto-neutron star
surface rapidly decreases after the supernova shockwave passes through
a Fe-core. This region is often called a hot bubble, in which
$Y_{\rm e}$ is set by a competition between different 
lepton capture processes on free nucleons.
At the beginning of the explosion, an excess of electron neutrinos over antineutrinos makes the matter
tend to be proton-rich (Qian and Woosley 1996).
Recent detailed one- and two-dimensional simulations have shown that 
some of these proton-rich matter is actually ejected.
In the later stages, the fluxes and 
spectral change of neutrinos make $Y_{\rm e}$ less than 0.5.
2D simulations by Janka et al. (2003) 
have shown that not only the proton-rich matter
but also some neutron-rich matter ($Y_{\rm e}$$\gsim$0.46-0.47) is ejected.
This is because the hot bubble is convective and some of the inner matter
can be dragged outside.

Although these previous calculations are for less massive normal
supernovae, the similar mechanism may work for more massive supernova.
Even if the explosion mechanism is completely different,
the inner matter may be carried outside along the jets in a
jet-like explosions.
Therefore, in the following we calculate nucleosynthesis in a deep region
of supernovae to estimate the total yield
when the same amount of matter below the conventional mass cut,
with $Y_{\rm e}$ arbitrarily changed with respect to the presupernova value, is ejected.   

In order to perform nucleosynthesis calculations,
we need histories of temperature and density for a given mass element.
Strictly speaking, the histories depend on explosion models and 
how the matter is carried outside and cannot be represented 
by a one-dimensional model.
To avoid complication, however, we carry out the same calculations with Section 3 but in the region below $M_{\rm cut}$ and arbitrarily changing the $Y_{\rm e}$ in the progenitor model according to Figure 4.
This approach helps simplifying the complicated problems and clarifying the
essence of physics.
With this assumption, the entropy of the low $Y_{\rm e}$ flow
is $s/k_{\rm b}$$\sim$3 for SNe and $\sim$15 for HNe,
which is similar value as the matter just above $M_{\rm cut}$.

It has been long discussed that if a normal SN produces
main r-process elements. This is because the SN has to eject quite high-entropy
neutrino driven wind.
As for the weak r-process, Wanajo et al. (2001), for example,
showed that if the entropy of the neutrino driven wind is not high enough for
the main r-process, a weak r-process-like pattern is obtained.
However, whether a SN can eject such kind of mater is
unpredictable because explosion simulations have not been succeeded.
Therefore, we consider an extreme case that 
no neutrino-driven high-entropy matter but the supernova-shocked matter 
is ejected. The ejection of high entropy matter 
are discussed in the last part of Section 5 briefly and will
be discussed in detail elsewhere.

\subsection{Parameter Dependences of ``Complete Si-Burning''}

In this subsection, we show parameter dependences of the products
of the complete-Si burning.
The complete Si-burning takes place in the shocked matter
attaining the maximum temperature of $\log_{10} T_{\rm max}$ $>$9.5
and thus no unburned Si is left after the complete Si-burning.
Since we are mainly interested in the ejection of low $Y_{\rm e}$ matter
from the beneath of a conventional mass-cut, $M_{\rm cut}$,
we change $Y_{\rm e}$ in the region from 0.40 to 0.50.
Although 
Janka et al. (2003) showed the small amount of ejection of matter with
$Y_{\rm e}$ $\sim$0.56-0.46 from the hot bubble region,
we consider $Y_{\rm e}$ as low as 0.40 because
the explosion mechanism is quite uncertain especially for
hypernovae. In this paper we do not consider $Y_{\rm e}$ $>$0.5 matter. 
Its effect is briefly discussed in Section 5.

We use the temperature and density trajectories of the models 1301,
2501 and 2520 as \S 3.
The difference from \S 3 is that we consider the deep region below
$M_{\rm cut}$.
We denote the mass coordinate of the inner and outer boundaries of the
region by $M_1$ and $M_2$, respectively, and take mass average in the region.
$M_2$ is defined by the location where $X$($^{28}$Si)$\sim$10$^{-3}$.
$M_1$ is chosen at a point near Fe core surface.
We note that the result is not sensitive to $M_1$
because density and temperature trajectories during nucleosynthesis
is almost the same around $M_1$.
The specific values taken are
($M_1$,$M_2$)=(1.41$M_\odot$,1.64$M_\odot$),(1.52$M_\odot$,1.92$M_\odot$) and 
(1.61$M_\odot$,2.69$M_\odot$) for model-1301, model-2501, and model-2520, respectively.
Figure 5 illustrates the $Y_{\rm e}$ profile for the three models
we have assumed.

 Nucleosynthesis of this region basically proceeds as $\alpha$-rich freezeout.
When the shock wave reaches the region, temperature rises rapidly and
heavy elements are decomposed mainly into $\alpha$-particles.
As the star expands and the temperature drops,
$\alpha$-rich freezeout takes place with roughly constant $Y_{\rm e}$.
The mass fraction of $\alpha$ at the maximum temperature and later stages
depends on the entropy, i.e., temperature and density.
More $\alpha$ is produced and heavier elements are synthesized
for higher entropy explosions (e.g., UN02).

\subsubsection{$Y_{\rm e}$ Dependence}

In Figures 6-8, abundance ratios, [X/Fe], of the complete Si-burning 
products integrated over $M_1$ and $M_2$ are shown.
The abundance ratios of heavy elements as a function of $Y_{\rm e}$
show non-monotonic behavior in Figure 6-8.
This behavior depends not only on the abundance of X, but also on that of Fe.
For example, in Figure 6 and 7, Fe to heavy elements ratios, 
such as Sr/Fe, appear
to be minimum for $Y_{\rm e}$ $\sim$0.45.
This is because for $Y_{\rm e}$ $\sim$0.45 the most abundant isotope
which becomes Fe is $^{56}$Fe and not $^{56}$Ni.
$^{56}$Fe is produced relatively a lot for $Y_{\rm e}$ $\sim$0.45.
Therefore, the reason for this non-monotonic behavior is not simple.
These figures also show that for $Y_{\rm e}$ $\gsim$ 0.49,
elements heavier than Zn are not efficiently produced
compared to Fe.
We provide some explanations of the result paying attention to
the abundance of X in the following paragraphs.

Absolute amounts of the synthesized weak r-process elements
are easily seen in Figure 9, which
shows the mass fractions of Sr, Y and Zr in the region.
This figure shows that the abundances of these elements are
relatively large for $Y_{\rm e}$$\lsim$0.48 and have a peak around
$Y_{\rm e}$$\sim$0.43-0.46.
For $Y_{\rm e}$$\sim$0.42-0.43, most abundant isotopes that become Sr, Y and Zr
are $^{88}$Kr, and $^{89}$Kr and $^{90}$Kr, respectively.
For $Y_{\rm e}$$\sim$0.45-0.46, these are $^{88}$Sr, $^{89}$Y and $^{90}$Zr, respectively.
Since these isotopes are all neutron-rich having $Y_{\rm e}$$\equiv$Z/A$\sim$0.41-0.45, 
the weak r-process elements are tend to be
produced most efficiently for this range of $Y_{\rm e}$.
We note that the exact processes complicatedly depend on entropy, $Y_{\rm e}$,
the properties of nuclear states, and the mass fraction of $\alpha$
particles during $\alpha$-rich freezeout.

The abundance peaks seem to locate at $Y_{\rm e}$$\sim$0.43 
for the normal energy models (1301 and 2501), and 
$Y_{\rm e}$$\sim$0.45-0.46 for model-2520.
It is interesting that the peak is located at a larger value
of $Y_{\rm e}$ for the higher energy model. The reason for this is
not simple because the yields depend on the complicated properties
of nuclear structures. For example,
in the $Y_{\rm e}$=0.40 case in Figure 8 (model-2520),
the synthesized amounts of the weak r-process
elements are small, but lighter elements, Ge to Kr, are quite abundant.
For this specific case
$^{82}$Ge, with $Y_{\rm e}$$\equiv$Z/A=0.390,
is quite abundant after the explosive synthesis. This decays
into $^{82}$Se and the synthesis of heavier weak r-process elements are
suppressed. 

\subsubsection{$M$ Dependence}

Since the density and temperature trajectories of 
the complete Si-burning region are not so different
for model-1301 and model-2501, 
the abundance of weak r-process elements in model-1301 and model-2501
are similar (Figure 6, 7 and 9).

\subsubsection{$E$ Dependence}

As seen in Figure 7, 8 and 9,
high $E$ enhances the weak r-elements especially 
from $Y_{\rm e}$ = 0.45 to 0.47.
Temperature of the complete Si-burning region in model-2520 is much higher
than that in the model-2501.
Therefore, an entropy of model-2520 is much higher than that of model-2501,
and much more $\alpha$-particles can be obtained in model-2520 than
in model-2501.
This is why more Sr, Y and Zr are produced in model-2520 than in
model-2501.

The abundances of Sr, Y and Zr are almost same in model-2501 and model-2520
when $Y_{\rm e}$=0.49 and 0.50.
When $Y_{\rm e}$ is 0.49 and 0.50, the elements produced also have 
$Y_{\rm e} \simeq $0.5.
Since heavy nucleus with $Y_{\rm e} \simeq $0.5 are 
less bound than those with neutron-rich (Hoffman et al. 1996),
 when $Y_{\rm e} \simeq$ 0.5,
heavier elements than Zn are not produced even though $E$ is high.

\subsection{Abundance Patterns of Whole Ejecta with Mass Ejection below $M_{\rm cut}$} 
Nucleosynthesis pattern 
in the complete Si-burning region with constant $Y_{\rm e}$ is shown in the previous subsection. 
In this subsection, assuming various $Y_{\rm e}$ distributions in region below $M_{\rm cut}$,
we present abundance patterns of whole ejecta with mass ejection below $M_{\rm cut}$.
For the matter below $M_{\rm cut}$,
the yields are averaged for 
the $Y_{\rm e}$ values ranged from $Y_{\rm e}^{\rm min}$ to $Y_{\rm e}^{\rm max}$.
For example, the Y$_{\rm e}$ distribution Y$_{\rm e}$ = 0.45-0.50 means
that the yields are the average of the six models with Y$_{\rm e}$ = 0.45, 0.46, 0.47, 0.48, 0.49 and 0.50.
In this subsection again we do not consider the matter with $Y_{\rm e}$$>$0.5.

The main parameters of our models are $\Delta M$ and $Y_{\rm e}^{\rm min}$.
$\Delta M$ is the mass of the ejected matter from the region below
$M_{\rm cut}$, and $Y_{\rm e}^{\rm min}$ is the lowest $Y_{\rm e}$
of the ejected matter.
$\Delta M$ of the matter below $M_{\rm cut}$ is added to the matter above $M_{\rm cut}$,
and they are assumed to be ejected to the outer space all together.
Small amounts of matter with low $Y_{\rm e}$
could be ejected in two-dimensional or jet-like explosion models 
(e.g., Janka et al. 2003).
As shown in the previous subsection,
large amounts of Sr, Y and Zr are obtained when $Y_{\rm e}$ = 0.43 in model-1301 and model-2501, and
when $Y_{\rm e}$ = 0.45 in model-2520. 
Therefore, we take $Y_{\rm e}^{\rm max}$ = 0.5, and $Y_{\rm e}^{\rm min}$
= 0.42 and 0.43 for model-1301 and model-2501, and
various values from $Y_{\rm e}^{\rm min}$ = 0.45 to 0.49 for model-2520.
Here, $\Delta M$ is selected to obtain 0 $\lsim$
[Zn/Fe] $\lsim$ 0.5.


The abundance patterns of the ejected matter for model-1301, model-2501 and 
model-2520 are shown in Figure 10, 11 and 12-13, respectively.
$Y_{\rm e}^{\rm min}$ and $\Delta M$ and some related numbers for models (a) are summarized in Table 2.
In the models
(a), $\Delta  M$ is set to obtain [Zn/Fe]$\simeq$0.5 for $Y_{\rm e}^{\rm min}$$\lsim$0.46.
For $Y_{\rm e}^{\rm min} \gsim$0.47 (model-2520), the same value of $\Delta M$ 
with the $Y_{\rm e}^{\rm min}$$\lsim$0.46 models is adopted
because [Zn/Fe] is lower than 0.5 even if all the matter below $M_{\rm cut}$ is ejected.
For the models (b) and (c), $\Delta M$ of the model (a) is divided by 3 and 10, respectively.

 As shown in Figure 10 and 11, [Sr,Y,Zr/Fe] are not improved to fit the observation
in model-1301 and model-2501.
On the contrary, [Sr,Y,Zr/Fe] in model-2520 are high enough to be ranged in 
$-1 <$ [Sr,Y,Zr/Fe] $<$ 1 when $Y_{\rm e}^{\rm min}$ is from 0.45 to
0.46 (Fig. 11).
Note that the abundance patterns of [Sr,Y,Zr/Fe] are different depending on $Y_{\rm e}^{\rm min}$.
When $Y_{\rm e}^{\rm min}$ $=$0.45, those ratios have the relation of
[Sr/Fe]$\sim$[Zr/Fe]$>$[Y/Fe].
When $Y_{\rm e}^{\rm min}$ $=$0.452, [Sr/Fe] $\simeq $ [Y/Fe] $<$ [Zr/Fe].
When $Y_{\rm e}^{\rm min}$ $=0.454-0.46$, [Sr/Fe] $<$ [Y/Fe] $<$ [Zr/Fe].

\subsection{Comparisons with Individual EMP Star}
In this subsection we compare our yields of model-2520 with abundance
patterns of EMP stars.
Among the 22 weak r-process stars mentioned in $\S$ 1,
15 stars have [Sr/Ba]$>$ 1 and [Y/Eu] $>$ 1.
We select Sr-rich stars with [Sr/Ba] $>$ 1, [Y/Eu] $>$ 1
and [Sr/Fe] $>$ 0: BS16477-003 ([Fe/H]=-3.36), CS22873-166 ([Fe/H]=-2.97)
, CS22897-008 ([Fe/H]=-3.41) and CS29518-051 ([Fe/H]=-2.78).
The observational data from He to Zn and beyond Zn are taken
from Cayrel et al. (2004) and Franc\c ois et al. (2007), respectively.
$Y_{\rm e}$ distribution and $\Delta M$ have been chosen in order to obtain 
the best fit to the observed abundances of Sr, Y and Zr.
In order to adjust light to heavy element ratios, such as O/Fe,
we also assume the following mixing-fallback process to take place in
HN model-2520 (see e.g., Tominaga et al. 2007 for detail):\\
1. Burned material below $M_{\rm mix}$(out) is uniformly mixed.\\
2. Afterward only a fraction ($f$) of the mixed material is ejected
with the matter above $M_{\rm mix}$(out). \\
We set $M_{\rm mix}$(out) = 3.60$M_\odot$ 
for the C-poor stars with [C/Mg]$<$0, such as CS22873-166 and CS29518-051,
and 5.76$M_\odot$ for C-normal stars with [C/Mg]$\sim$0, such as BS16477-003 and CS22897-008.
$M_r$ = 3.60$M_\odot$ is a point above which X(Ca)$<$X(Mg), 
and $M_r$ = 5.76$M_\odot$ is a point 
in the O-rich layer. 
$Y_{\rm e}^{\rm min}$, $\Delta M$, $f$, $M_{\rm mix}$(out) and 
the other values for each comparison are summarized in Table 3.
Figure 14 shows comparisons between the yields of our mixing-fallback models and 
the abundance patterns of EMP stars.
In addition to the abundance ratios of the elements heavier than Si,
[C/Fe], [Mg/Fe] and [Al/Fe] show reasonable agreements with the observations.
The nucleosynthesis yields in the ejecta for selected isotopes
at the time around 150 seconds after the
explosion are also given in Tables 4 to 7. To obtain these tables, the isotopes
with their half-lives less than 30 days except $^{56}$Ni
are radioactively decayed.

\section{CONCLUSIONS AND DISCUSSIONS}
In this paper we assume uncertainty of $Y_{\rm e}$ in the deep regions below $M_{\rm cut}$
and mass ejection from the regions for three models model-1301, model-2501
and model-2520.
Among those models, we obtain high [Sr,Y,Zr/Fe] (ranged from -1 to 1)
only in the model-2520.
The ``hypernova'' model-2520
can reproduce the observational data of Sr, Y and Zr
in addition to the elements from C to Zn.

We also find that the weak r-process elements are not contained in the 
``normal'' supernova models 1301 $\&$ 2501, even though low $Y_{\rm e}$
($\gsim$0.40) matter is ejected.
In the normal supernova models, however, intermediate mass elements from Ga to Rb may be
abundantly ejected (Figure 10 and 11).
It is interesting to note that there have been no observational evidences that
Ga-Rb-rich stars exist in EMP stars.
It is possible that normal SNe do not eject sufficient amounts of
low $Y_{\rm e}$ matter or that we have observationally overlooked
such Ga-Rb-rich stars.

In comparisons with 4 EMP stars in Figure 14,
the ratios of some elements, i.e., 
[Na/Fe], [K/Fe], [Sc/Fe], [Mn/Fe] and [Co/Fe]
show deficiencies from the observation as our previous finding
(see e.g., UN05; Tominaga et al. 2007).
In Tominaga et al. (2007), possible solutions are discussed as follows:
Na is mostly synthesized in the C-shell burning, and
the produced amount of Na depends on the overshooting at the edge of the convective
C-burning shell (Iwamoto et al. 2005).
Since no overshooting is included in the present presupernova
evolution models, the inclusion of the overshooting could enhance
the Na abundance.
[K/Fe] is slightly enhanced by the ``low-density'' modification
(UN05; Tominaga et al. 2007)
and Iwamoto et al. (2006) suggests that the matter with large
$Y_{\rm e}$ ($>$0.5) can produce enough K.
[Sc/Fe] and [Ti/Fe] can be enhanced by nucleosynthesis in high-entropy environments
(a low-density modification, UN05; Tominaga et al. 2007) or in a jet-like explosion 
(Nagataki et al. 2003; Maeda $\&$ Nomoto 2003; Tominaga 2009),
and further an enhancement of [Sc/Fe] can be realized if $Y_{\rm e}$ ($>$0.5)
(Pruet et al. 2004a, 2005; Fr\"ohlich et al. 2006b; 
Iwamoto et al. 2006).
[Co/Fe] and [Mn/Fe] can be improved by the $Y_{\rm e}$ modification
in the Si-burning region (UN05; Tominaga et al. 2007)
and [Mn/Fe] can also be enhanced by a neutrino process (Woosley $\&$
Weaver 1995; Yoshida et al. 2008).

Many of the solutions discussed above include the the ejection of proton-rich
($Y_{\rm e}$$>$0.5) ``complete Si-burning'' matter.
This does not contradict with our assumption that 
the neutron-rich ($Y_{\rm e}$$<$0.5)  ``complete Si-burning'' 
matter is ejected.
This is because both $Y_{\rm e}$$>$0.5 matter and $Y_{\rm e}$$<$0.5
matter could be ejected simultaneously from the ``hot-bubble region''
in the multi-dimensional simulations (e.g., Janka et al. 2003; Pruet et al. 2005).
Although we do not include the proton-rich matter, 
the inclusion of the matter does not change the present results
because the contributions from the proton-rich matter can merely be added 
to the present results.
The nucleosynthesis in the proton-rich matter as well as the neutrino process
will be considered elsewhere.

There also remains a possible problem in elements Mo, Ru and Rh.
The observational data of those elements is obtained in only two 
EMP stars HD122563 and HD88609 (see Honda et al. 2007).
The abundance patterns of them show continuously decreasing trends
compared with the main r-process as a function of atomic number,
from Sr to Yb (Z=38$-$70).
 Our models have [Mo/Fe]$\sim-$5, [Ru/Fe]$\sim-$2 and [Rh/Fe]$\sim-$3,
while HD122563 has [Mo/Fe]=$-$0.02, [Ru/Fe]=0.07 and [Rh/Fe]$<$0.45,
and HD88609 has [Mo/Fe]=0.15, [Ru/Fe]=0.32 and [Rh/Fe]$<$0.70.
Further observation will be needed to investigate whether high ratios of
Mo, Ru and Rh
are typical in the weak r-process stars.
Since there two stars are relatively metal rich,
[Fe/H] $\sim -3.0$ (HD88609) and [Fe/H] $\sim -2.7$ (HD122563),
their abundance patterns may be contaminated by several SNe,
r-process and s-process.

There may be a solution in a high-entropy matter ejection.
Pruet et al. (2006) investigated the contribution of the proton-rich
high-entropy winds
using the two-dimensional
15$M_\odot$ core collapse model of Janka et al. (2003).
The origin of the so-called p-process nuclei from A=92 to 126
is an unsolved riddles of nuclear astrophysics,
but they found synthesis of p-rich nuclei up to $^{102}$Pd in
the proton-rich wind, although their calculations do not show an efficient
production of $^{92}$Mo.
$Y_{\rm e}$ of proton-rich neutrino wind in Pruet et al. (2006)
is ranged from 0.539 to 0.558, and
entropy ($s$/$k_b$) is from 54.8 to 76.9.
The entropy in the supernova shock model  ($s$/$k_b$$\lsim$15) is
much smaller.
Since the properties of the neutrino driven wind is uncertain,
the nucleosynthesis in the proton-rich wind is certainly interesting, especially if there are
no other possibilities.

The $Y_{\rm e}$ below the mass cut is very sensitive to the rates
for the neutrino and positron captures on neutrons
and for the inverse captures on protons.
Unfortunately the actual amount of neutrino flux depends on the unknown explosion
mechanisms.
Therefore, the $Y_{\rm e}$ of ejecta for a specific model needs to be 
calculated in the future works.

In Figure 15 we show [Sr/Fe] vs. [Zn/Fe]
because [Zn/Fe] is a rough barometer of
the SN explosion energy (e.g., UN02, UN05).
The implications obtained from this figure are as follows.
We have shown that the weak r-process elements can
be produced without introducing extra higher-entropy matter in high $E$
SN models.
Previous work (UN02) suggests that the abundance of EMP stars
with high [Zn/Fe] are reproduced by high $E$ SN models.
The apparent no-correlation between [Sr/Fe] and [Zn/Fe] 
means that, if our interpretation is correct,
only a portion of HNe eject a large amount of Sr but
the rest of HNe eject a small amount of Sr.
In other words high $E$ is just a necessary condition to eject
the weak r-process elements and other factors determine the ejected mass
of the weak r-process elements.

Our results show that normal $E$ models do not produce
large amount of Sr, Y, and Zr.
However we should note that this is not the case if 
a normal SN ejects somehow higher-entropy matter than
the supernova shock. Figure 15 does not deny such
a possibility because
[Zn/Fe]$\sim$0 stars may be reproduced by a normal
SNe or the mixture of several SNe (Tominaga et al. 2007).
This figure show that all [Zn/Fe]$\sim$0 stars show high [Sr/Fe].
A possible interpretation is that the actual normal SNe can produce
[Sr/Fe]$\sim$ 0.
If this is the case, our results imply that a normal SN can be ejecting a higher
entropy matter than the supernova shock,
that is likely the neutrino driven wind.
This fact may be used to constrain the neutrino driven wind of a normal SN,
though we have to handle the $Y_{\rm e} >$0.5 matter before
qualitatively constraining the model,
because the high-entropy proton-rich
matter may also produce the weak r-process elements 
as shown in Pruet et al. (2006).

\acknowledgments

 We would like to thank T. Yoshida, W. Aoki and 
S. Wanajo for useful comments and discussions.
This work has been supported in part by the grants-in-aid
for Scientific Research (19840010) from the MEXT of Japan.

\clearpage

\begin{deluxetable}{llcc}
\tabletypesize{\scriptsize}
\tablecaption{Isotopes included in the nuclear reaction network
\label{tbl-1}}
\tablewidth{0pt}
\tablehead{
\colhead{Isotope} & \colhead{A} & \colhead{Isotope} & \colhead{A}}
\startdata
n & 1 & V & 44-60 \cr
H & 1-3 & Cr & 46-63 \cr
He & 3-4 & Mn & 48-65 \cr
Li & 6-7 & Fe & 50-68  \cr
Be & 7-9 & Co & 51-71  \cr
B & 8-13 & Ni & 54-73 \cr
C & 11-15 & Cu & 56-76 \cr
N & 13-18 & Zn & 59-78 \cr
O & 14-21 & Ga & 60-81 \cr
F & 17-23 & Ge & 59-84 \cr
Ne & 18-26 & As & 64-86 \cr
Na & 21-28 & Se & 65-89 \cr
Mg & 22-31 & Br & 68-92 \cr
Al & 25-34 & Kr & 66-94 \cr
Si & 26-36 & Rb & 72-97\cr
P & 27-39 & Sr & 69-100\cr
S & 30-42 & Y & 76-102 \cr
Cl & 32-44 & Zr & 74-105\cr
Ar & 34-47 & Nb & 80-107 \cr
K & 36-50 & Mo & 79-110\cr
Ca & 38-52 & Tc & 85-113\cr
Sc & 40-55 & Ru & 84-115\cr
Ti & 42-57 & Rh & 89-118\cr
   &       & Pd & 89-121\cr
\enddata
\end{deluxetable}

\clearpage

\begin{deluxetable}{rlcccccccc}
\tabletypesize{\scriptsize}
\tablecaption{
\label{tbl-1}}
\tablewidth{0pt}
\tablehead{
\colhead{model} &
\colhead{$Y_{\rm e}^{\rm min}$} & \colhead{$\Delta M$} &  
\colhead{[Sr/Fe]} & \colhead{[Y/Fe]} & \colhead{[Zr/Fe]} & \colhead{$M$(Sr)}
  & \colhead{$M$(Y)} & \colhead{$M$(Zr)} & \colhead{$M$($^{56}$Ni)}}
\startdata
1301(a) &0.42 & 1.65E-03 & -2.38 & -3.38 & -3.20 & 1.02E-08 & 2.10E-10 & 8.02E-10 & 5.93E-02\cr
1301(a) &0.43 & 1.65E-03 & -2.37 & -3.47 & -3.25 & 1.05E-08 & 1.69E-10 & 7.16E-10 & 5.93E-02\cr
\hline
2501(a) &0.42 &2.02E-02 &            -1.99 & -3.18 & -3.12 & 9.12E-08 & 1.21E-09 & 3.55E-09 & 2.19E-01\cr
2501(a) &0.43 & 2.02E-02&            -1.99 & -3.33 & -3.20 & 9.26E-08 & 8.62E-10 & 2.92E-09 & 2.19E-01\cr
\hline
2520(a) &0.45 &1.08E-02&  0.54 & 0.22 &  0.43 &  6.56E-05 & 6.40E-06 & 2.59E-05 & 4.60E-01\cr
2520(a) &0.452& 1.08E-02 &              0.38 & 0.36 & 0.52 & 4.51E-05 & 8.70E-06 & 3.20E-05  & 4.60E-01\cr
2520(a) &0.454&1.08E-02  &            0.15 & 0.35 & 0.60  & 2.65E-05 & 8.53E-06 & 3.68E-05 & 4.60E-01\cr
2520(a) &0.456& 1.08E-02 &            -0.11 & 0.26 & 0.65 & 1.47E-05 & 7.00E-06 & 4.28E-05 & 4.60E-01\cr
2520(a) &0.458& 1.08E-02 &            -0.33 & 0.15 & 0.66 &  8.73E-06 & 5.47E-06 & 4.44E-05 & 4.60E-01\cr
2520(a) & 0.46& 1.08E-02 &            -0.72 & -0.17 & 0.46 & 3.57E-06 & 2.61E-06 & 2.77E-05 & 4.61E-01\cr
2520(a) & 0.47&1.08E-02 &            -2.24 & -2.32 & -0.51 & 1.07E-07 & 1.84E-08 & 2.98E-06& 4.61E-01\cr
2520(a) & 0.48&1.08E-02  &            -3.41 & -4.13 & -3.37 & 7.32E-09 & 2.84E-10 & 4.09E-09& 4.62E-01\cr
2520(a) & 0.49&1.08E-02  &            -3.67 & -4.51 & -3.67 & 4.01E-09 & 1.20E-10 & 2.07E-09& 4.64E-01\cr
\enddata

\tablecomments{Each column shows model name, $Y_{\rm e}^{\rm min}$, $\Delta M$,
[Sr/Fe], [Y/Fe], [Zr/Fe] and ejected masses of Sr, Y, Zr and $^{56}$Ni
for the models (a) in the Figures 10 - 13. In the models
(a), $\Delta  M$ is set to obtain [Zn/Fe]$\simeq$0.5 for $Y_{\rm e}^{\rm min} \lsim$0.46.
For $Y_{\rm e}^{\rm min} \gsim$0.47 (model-2520), the same value of $\Delta M$ 
with the $Y_{\rm e}^{\rm min} \lsim $0.46 models is adopted
because [Zn/Fe] is lower than 0.5 even if all the matter below $M_{\rm cut}$ is ejected.
}
\end{deluxetable}

\clearpage

\begin{deluxetable}{llccccccccc}
\tabletypesize{\scriptsize}
\tablecaption{
\label{tbl-1}}
\tablewidth{0pt}
\tablehead{
\colhead{model} &
\colhead{$Y_{\rm e}^{\rm min}$} & \colhead{$f \times \Delta M$}&  
\colhead{$f$} & \colhead{$M_{\rm mix}$(out)}
& \colhead{$M$(Sr)} & \colhead{$M$(Y)} & \colhead{$M$(Zr)} & \colhead{$M$($^{56}$Ni)} & \colhead{star}}
\startdata
2520-1 & 0.452 & 2.70E-04 &  0.05 & 5.76 & 1.13E-06 & 2.17E-07 & 8.00E-07&2.30E-02  &BS16477-003\cr
2520-2 & 0.45 &5.90E-04  & 0.05 & 5.76 &3.61E-06 & 3.52E-07 & 1.43E-06 &2.30E-02       &CS22897-008\cr
2520-3 & 0.45 & 1.08E-03  & 0.20 & 3.60&6.56E-06 & 6.40E-07 & 2.60E-06 &9.19E-02       &CS22873-166\cr
2520-4 & 0.45 & 1.35E-03 & 0.25 & 3.60 &8.19E-06&8.00E-07&3.24E-06& 1.15E-01         &CS29518-051\cr
\enddata
\tablecomments{The numbers shown are model, $Y_{\rm e}^{\rm min}$, $f \times \Delta M$,
$f$, $M_{\rm mix}$(out), $M$(Sr), $M$(Y), $M$(Zr), $M$($^{56}$Ni) and star names
for the models in Figure 14. The masses are in the units of $M_\odot$.
In our definition $f \times \Delta M$ is the actual ejected mass coming from
the below the mass-cut.
}
\end{deluxetable}

\clearpage

\begin{deluxetable}{llccccccccc}
\tabletypesize{\scriptsize}
\tablecaption{Yields (in $M_\odot$) for the model 2520-1 
\label{tbl-1}}
\tablewidth{0pt}
\tablehead{
\colhead{Isotope} 
& \colhead{mass} 
& \colhead{Isotope} 
& \colhead{mass}
& \colhead{Isotope}
& \colhead{mass}
& \colhead{Isotope}
& \colhead{mass}
}
\startdata
 $^{ 12}$C & 1.755E-01 &$^{ 46}$Sc& 4.216E-13 &$^{ 73}$Ge& 3.551E-10 & $^{ 92}$Mo& 4.850E-09 \cr
 $^{ 13}$C & 5.045E-08 &$^{ 44}$Ti& 1.502E-05 &$^{ 74}$Ge& 8.559E-09 & $^{ 93}$Mo& 9.053E-11 \cr
 $^{ 14}$C & 1.369E-10 &$^{ 46}$Ti& 2.482E-07 &$^{ 76}$Ge& 2.588E-12 & $^{ 94}$Mo& 2.144E-10 \cr
 $^{ 14}$N & 6.300E-04 &$^{ 47}$Ti& 9.284E-07 &$^{ 73}$As& 4.569E-09 & $^{ 95}$Mo& 5.496E-13 \cr
 $^{ 15}$N & 4.046E-07 &$^{ 48}$Ti& 4.214E-05 &$^{ 75}$As& 3.377E-09 & $^{ 96}$Mo& 9.662E-13 \cr
 $^{ 16}$O & 4.086E-01 &$^{ 49}$Ti& 3.174E-10 &$^{ 74}$Se& 3.920E-08 & $^{ 97}$Mo& 2.373E-13 \cr
 $^{ 17}$O & 1.478E-06 &$^{ 50}$Ti& 2.678E-07 &$^{ 75}$Se& 1.748E-09 & $^{ 98}$Mo& 9.873E-14 \cr
 $^{ 18}$O & 5.506E-07 &$^{ 49}$V & 4.535E-07 &$^{ 76}$Se& 1.596E-07 & $^{100}$Mo& 6.059E-15 \cr
 $^{ 19}$F & 8.379E-08 &$^{ 50}$V & 1.184E-11 &$^{ 77}$Se& 1.744E-09 & $^{ 95}$Tc& 8.480E-11 \cr
 $^{ 20}$Ne& 6.517E-02 &$^{ 51}$V & 2.164E-06 &$^{ 78}$Se& 4.886E-08 & $^{ 97}$Tc& 5.534E-11 \cr
 $^{ 21}$Ne& 1.754E-06 &$^{ 50}$Cr& 4.088E-07 &$^{ 79}$Se& 7.851E-11 & $^{ 98}$Tc& 1.019E-13 \cr
 $^{ 22}$Ne& 7.150E-07 &$^{ 52}$Cr& 4.455E-04 &$^{ 80}$Se& 6.644E-10 & $^{ 99}$Tc& 7.414E-14 \cr
 $^{ 22}$Na& 1.287E-06 &$^{ 53}$Cr& 9.406E-08 &$^{ 82}$Se& 2.263E-13 & $^{ 96}$Ru& 2.458E-10 \cr
 $^{ 23}$Na& 1.496E-05 &$^{ 54}$Cr& 1.555E-06 &$^{ 79}$Br& 4.948E-09 & $^{ 98}$Ru& 1.099E-10 \cr
 $^{ 24}$Mg& 3.295E-02 &$^{ 53}$Mn& 1.134E-05 &$^{ 81}$Br& 2.286E-09 & $^{ 99}$Ru& 5.395E-11 \cr
 $^{ 25}$Mg& 8.304E-05 &$^{ 54}$Mn& 1.981E-09 &$^{ 78}$Kr& 9.330E-09 & $^{100}$Ru& 3.006E-10 \cr
 $^{ 26}$Mg& 5.831E-06 &$^{ 55}$Mn& 3.709E-07 &$^{ 80}$Kr& 2.412E-08 & $^{101}$Ru& 9.298E-14 \cr
 $^{ 26}$Al& 2.346E-06 &$^{ 54}$Fe& 4.782E-06 &$^{ 81}$Kr& 5.302E-10 & $^{102}$Ru& 5.723E-14 \cr
 $^{ 27}$Al& 2.605E-04 &$^{ 55}$Fe& 5.501E-06 &$^{ 82}$Kr& 3.516E-08 & $^{103}$Ru& 1.393E-13 \cr
 $^{ 28}$Si& 4.050E-02 &$^{ 56}$Fe& 5.754E-06 &$^{ 83}$Kr& 5.357E-10 & $^{104}$Ru& 2.455E-14 \cr
 $^{ 29}$Si& 9.614E-05 &$^{ 57}$Fe& 2.071E-07 &$^{ 84}$Kr& 1.470E-08 & $^{106}$Ru& 1.458E-15 \cr
 $^{ 30}$Si& 3.956E-05 &$^{ 58}$Fe& 5.094E-06 &$^{ 85}$Kr& 3.664E-11 & $^{101}$Rh& 4.509E-11 \cr
 $^{ 32}$Si& 2.587E-13 &$^{ 59}$Fe& 8.189E-10 &$^{ 86}$Kr& 2.902E-10 & $^{102}$Rh& 2.008E-14 \cr
 $^{ 31}$P & 1.844E-05 &$^{ 60}$Fe& 7.672E-10 &$^{ 83}$Rb& 1.383E-09 & $^{103}$Rh& 9.428E-12 \cr
 $^{ 32}$S & 1.471E-02 &$^{ 56}$Co& 4.064E-08 &$^{ 84}$Rb& 1.082E-10 \cr
 $^{ 33}$S & 1.725E-05 &$^{ 57}$Co& 3.451E-04 &$^{ 85}$Rb& 3.471E-09 \cr
 $^{ 34}$S & 3.206E-06 &$^{ 58}$Co& 2.239E-09 &$^{ 87}$Rb& 8.623E-09 \cr
 $^{ 35}$S & 4.812E-11 &$^{ 59}$Co& 1.883E-07 &$^{ 84}$Sr& 1.239E-09 \cr
 $^{ 36}$S & 3.057E-12 &$^{ 60}$Co& 1.308E-09 &$^{ 85}$Sr& 1.931E-10 \cr
 $^{ 35}$Cl& 1.906E-06 &$^{ 56}$Ni& 2.297E-02 &$^{ 86}$Sr& 3.750E-09 \cr
 $^{ 36}$Cl& 1.244E-09 &$^{ 58}$Ni& 1.897E-04 &$^{ 87}$Sr& 1.490E-09 \cr
 $^{ 37}$Cl& 1.131E-09 &$^{ 59}$Ni& 5.428E-05 &$^{ 88}$Sr& 1.121E-06 \cr
 $^{ 36}$Ar& 2.644E-03 &$^{ 60}$Ni& 5.189E-04 &$^{ 89}$Sr& 3.053E-11 \cr
 $^{ 37}$Ar& 1.193E-06 &$^{ 61}$Ni& 9.478E-06 &$^{ 90}$Sr& 9.241E-14 \cr
 $^{ 38}$Ar& 6.251E-07 &$^{ 62}$Ni& 4.693E-05 &$^{ 88}$Y & 7.221E-11 \cr
 $^{ 39}$Ar& 1.520E-12 &$^{ 63}$Ni& 2.101E-08 &$^{ 89}$Y & 2.174E-07 \cr
 $^{ 40}$Ar& 7.346E-13 &$^{ 64}$Ni& 2.509E-07 &$^{ 91}$Y & 3.240E-13 \cr
 $^{ 42}$Ar& 3.761E-17 &$^{ 63}$Cu& 2.266E-06 &$^{ 88}$Zr& 2.081E-10 \cr
 $^{ 39}$K & 6.151E-07 &$^{ 65}$Cu& 1.583E-07 &$^{ 90}$Zr& 7.952E-07 \cr
 $^{ 40}$K & 3.358E-11 &$^{ 64}$Zn& 6.511E-05 &$^{ 91}$Zr& 9.083E-10 \cr
 $^{ 41}$K & 1.288E-11 &$^{ 65}$Zn& 2.625E-07 &$^{ 92}$Zr& 4.393E-11 \cr
 $^{ 40}$Ca& 2.605E-03 &$^{ 66}$Zn& 1.272E-05 &$^{ 93}$Zr& 4.954E-14 \cr
 $^{ 41}$Ca& 2.355E-07 &$^{ 67}$Zn& 7.894E-08 &$^{ 94}$Zr& 2.988E-14 \cr
 $^{ 42}$Ca& 1.772E-08 &$^{ 68}$Zn& 3.639E-07 &$^{ 95}$Zr& 7.698E-14 \cr
 $^{ 43}$Ca& 1.459E-08 &$^{ 70}$Zn& 4.206E-11 &$^{ 96}$Zr& 1.295E-14 \cr
 $^{ 44}$Ca& 3.043E-11 &$^{ 69}$Ga& 5.433E-08 &$^{ 91}$Nb& 3.611E-09 \cr
 $^{ 45}$Ca& 3.966E-14 &$^{ 71}$Ga& 4.531E-09 &$^{ 92}$Nb& 1.732E-11 \cr
 $^{ 46}$Ca& 3.965E-12 &$^{ 68}$Ge& 1.131E-06 &$^{ 93}$Nb& 2.300E-12 \cr
 $^{ 48}$Ca& 2.440E-12 &$^{ 70}$Ge& 1.089E-06 &$^{ 94}$Nb& 6.665E-14 \cr
 $^{ 45}$Sc& 1.662E-08 &$^{ 72}$Ge& 4.058E-07 &$^{ 95}$Nb& 2.658E-13 \cr
\enddata
\tablecomments{Nucleosynthesis yields in the ejecta for the 
model 2520-1 (see Table 3 for the parameters of the model) evaluated at the
time about 150 seconds after the explosion. To obtain this table, the isotopes  
with their half-lives less than 30 days (except $^{56}$Ni) 
are radioactively decayed.
}
\end{deluxetable}

\clearpage

\begin{deluxetable}{llccccccccc}
\tabletypesize{\scriptsize}
\tablecaption{Yields (in $M_\odot$) for the model 2520-2 
\label{tbl-1}}
\tablewidth{0pt}
\tablehead{
\colhead{Isotope} 
& \colhead{mass} 
& \colhead{Isotope} 
& \colhead{mass}
& \colhead{Isotope} 
& \colhead{mass}
& \colhead{Isotope} 
& \colhead{mass}
}
\startdata $^{ 12}$C & 1.755E-01 &$^{ 46}$Sc& 9.068E-13 &$^{ 73}$Ge& 1.385E-09 & $^{ 92}$Mo& 1.046E-08 \cr
 $^{ 13}$C & 5.045E-08 &$^{ 44}$Ti& 1.508E-05 &$^{ 74}$Ge& 6.408E-08 & $^{ 93}$Mo& 1.500E-10 \cr
 $^{ 14}$C & 1.369E-10 &$^{ 46}$Ti& 2.600E-07 &$^{ 76}$Ge& 9.684E-11 & $^{ 94}$Mo& 2.518E-10 \cr
 $^{ 14}$N & 6.300E-04 &$^{ 47}$Ti& 9.410E-07 &$^{ 73}$As& 8.465E-09 & $^{ 95}$Mo& 6.159E-13 \cr
 $^{ 15}$N & 4.046E-07 &$^{ 48}$Ti& 4.224E-05 &$^{ 75}$As& 1.002E-08 & $^{ 96}$Mo& 9.878E-13 \cr
 $^{ 16}$O & 4.086E-01 &$^{ 49}$Ti& 1.117E-09 &$^{ 74}$Se& 8.518E-08 & $^{ 97}$Mo& 2.383E-13 \cr
 $^{ 17}$O & 1.478E-06 &$^{ 50}$Ti& 1.168E-06 &$^{ 75}$Se& 3.689E-09 & $^{ 98}$Mo& 9.912E-14 \cr
 $^{ 18}$O & 5.506E-07 &$^{ 49}$V & 4.614E-07 &$^{ 76}$Se& 2.947E-07 & $^{100}$Mo& 6.227E-15 \cr
 $^{ 19}$F & 8.379E-08 &$^{ 50}$V & 2.573E-11 &$^{ 77}$Se& 4.212E-09 & $^{ 95}$Tc& 8.516E-11 \cr
 $^{ 20}$Ne& 6.517E-02 &$^{ 51}$V & 2.287E-06 &$^{ 78}$Se& 1.252E-07 & $^{ 97}$Tc& 5.567E-11 \cr
 $^{ 21}$Ne& 1.754E-06 &$^{ 50}$Cr& 4.174E-07 &$^{ 79}$Se& 4.423E-10 & $^{ 98}$Tc& 1.022E-13 \cr
 $^{ 22}$Ne& 7.150E-07 &$^{ 52}$Cr& 4.468E-04 &$^{ 80}$Se& 7.106E-09 & $^{ 99}$Tc& 7.460E-14 \cr
 $^{ 22}$Na& 1.287E-06 &$^{ 53}$Cr& 2.320E-07 &$^{ 82}$Se& 1.243E-11 & $^{ 96}$Ru& 2.472E-10 \cr
 $^{ 23}$Na& 1.496E-05 &$^{ 54}$Cr& 5.295E-06 &$^{ 79}$Br& 8.292E-09 & $^{ 98}$Ru& 1.105E-10 \cr
 $^{ 24}$Mg& 3.295E-02 &$^{ 53}$Mn& 1.135E-05 &$^{ 81}$Br& 8.530E-09 & $^{ 99}$Ru& 5.432E-11 \cr
 $^{ 25}$Mg& 8.304E-05 &$^{ 54}$Mn& 3.734E-09 &$^{ 78}$Kr& 1.820E-08 & $^{100}$Ru& 3.021E-10 \cr
 $^{ 26}$Mg& 5.831E-06 &$^{ 55}$Mn& 8.530E-07 &$^{ 80}$Kr& 5.220E-08 & $^{101}$Ru& 9.375E-14 \cr
 $^{ 26}$Al& 2.347E-06 &$^{ 54}$Fe& 4.860E-06 &$^{ 81}$Kr& 1.098E-09 & $^{102}$Ru& 5.789E-14 \cr
 $^{ 27}$Al& 2.605E-04 &$^{ 55}$Fe& 5.526E-06 &$^{ 82}$Kr& 6.539E-08 & $^{103}$Ru& 1.403E-13 \cr
 $^{ 28}$Si& 4.050E-02 &$^{ 56}$Fe& 1.081E-05 &$^{ 83}$Kr& 2.574E-09 & $^{104}$Ru& 2.507E-14 \cr
 $^{ 29}$Si& 9.614E-05 &$^{ 57}$Fe& 4.053E-07 &$^{ 84}$Kr& 4.073E-08 & $^{106}$Ru& 2.162E-15 \cr
 $^{ 30}$Si& 3.956E-05 &$^{ 58}$Fe& 1.482E-05 &$^{ 85}$Kr& 2.244E-10 & $^{101}$Rh& 4.534E-11 \cr
 $^{ 32}$Si& 2.758E-13 &$^{ 59}$Fe& 4.922E-09 &$^{ 86}$Kr& 3.441E-09 & $^{102}$Rh& 2.011E-14 \cr
 $^{ 31}$P & 1.844E-05 &$^{ 60}$Fe& 8.019E-09 &$^{ 83}$Rb& 2.995E-09 & $^{103}$Rh& 9.514E-12 \cr
 $^{ 32}$S & 1.471E-02 &$^{ 56}$Co& 4.292E-08 &$^{ 84}$Rb& 2.291E-10 \cr
 $^{ 33}$S & 1.725E-05 &$^{ 57}$Co& 3.476E-04 &$^{ 85}$Rb& 6.494E-09 \cr
 $^{ 34}$S & 3.210E-06 &$^{ 58}$Co& 4.357E-09 &$^{ 87}$Rb& 2.652E-08 \cr
 $^{ 35}$S & 4.816E-11 &$^{ 59}$Co& 3.613E-07 &$^{ 84}$Sr& 2.651E-09 \cr
 $^{ 36}$S & 6.989E-12 &$^{ 60}$Co& 4.607E-09 &$^{ 85}$Sr& 4.205E-10 \cr
 $^{ 35}$Cl& 1.913E-06 &$^{ 56}$Ni& 2.302E-02 &$^{ 86}$Sr& 6.693E-09 \cr
 $^{ 36}$Cl& 1.246E-09 &$^{ 58}$Ni& 2.436E-04 &$^{ 87}$Sr& 2.464E-09 \cr
 $^{ 37}$Cl& 1.132E-09 &$^{ 59}$Ni& 5.601E-05 &$^{ 88}$Sr& 3.593E-06 \cr
 $^{ 36}$Ar& 2.644E-03 &$^{ 60}$Ni& 5.788E-04 &$^{ 89}$Sr& 2.271E-10 \cr
 $^{ 37}$Ar& 1.194E-06 &$^{ 61}$Ni& 1.067E-05 &$^{ 90}$Sr& 1.103E-12 \cr
 $^{ 38}$Ar& 6.295E-07 &$^{ 62}$Ni& 9.771E-05 &$^{ 88}$Y & 1.393E-10 \cr
 $^{ 39}$Ar& 2.135E-12 &$^{ 63}$Ni& 1.066E-07 &$^{ 89}$Y & 3.515E-07 \cr
 $^{ 40}$Ar& 1.633E-12 &$^{ 64}$Ni& 2.265E-06 &$^{ 91}$Y & 1.904E-12 \cr
 $^{ 42}$Ar& 1.188E-16 &$^{ 63}$Cu& 3.344E-06 &$^{ 88}$Zr& 3.741E-10 \cr
 $^{ 39}$K & 6.229E-07 &$^{ 65}$Cu& 5.006E-07 &$^{ 90}$Zr& 1.418E-06 \cr
 $^{ 40}$K & 3.891E-11 &$^{ 64}$Zn& 8.389E-05 &$^{ 91}$Zr& 1.582E-09 \cr
 $^{ 41}$K & 2.182E-11 &$^{ 65}$Zn& 4.592E-07 &$^{ 92}$Zr& 1.178E-10 \cr
 $^{ 40}$Ca& 2.605E-03 &$^{ 66}$Zn& 2.364E-05 &$^{ 93}$Zr& 1.000E-13 \cr
 $^{ 41}$Ca& 2.359E-07 &$^{ 67}$Zn& 1.572E-07 &$^{ 94}$Zr& 3.159E-14 \cr
 $^{ 42}$Ca& 2.588E-08 &$^{ 68}$Zn& 1.622E-06 &$^{ 95}$Zr& 7.723E-14 \cr
 $^{ 43}$Ca& 1.619E-08 &$^{ 70}$Zn& 9.605E-10 &$^{ 96}$Zr& 1.357E-14 \cr
 $^{ 44}$Ca& 6.262E-11 &$^{ 69}$Ga& 1.067E-07 &$^{ 91}$Nb& 7.695E-09 \cr
 $^{ 45}$Ca& 1.711E-13 &$^{ 71}$Ga& 1.207E-08 &$^{ 92}$Nb& 3.452E-11 \cr
 $^{ 46}$Ca& 2.337E-11 &$^{ 68}$Ge& 1.900E-06 &$^{ 93}$Nb& 3.862E-12 \cr
 $^{ 48}$Ca& 1.134E-11 &$^{ 70}$Ge& 2.229E-06 &$^{ 94}$Nb& 6.996E-14 \cr
 $^{ 45}$Sc& 1.798E-08 &$^{ 72}$Ge& 8.084E-07 &$^{ 95}$Nb& 2.675E-13 \cr
\enddata
\end{deluxetable}

\clearpage

\begin{deluxetable}{llccccccccc}
\tabletypesize{\scriptsize}
\tablecaption{Yields (in $M_\odot$) for the model 2520-3 
\label{tbl-1}}
\tablewidth{0pt}
\tablehead{
\colhead{Isotope} 
& \colhead{mass} 
& \colhead{Isotope} 
& \colhead{mass}
& \colhead{Isotope} 
& \colhead{mass}
& \colhead{Isotope} 
& \colhead{mass}
}
\startdata
 $^{ 12}$C & 2.417E-01 &$^{ 46}$Sc& 1.775E-12 &$^{ 73}$Ge& 2.519E-09 & $^{ 92}$Mo& 1.938E-08 \cr
 $^{ 13}$C & 7.670E-08 &$^{ 44}$Ti& 6.014E-05 &$^{ 74}$Ge& 1.165E-07 & $^{ 93}$Mo& 3.594E-10 \cr
 $^{ 14}$C & 3.219E-10 &$^{ 46}$Ti& 1.028E-06 &$^{ 76}$Ge& 1.761E-10 & $^{ 94}$Mo& 8.444E-10 \cr
 $^{ 14}$N & 6.309E-04 &$^{ 47}$Ti& 3.714E-06 &$^{ 73}$As& 1.631E-08 & $^{ 95}$Mo& 2.131E-12 \cr
 $^{ 15}$N & 5.305E-07 &$^{ 48}$Ti& 1.686E-04 &$^{ 75}$As& 1.821E-08 & $^{ 96}$Mo& 3.853E-12 \cr
 $^{ 16}$O & 2.043E+00 &$^{ 49}$Ti& 2.032E-09 &$^{ 74}$Se& 1.557E-07 & $^{ 97}$Mo& 9.491E-13 \cr
 $^{ 17}$O & 1.487E-06 &$^{ 50}$Ti& 2.123E-06 &$^{ 75}$Se& 6.738E-09 & $^{ 98}$Mo& 3.949E-13 \cr
 $^{ 18}$O & 5.541E-07 &$^{ 49}$V & 1.819E-06 &$^{ 76}$Se& 5.390E-07 & $^{100}$Mo& 2.424E-14 \cr
 $^{ 19}$F & 8.399E-08 &$^{ 50}$V & 4.735E-11 &$^{ 77}$Se& 7.826E-09 & $^{ 95}$Tc& 3.392E-10 \cr
 $^{ 20}$Ne& 1.646E-01 &$^{ 51}$V & 8.739E-06 &$^{ 78}$Se& 2.277E-07 & $^{ 97}$Tc& 2.213E-10 \cr
 $^{ 21}$Ne& 8.490E-06 &$^{ 50}$Cr& 1.732E-06 &$^{ 79}$Se& 8.043E-10 & $^{ 98}$Tc& 4.076E-13 \cr
 $^{ 22}$Ne& 1.462E-06 &$^{ 52}$Cr& 1.782E-03 &$^{ 80}$Se& 1.292E-08 & $^{ 99}$Tc& 2.966E-13 \cr
 $^{ 22}$Na& 5.343E-06 &$^{ 53}$Cr& 4.218E-07 &$^{ 82}$Se& 2.261E-11 & $^{ 96}$Ru& 9.833E-10 \cr
 $^{ 23}$Na& 2.018E-04 &$^{ 54}$Cr& 9.627E-06 &$^{ 79}$Br& 1.520E-08 & $^{ 98}$Ru& 4.396E-10 \cr
 $^{ 24}$Mg& 1.477E-01 &$^{ 53}$Mn& 4.536E-05 &$^{ 81}$Br& 1.551E-08 & $^{ 99}$Ru& 2.158E-10 \cr
 $^{ 25}$Mg& 1.102E-04 &$^{ 54}$Mn& 6.796E-09 &$^{ 78}$Kr& 3.737E-08 & $^{100}$Ru& 1.202E-09 \cr
 $^{ 26}$Mg& 2.944E-05 &$^{ 55}$Mn& 1.551E-06 &$^{ 80}$Kr& 9.522E-08 & $^{101}$Ru& 3.719E-13 \cr
 $^{ 26}$Al& 3.404E-06 &$^{ 54}$Fe& 1.958E-05 &$^{ 81}$Kr& 2.033E-09 & $^{102}$Ru& 2.289E-13 \cr
 $^{ 27}$Al& 1.036E-03 &$^{ 55}$Fe& 2.202E-05 &$^{ 82}$Kr& 1.194E-07 & $^{103}$Ru& 5.570E-13 \cr
 $^{ 28}$Si& 2.094E-01 &$^{ 56}$Fe& 1.968E-05 &$^{ 83}$Kr& 4.680E-09 & $^{104}$Ru& 9.821E-14 \cr
 $^{ 29}$Si& 5.847E-04 &$^{ 57}$Fe& 7.369E-07 &$^{ 84}$Kr& 7.406E-08 & $^{106}$Ru& 5.831E-15 \cr
 $^{ 30}$Si& 9.516E-05 &$^{ 58}$Fe& 2.695E-05 &$^{ 85}$Kr& 4.080E-10 & $^{101}$Rh& 1.804E-10 \cr
 $^{ 32}$Si& 3.666E-13 &$^{ 59}$Fe& 8.949E-09 &$^{ 86}$Kr& 6.256E-09 & $^{102}$Rh& 8.031E-14 \cr
 $^{ 31}$P & 6.653E-05 &$^{ 60}$Fe& 1.458E-08 &$^{ 83}$Rb& 5.462E-09 & $^{103}$Rh& 3.771E-11 \cr
 $^{ 32}$S & 7.671E-02 &$^{ 56}$Co& 1.632E-07 &$^{ 84}$Rb& 4.166E-10 \cr
 $^{ 33}$S & 1.964E-04 &$^{ 57}$Co& 1.380E-03 &$^{ 85}$Rb& 1.181E-08 \cr
 $^{ 34}$S & 2.550E-05 &$^{ 58}$Co& 7.945E-09 &$^{ 87}$Rb& 4.823E-08 \cr
 $^{ 35}$S & 5.278E-10 &$^{ 59}$Co& 6.570E-07 &$^{ 84}$Sr& 4.984E-09 \cr
 $^{ 36}$S & 5.323E-11 &$^{ 60}$Co& 8.377E-09 &$^{ 85}$Sr& 7.702E-10 \cr
 $^{ 35}$Cl& 1.503E-05 &$^{ 56}$Ni& 9.188E-02 &$^{ 86}$Sr& 1.225E-08 \cr
 $^{ 36}$Cl& 2.379E-08 &$^{ 58}$Ni& 7.591E-04 &$^{ 87}$Sr& 4.552E-09 \cr
 $^{ 37}$Cl& 2.221E-08 &$^{ 59}$Ni& 2.171E-04 &$^{ 88}$Sr& 6.533E-06 \cr
 $^{ 36}$Ar& 1.234E-02 &$^{ 60}$Ni& 2.070E-03 &$^{ 89}$Sr& 4.130E-10 \cr
 $^{ 37}$Ar& 1.576E-05 &$^{ 61}$Ni& 3.764E-05 &$^{ 90}$Sr& 2.066E-12 \cr
 $^{ 38}$Ar& 7.734E-06 &$^{ 62}$Ni& 1.880E-04 &$^{ 88}$Y & 2.534E-10 \cr
 $^{ 39}$Ar& 2.101E-11 &$^{ 63}$Ni& 1.939E-07 &$^{ 89}$Y & 6.392E-07 \cr
 $^{ 40}$Ar& 3.820E-12 &$^{ 64}$Ni& 4.118E-06 &$^{ 91}$Y & 3.541E-12 \cr
 $^{ 42}$Ar& 5.836E-16 &$^{ 63}$Cu& 8.746E-06 &$^{ 88}$Zr& 8.341E-10 \cr
 $^{ 39}$K & 4.466E-06 &$^{ 65}$Cu& 9.101E-07 &$^{ 90}$Zr& 2.578E-06 \cr
 $^{ 40}$K & 5.160E-10 &$^{ 64}$Zn& 2.602E-04 &$^{ 91}$Zr& 2.876E-09 \cr
 $^{ 41}$K & 1.293E-10 &$^{ 65}$Zn& 1.016E-06 &$^{ 92}$Zr& 2.143E-10 \cr
 $^{ 40}$Ca& 1.075E-02 &$^{ 66}$Zn& 4.431E-05 &$^{ 93}$Zr& 2.681E-13 \cr
 $^{ 41}$Ca& 1.993E-06 &$^{ 67}$Zn& 3.278E-07 &$^{ 94}$Zr& 1.224E-13 \cr
 $^{ 42}$Ca& 1.826E-07 &$^{ 68}$Zn& 2.972E-06 &$^{ 95}$Zr& 3.079E-13 \cr
 $^{ 43}$Ca& 5.851E-08 &$^{ 70}$Zn& 1.747E-09 &$^{ 96}$Zr& 5.192E-14 \cr
 $^{ 44}$Ca& 1.883E-10 &$^{ 69}$Ga& 1.988E-07 &$^{ 91}$Nb& 1.415E-08 \cr
 $^{ 45}$Ca& 3.175E-13 &$^{ 71}$Ga& 2.199E-08 &$^{ 92}$Nb& 6.286E-11 \cr
 $^{ 46}$Ca& 4.255E-11 &$^{ 68}$Ge& 4.524E-06 &$^{ 93}$Nb& 7.268E-12 \cr
 $^{ 48}$Ca& 2.025E-11 &$^{ 70}$Ge& 4.055E-06 &$^{ 94}$Nb& 2.663E-13 \cr
 $^{ 45}$Sc& 7.514E-08 &$^{ 72}$Ge& 1.481E-06 &$^{ 95}$Nb& 1.064E-12 \cr
\enddata
\end{deluxetable}

\clearpage

\begin{deluxetable}{llccccccccc}
\tabletypesize{\scriptsize}
\tablecaption{Yields (in $M_\odot$) for the model 2520-4 
\label{tbl-1}}
\tablewidth{0pt}
\tablehead{
\colhead{Isotope} 
& \colhead{mass} 
& \colhead{Isotope} 
& \colhead{mass}
& \colhead{Isotope} 
& \colhead{mass}
& \colhead{Isotope} 
& \colhead{mass}
}
\startdata
 $^{ 12}$C & 2.417E-01 &$^{ 46}$Sc& 2.184E-12 &$^{ 73}$Ge& 3.149E-09 & $^{ 92}$Mo& 2.423E-08 \cr
 $^{ 13}$C & 7.670E-08 &$^{ 44}$Ti& 7.515E-05 &$^{ 74}$Ge& 1.456E-07 & $^{ 93}$Mo& 4.492E-10 \cr
 $^{ 14}$C & 3.219E-10 &$^{ 46}$Ti& 1.274E-06 &$^{ 76}$Ge& 2.201E-10 & $^{ 94}$Mo& 1.056E-09 \cr
 $^{ 14}$N & 6.309E-04 &$^{ 47}$Ti& 4.642E-06 &$^{ 73}$As& 2.039E-08 & $^{ 95}$Mo& 2.664E-12 \cr
 $^{ 15}$N & 5.306E-07 &$^{ 48}$Ti& 2.107E-04 &$^{ 75}$As& 2.277E-08 & $^{ 96}$Mo& 4.817E-12 \cr
 $^{ 16}$O & 2.046E+00 &$^{ 49}$Ti& 2.540E-09 &$^{ 74}$Se& 1.946E-07 & $^{ 97}$Mo& 1.186E-12 \cr
 $^{ 17}$O & 1.487E-06 &$^{ 50}$Ti& 2.654E-06 &$^{ 75}$Se& 8.422E-09 & $^{ 98}$Mo& 4.936E-13 \cr
 $^{ 18}$O & 5.541E-07 &$^{ 49}$V & 2.272E-06 &$^{ 76}$Se& 6.737E-07 & $^{100}$Mo& 3.030E-14 \cr
 $^{ 19}$F & 8.399E-08 &$^{ 50}$V & 5.902E-11 &$^{ 77}$Se& 9.783E-09 & $^{ 95}$Tc& 4.240E-10 \cr
 $^{ 20}$Ne& 1.646E-01 &$^{ 51}$V & 1.092E-05 &$^{ 78}$Se& 2.846E-07 & $^{ 97}$Tc& 2.767E-10 \cr
 $^{ 21}$Ne& 8.490E-06 &$^{ 50}$Cr& 2.135E-06 &$^{ 79}$Se& 1.005E-09 & $^{ 98}$Tc& 5.095E-13 \cr
 $^{ 22}$Ne& 1.462E-06 &$^{ 52}$Cr& 2.227E-03 &$^{ 80}$Se& 1.615E-08 & $^{ 99}$Tc& 3.707E-13 \cr
 $^{ 22}$Na& 5.343E-06 &$^{ 53}$Cr& 5.273E-07 &$^{ 82}$Se& 2.826E-11 & $^{ 96}$Ru& 1.229E-09 \cr
 $^{ 23}$Na& 2.018E-04 &$^{ 54}$Cr& 1.203E-05 &$^{ 79}$Br& 1.901E-08 & $^{ 98}$Ru& 5.495E-10 \cr
 $^{ 24}$Mg& 1.477E-01 &$^{ 53}$Mn& 5.670E-05 &$^{ 81}$Br& 1.939E-08 & $^{ 99}$Ru& 2.697E-10 \cr
 $^{ 25}$Mg& 1.102E-04 &$^{ 54}$Mn& 8.493E-09 &$^{ 78}$Kr& 4.669E-08 & $^{100}$Ru& 1.503E-09 \cr
 $^{ 26}$Mg& 2.944E-05 &$^{ 55}$Mn& 1.939E-06 &$^{ 80}$Kr& 1.190E-07 & $^{101}$Ru& 4.649E-13 \cr
 $^{ 26}$Al& 3.417E-06 &$^{ 54}$Fe& 2.434E-05 &$^{ 81}$Kr& 2.541E-09 & $^{102}$Ru& 2.861E-13 \cr
 $^{ 27}$Al& 1.036E-03 &$^{ 55}$Fe& 2.752E-05 &$^{ 82}$Kr& 1.493E-07 & $^{103}$Ru& 6.963E-13 \cr
 $^{ 28}$Si& 2.261E-01 &$^{ 56}$Fe& 2.459E-05 &$^{ 83}$Kr& 5.850E-09 & $^{104}$Ru& 1.228E-13 \cr
 $^{ 29}$Si& 5.859E-04 &$^{ 57}$Fe& 9.211E-07 &$^{ 84}$Kr& 9.257E-08 & $^{106}$Ru& 7.289E-15 \cr
 $^{ 30}$Si& 9.565E-05 &$^{ 58}$Fe& 3.369E-05 &$^{ 85}$Kr& 5.100E-10 & $^{101}$Rh& 2.255E-10 \cr
 $^{ 32}$Si& 3.753E-13 &$^{ 59}$Fe& 1.119E-08 &$^{ 86}$Kr& 7.820E-09 & $^{102}$Rh& 1.004E-13 \cr
 $^{ 31}$P & 6.763E-05 &$^{ 60}$Fe& 1.823E-08 &$^{ 83}$Rb& 6.827E-09 & $^{103}$Rh& 4.714E-11 \cr
 $^{ 32}$S & 8.853E-02 &$^{ 56}$Co& 2.038E-07 &$^{ 84}$Rb& 5.207E-10 \cr
 $^{ 33}$S & 1.983E-04 &$^{ 57}$Co& 1.725E-03 &$^{ 85}$Rb& 1.477E-08 \cr
 $^{ 34}$S & 2.648E-05 &$^{ 58}$Co& 9.927E-09 &$^{ 87}$Rb& 6.028E-08 \cr
 $^{ 35}$S & 5.279E-10 &$^{ 59}$Co& 8.212E-07 &$^{ 84}$Sr& 6.221E-09 \cr
 $^{ 36}$S & 5.534E-11 &$^{ 60}$Co& 1.047E-08 &$^{ 85}$Sr& 9.622E-10 \cr
 $^{ 35}$Cl& 1.610E-05 &$^{ 56}$Ni& 1.149E-01 &$^{ 86}$Sr& 1.531E-08 \cr
 $^{ 36}$Cl& 2.384E-08 &$^{ 58}$Ni& 9.488E-04 &$^{ 87}$Sr& 5.690E-09 \cr
 $^{ 37}$Cl& 2.223E-08 &$^{ 59}$Ni& 2.714E-04 &$^{ 88}$Sr& 8.166E-06 \cr
 $^{ 36}$Ar& 1.486E-02 &$^{ 60}$Ni& 2.587E-03 &$^{ 89}$Sr& 5.163E-10 \cr
 $^{ 37}$Ar& 1.626E-05 &$^{ 61}$Ni& 4.706E-05 &$^{ 90}$Sr& 2.583E-12 \cr
 $^{ 38}$Ar& 8.030E-06 &$^{ 62}$Ni& 2.350E-04 &$^{ 88}$Y & 3.168E-10 \cr
 $^{ 39}$Ar& 2.157E-11 &$^{ 63}$Ni& 2.424E-07 &$^{ 89}$Y & 7.990E-07 \cr
 $^{ 40}$Ar& 4.532E-12 &$^{ 64}$Ni& 5.147E-06 &$^{ 91}$Y & 4.427E-12 \cr
 $^{ 42}$Ar& 6.299E-16 &$^{ 63}$Cu& 1.093E-05 &$^{ 88}$Zr& 1.042E-09 \cr
 $^{ 39}$K & 4.955E-06 &$^{ 65}$Cu& 1.138E-06 &$^{ 90}$Zr& 3.222E-06 \cr
 $^{ 40}$K & 5.257E-10 &$^{ 64}$Zn& 3.253E-04 &$^{ 91}$Zr& 3.594E-09 \cr
 $^{ 41}$K & 1.372E-10 &$^{ 65}$Zn& 1.271E-06 &$^{ 92}$Zr& 2.679E-10 \cr
 $^{ 40}$Ca& 1.334E-02 &$^{ 66}$Zn& 5.539E-05 &$^{ 93}$Zr& 3.352E-13 \cr
 $^{ 41}$Ca& 2.163E-06 &$^{ 67}$Zn& 4.097E-07 &$^{ 94}$Zr& 1.527E-13 \cr
 $^{ 42}$Ca& 1.934E-07 &$^{ 68}$Zn& 3.715E-06 &$^{ 95}$Zr& 3.849E-13 \cr
 $^{ 43}$Ca& 7.309E-08 &$^{ 70}$Zn& 2.183E-09 &$^{ 96}$Zr& 6.485E-14 \cr
 $^{ 44}$Ca& 2.157E-10 &$^{ 69}$Ga& 2.485E-07 &$^{ 91}$Nb& 1.768E-08 \cr
 $^{ 45}$Ca& 3.950E-13 &$^{ 71}$Ga& 2.748E-08 &$^{ 92}$Nb& 7.857E-11 \cr
 $^{ 46}$Ca& 5.316E-11 &$^{ 68}$Ge& 5.655E-06 &$^{ 93}$Nb& 9.085E-12 \cr
 $^{ 48}$Ca& 2.474E-11 &$^{ 70}$Ge& 5.068E-06 &$^{ 94}$Nb& 3.329E-13 \cr
 $^{ 45}$Sc& 9.122E-08 &$^{ 72}$Ge& 1.851E-06 &$^{ 95}$Nb& 1.329E-12 \cr
\enddata
\end{deluxetable}

\clearpage

\begin{figure}

\plotone{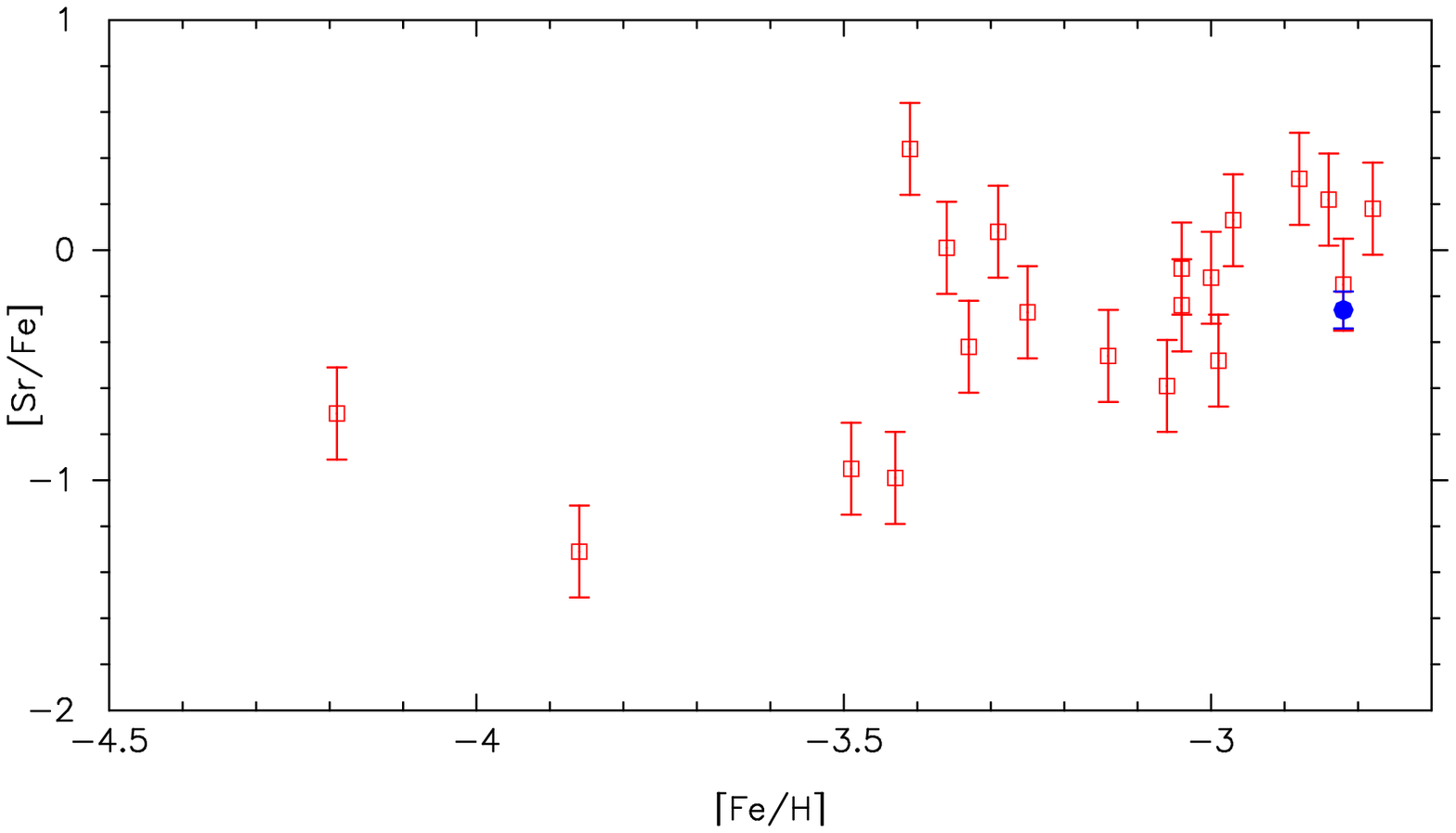}

\plotone{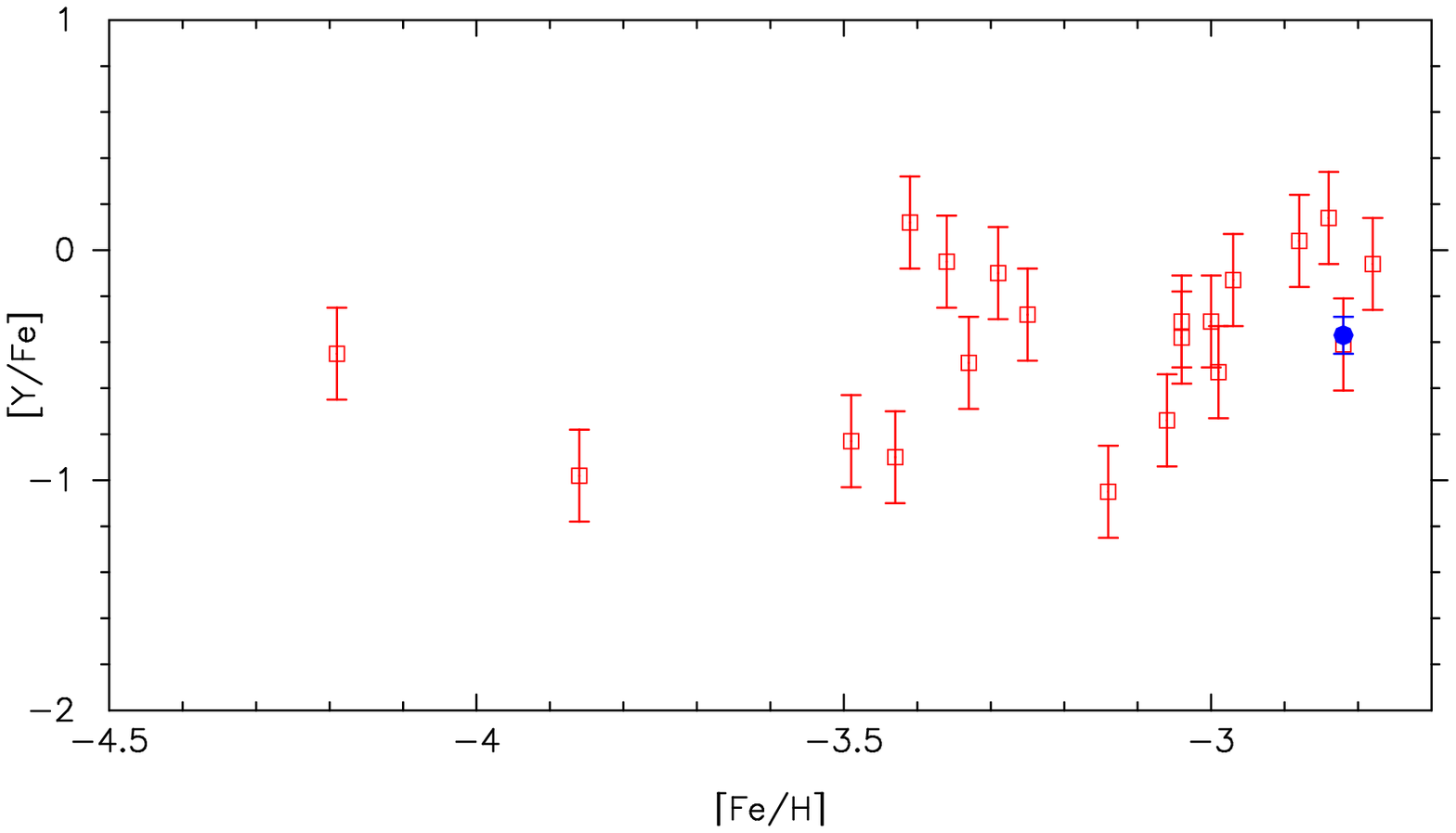}

\plotone{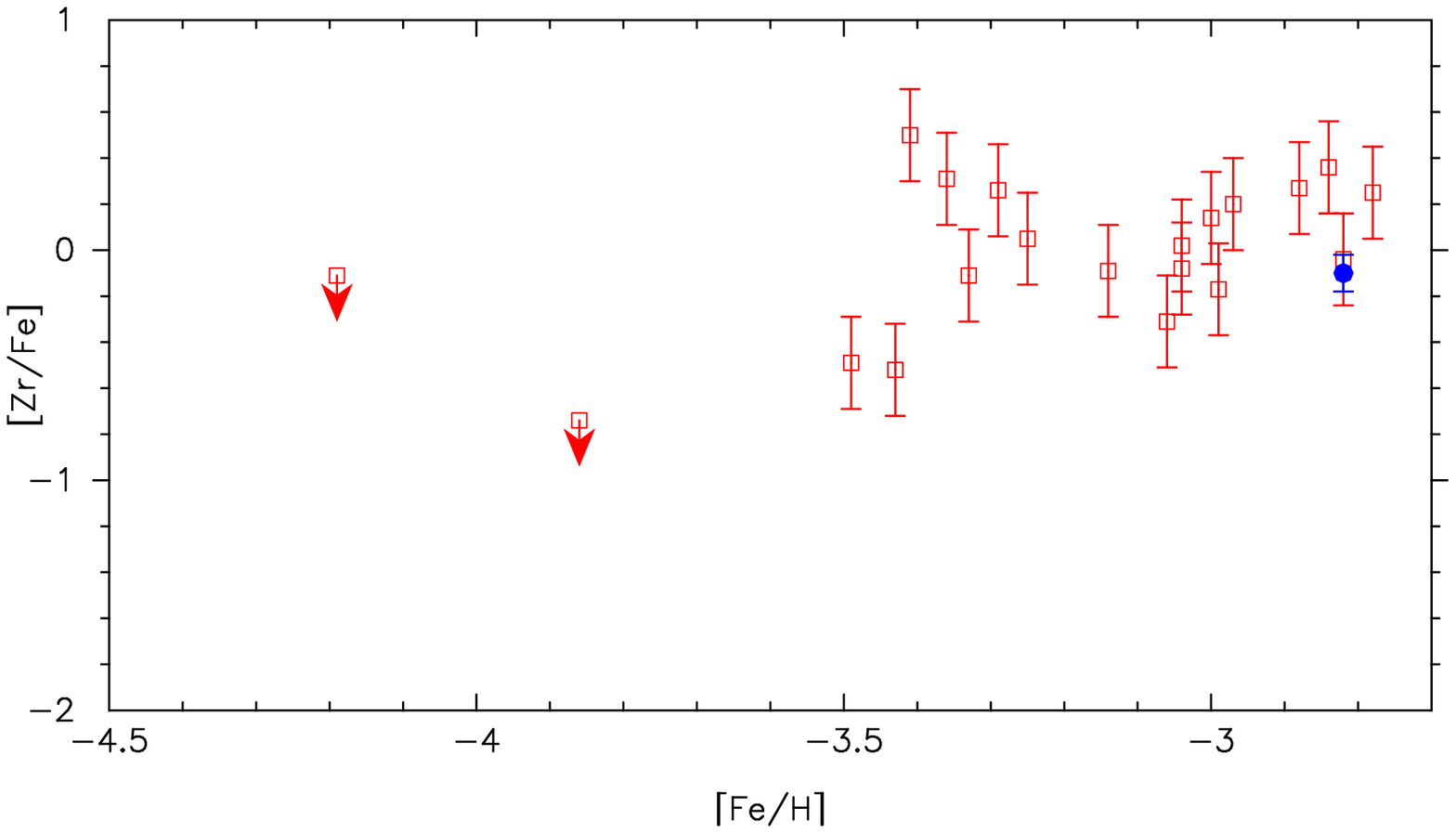}

\caption{[Sr/Fe] vs. [Fe/H], [Y/Fe] vs. [Fe/H], and [Zr/Fe] vs. [Fe/H]
of ``weak r-process stars''.
The solid circle point indicates HD122563 from Honda et al. (2006).
Other data are taken from Franc\c ois et al. (2007).}
\end{figure}

\clearpage

\begin{figure}

\plotone{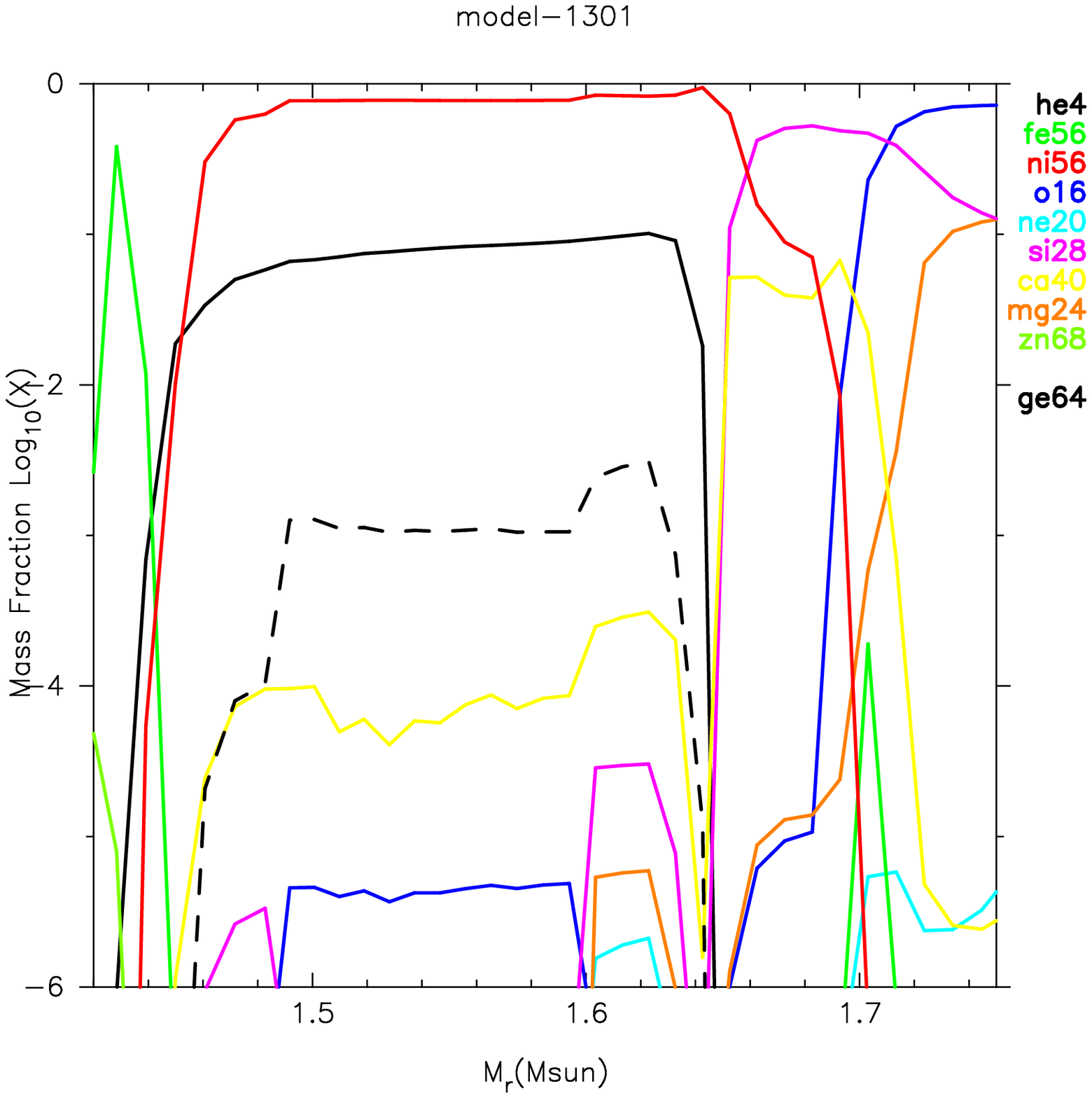}

\plotone{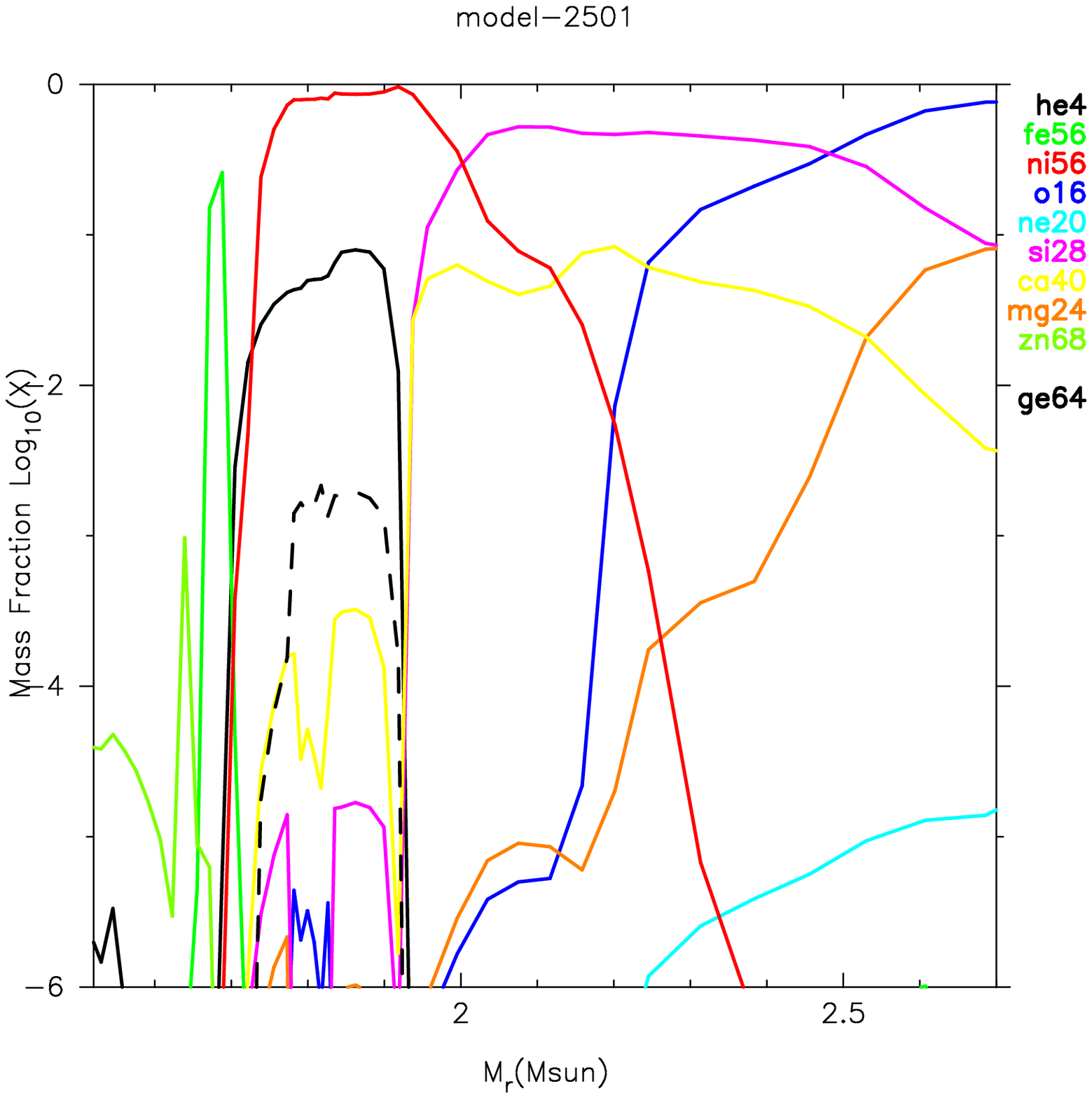}


\end{figure}

\begin{figure}




\plotone{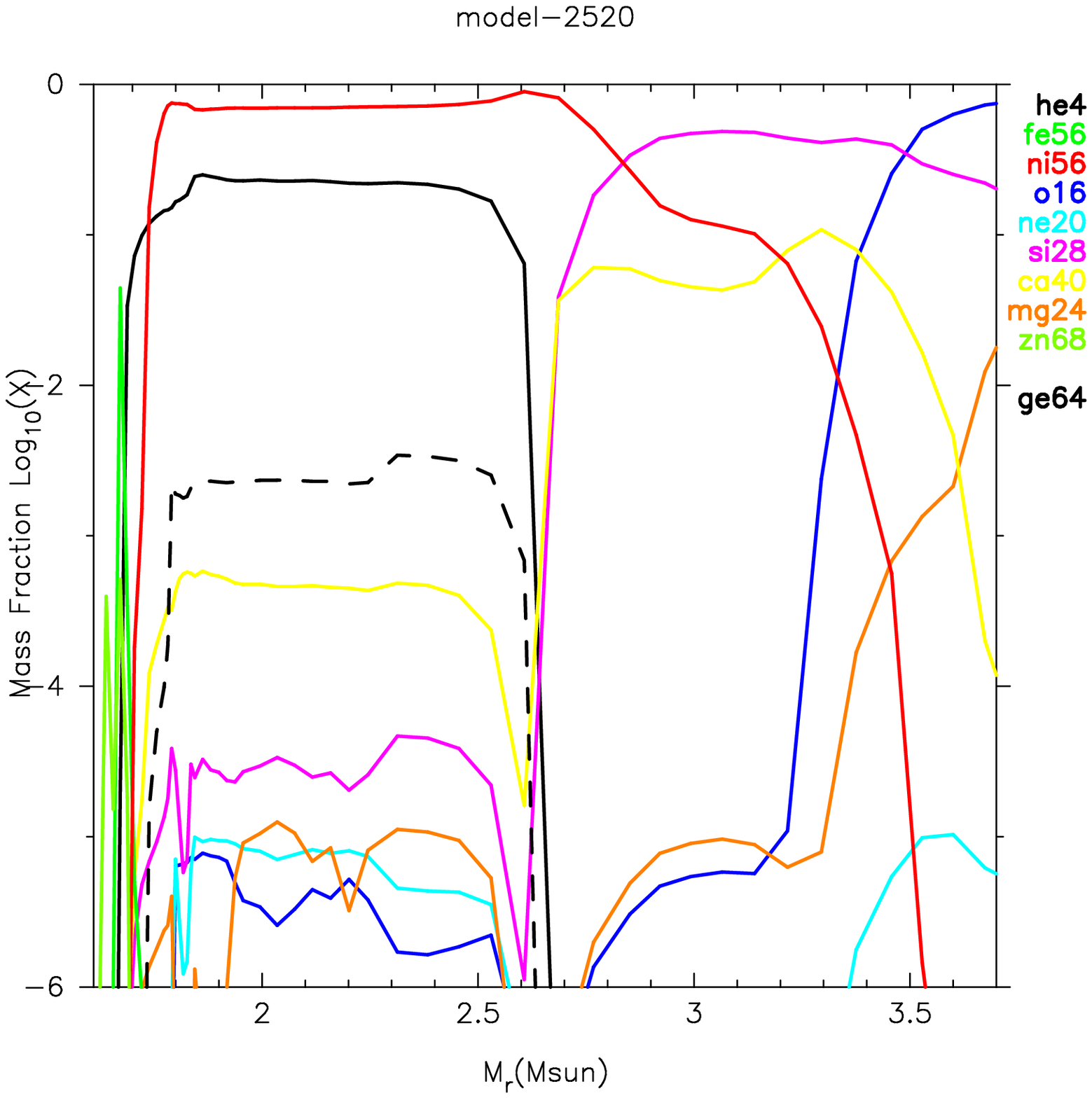}

\caption{Abundance distributions for selected species
after explosion: (a) model-1301, (b) model-2501, (c) model-2520.
The mass fraction $X$ for 
$^{56}$Ni, $^4$He, $^{28}$Si, $^{16}$O, $^{40}$Ca, 
$^{24}$Mg, and $^{64}$Ge are represented by   
solid, dashed, dotted, dash-dotted,
dashed, three dotted, thick solid, and 
thick dashed lines, respectively.
}
\end{figure}

\clearpage

\begin{figure}

\includegraphics[width=80mm]{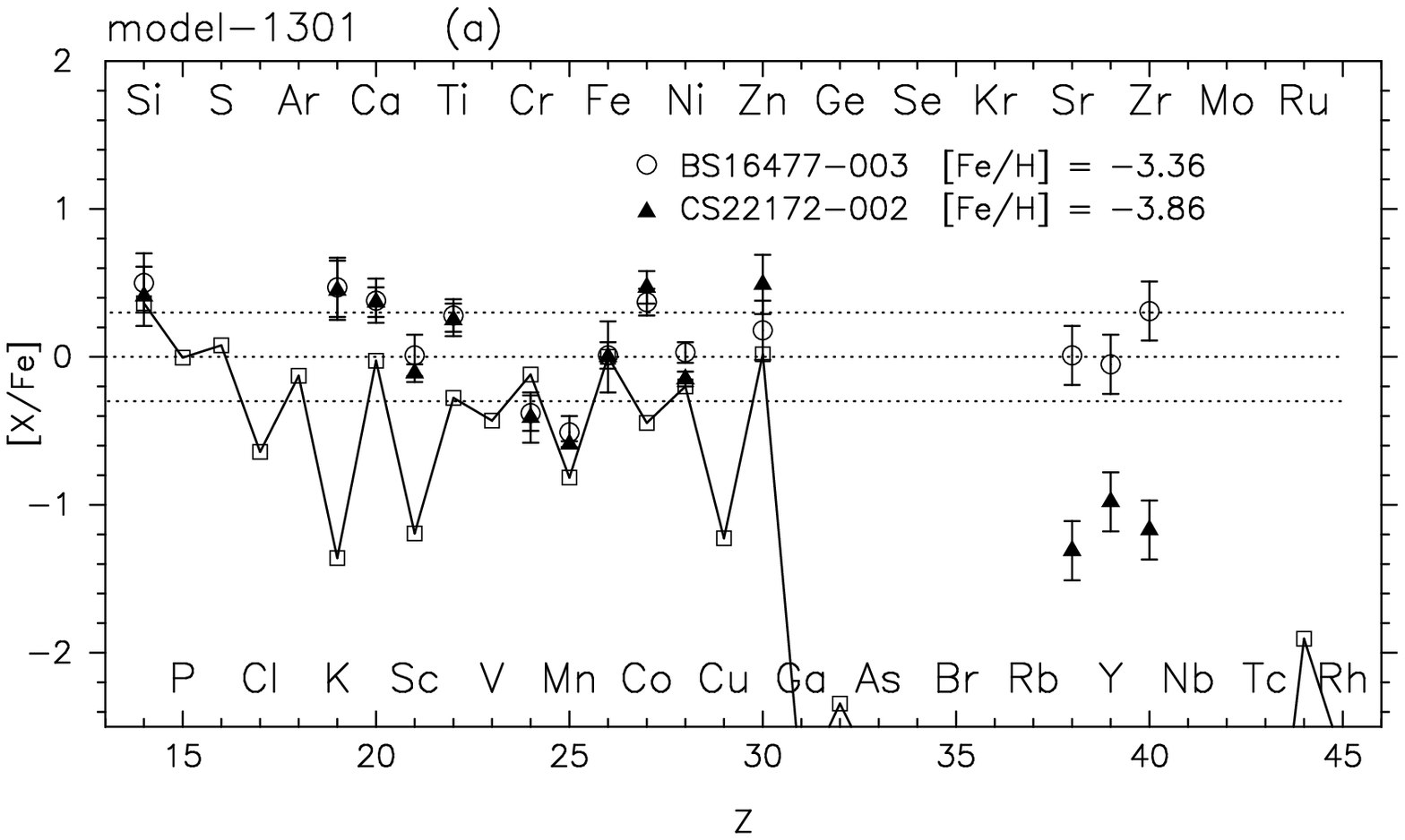}
\includegraphics[width=80mm]{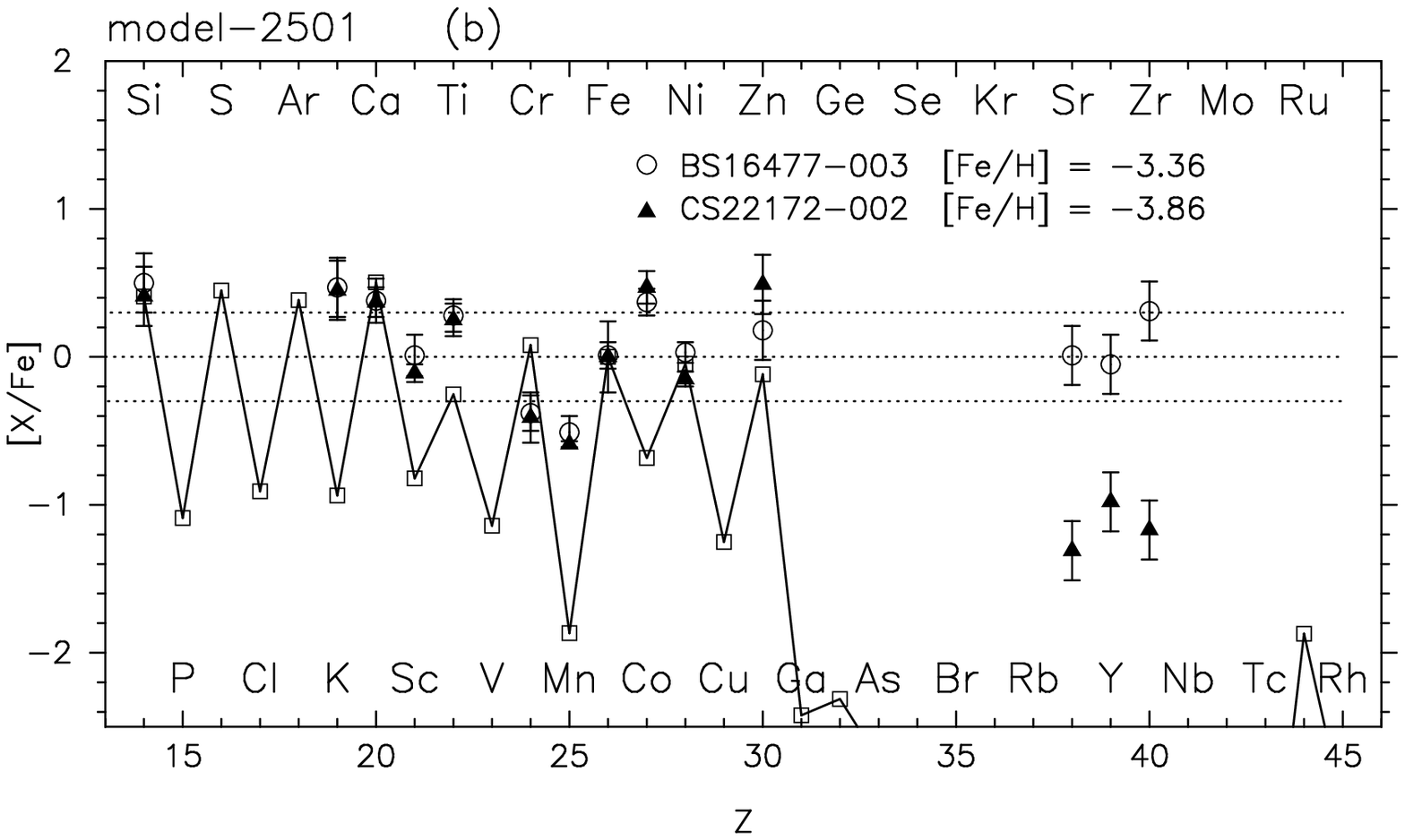}
\includegraphics[width=80mm]{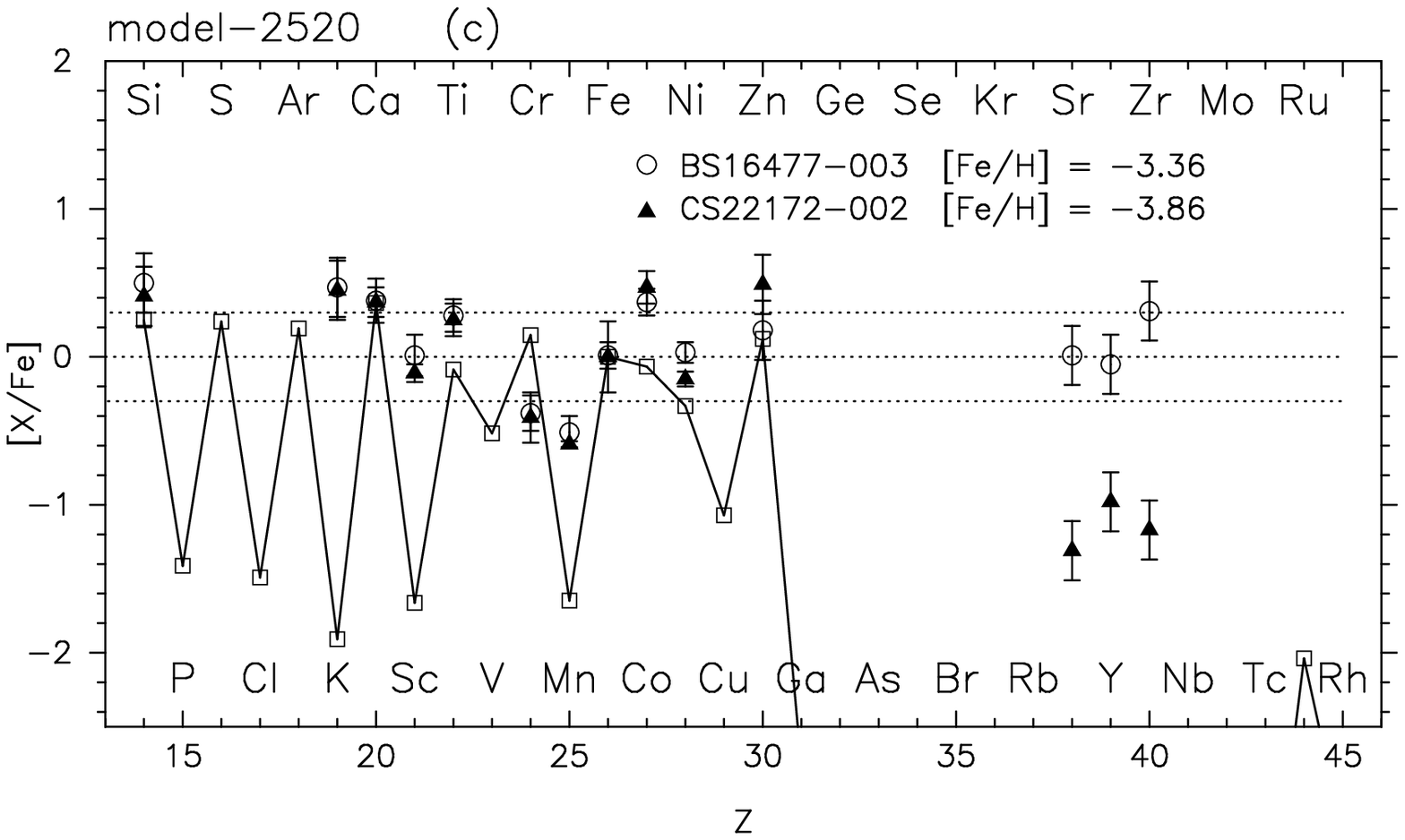}

\caption{Comparisons between our models with conventional mass-cut
and the abundance patterns of weak r-process stars.
Panels (a), (b), and (c) are model-1301, model-2501
and model-2520, respectively.
Mass-cuts are $M_{\rm cut}$=1.59$M_\odot$, 1.76$M_\odot$ and 2.31$M_\odot$ 
for model-1301, model-2501 and model-2520, respectively.
BS16477-003 is the weak r-process star with the highest abundance of Sr, Y and Zr, 
and CS22172-002 is the weak r-process star with the lowest abundance of 
these elements.}
\end{figure}

\begin{figure}

\includegraphics[width=100mm]{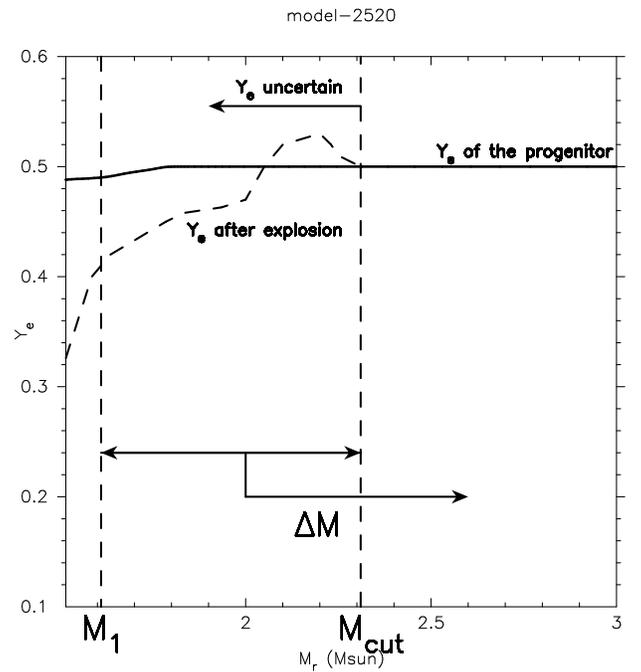}

\caption{Illustration of $M_1$, $M_{\rm cut}$, 
$\Delta M$, and $Y_{\rm e}$ distribution before and
after explosion described in Section 4.1.
}
\end{figure}

\clearpage
\begin{figure}


\includegraphics[width=80mm]{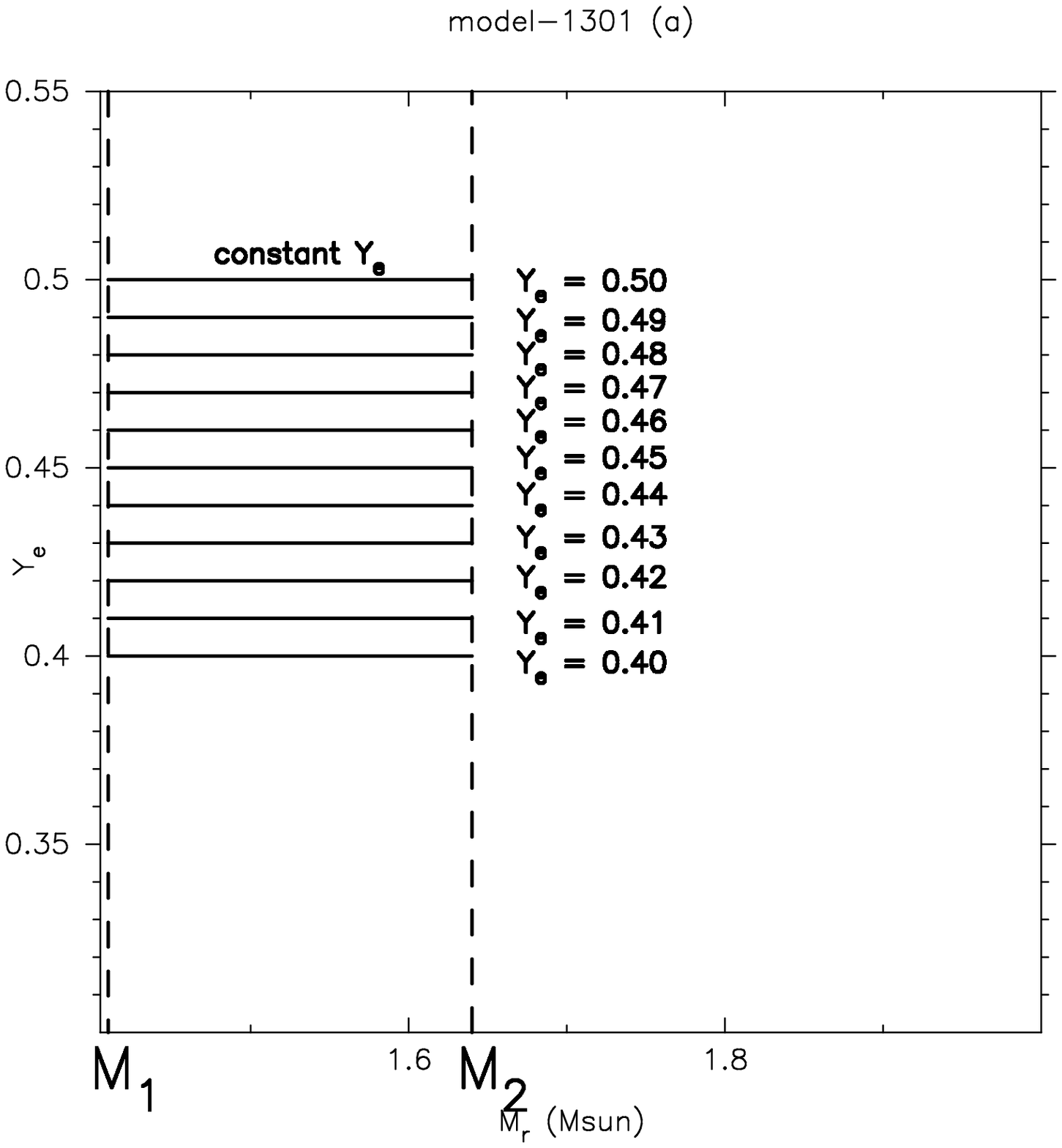}
\includegraphics[width=80mm]{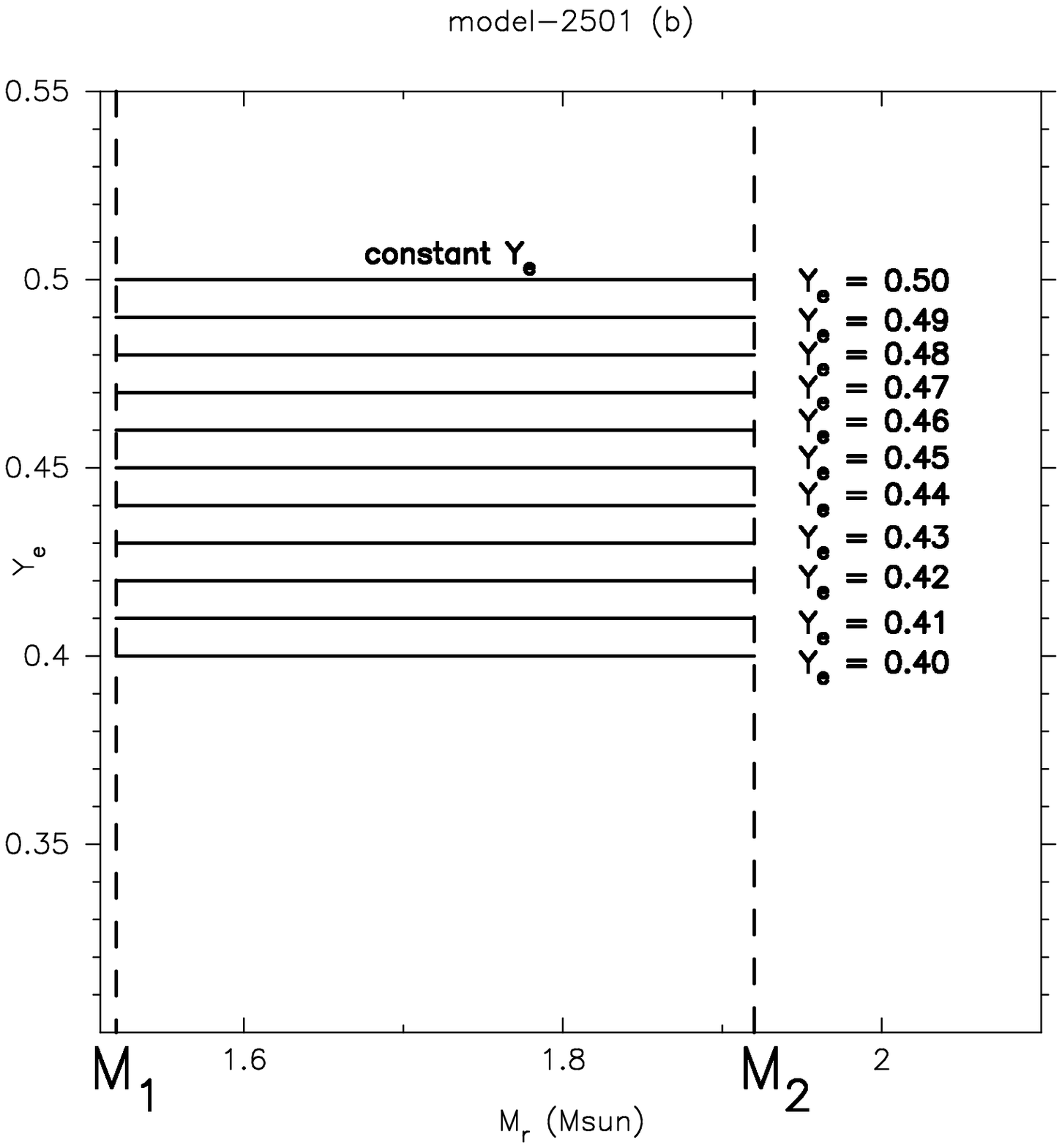}

\end{figure}

\begin{figure}


\includegraphics[width=80mm]{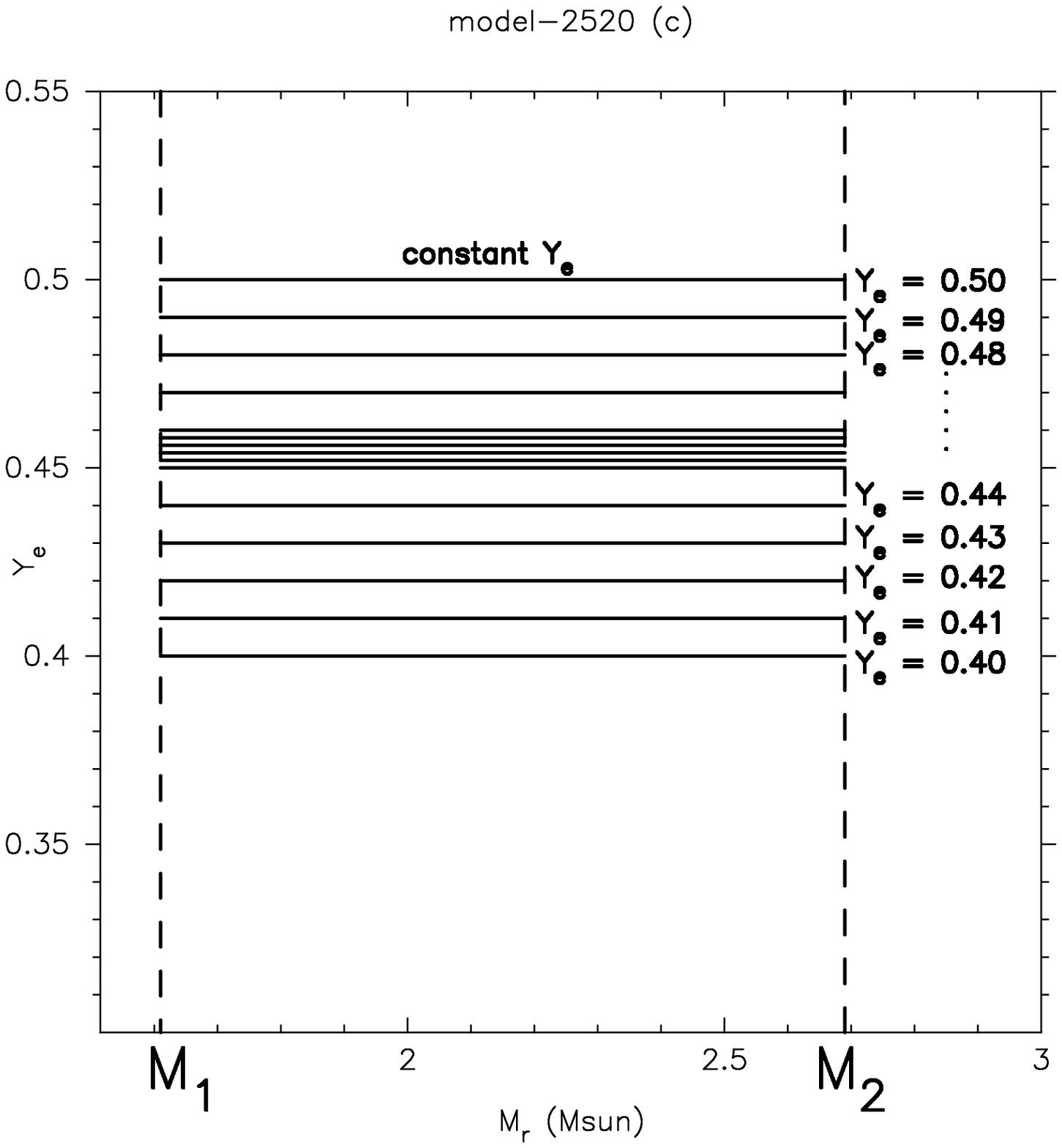}

\caption{Illustrations of $M_1$, $M_2$, and $Y_{\rm e}$ in Section 4.2.
Panels (a), (b) and (c) are model-1301, model-2501 and model-2520, respectively.
$M_1$ and $M_2$ are the inner and outer boundaries of the calculated region,
i.e., the complete Si-burning region, respectively.
We carry out the nucleosynthesis calculation in the region $M_1$$<$$M$$<$$M_2$ 
with constant $Y_{\rm e}$ from 0.40 to 0.50 as shown in solid lines.}
\end{figure}

\clearpage

\begin{figure}

\includegraphics[width=80mm]{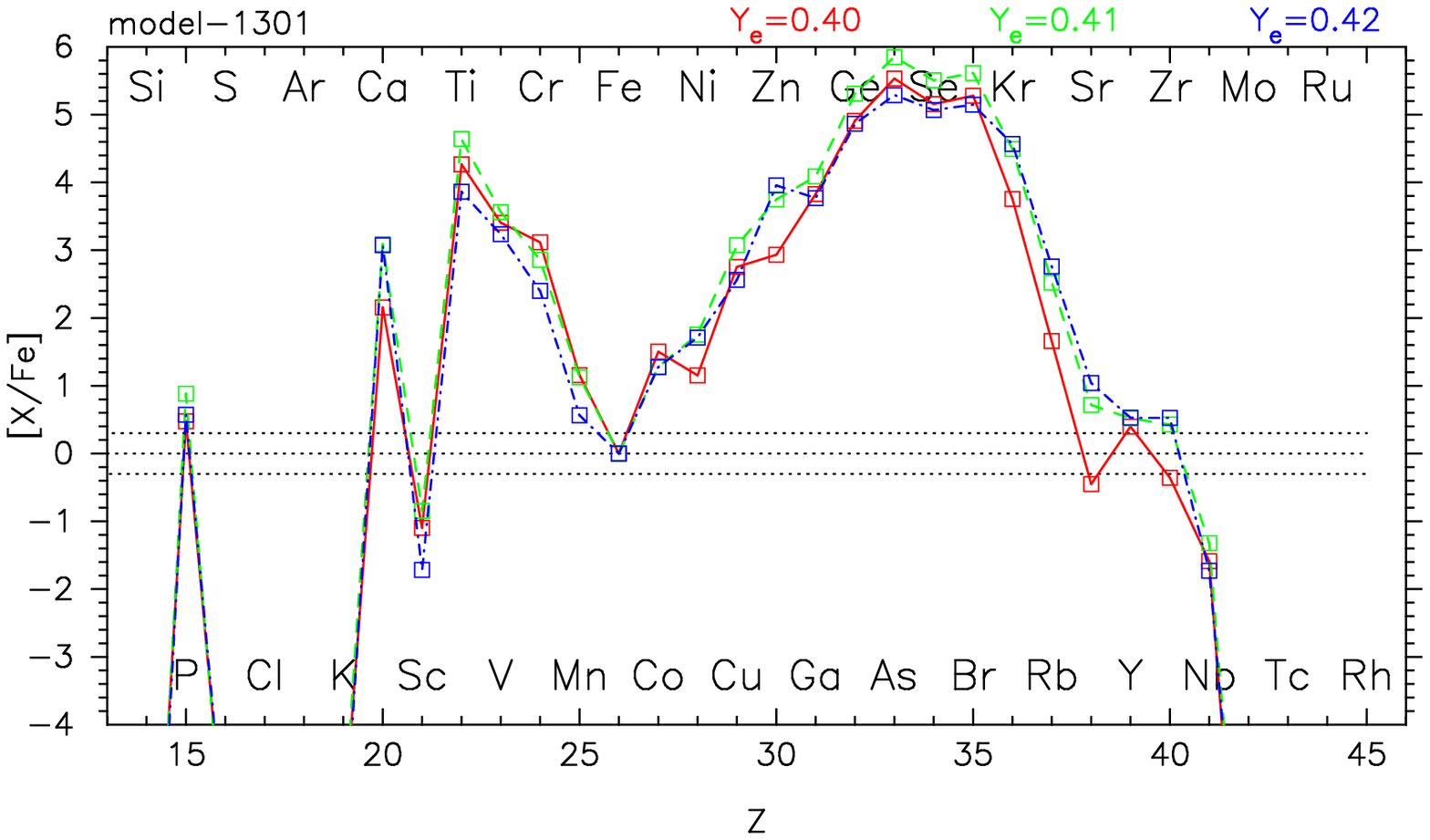}
\includegraphics[width=80mm]{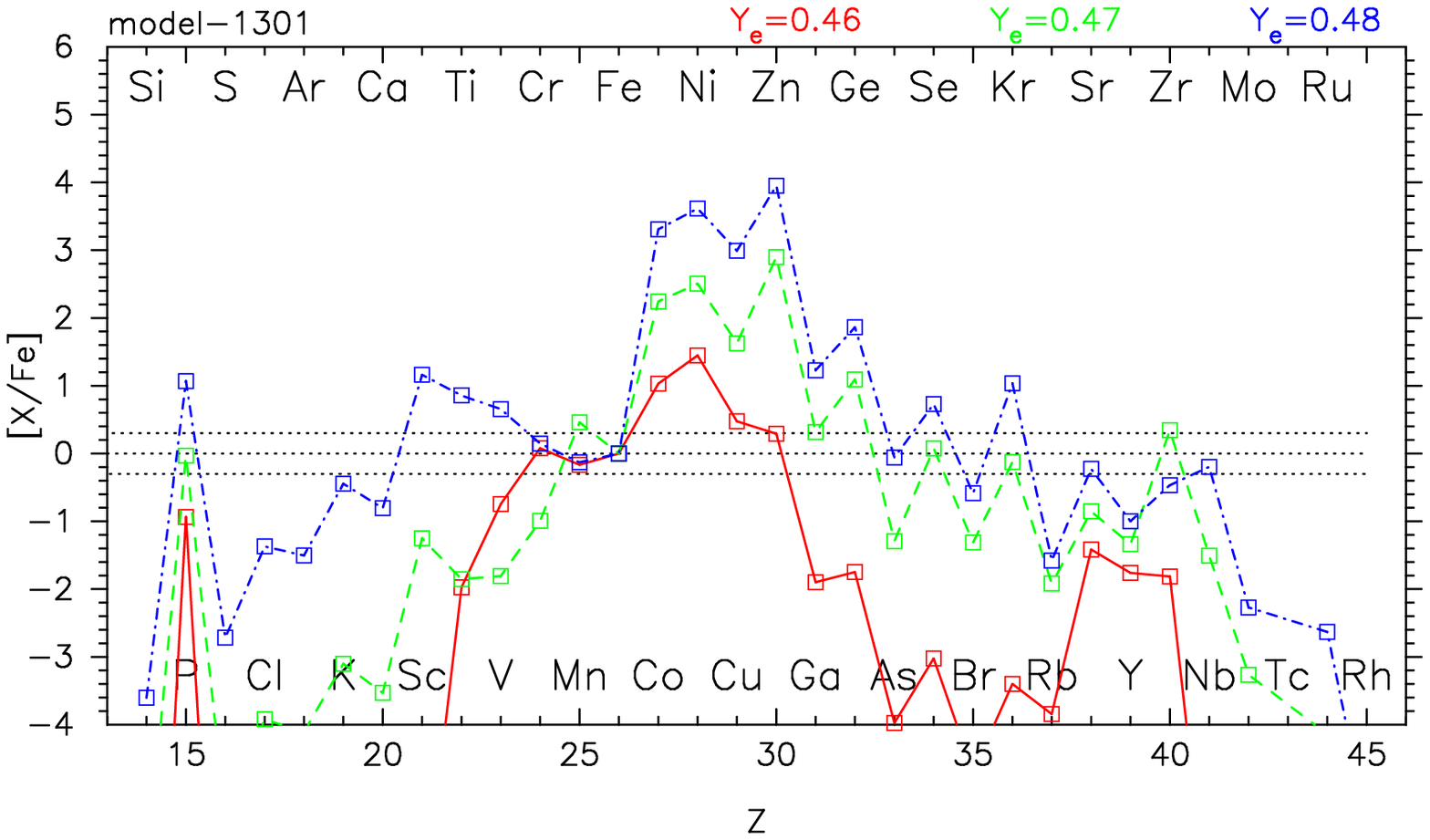}
\includegraphics[width=80mm]{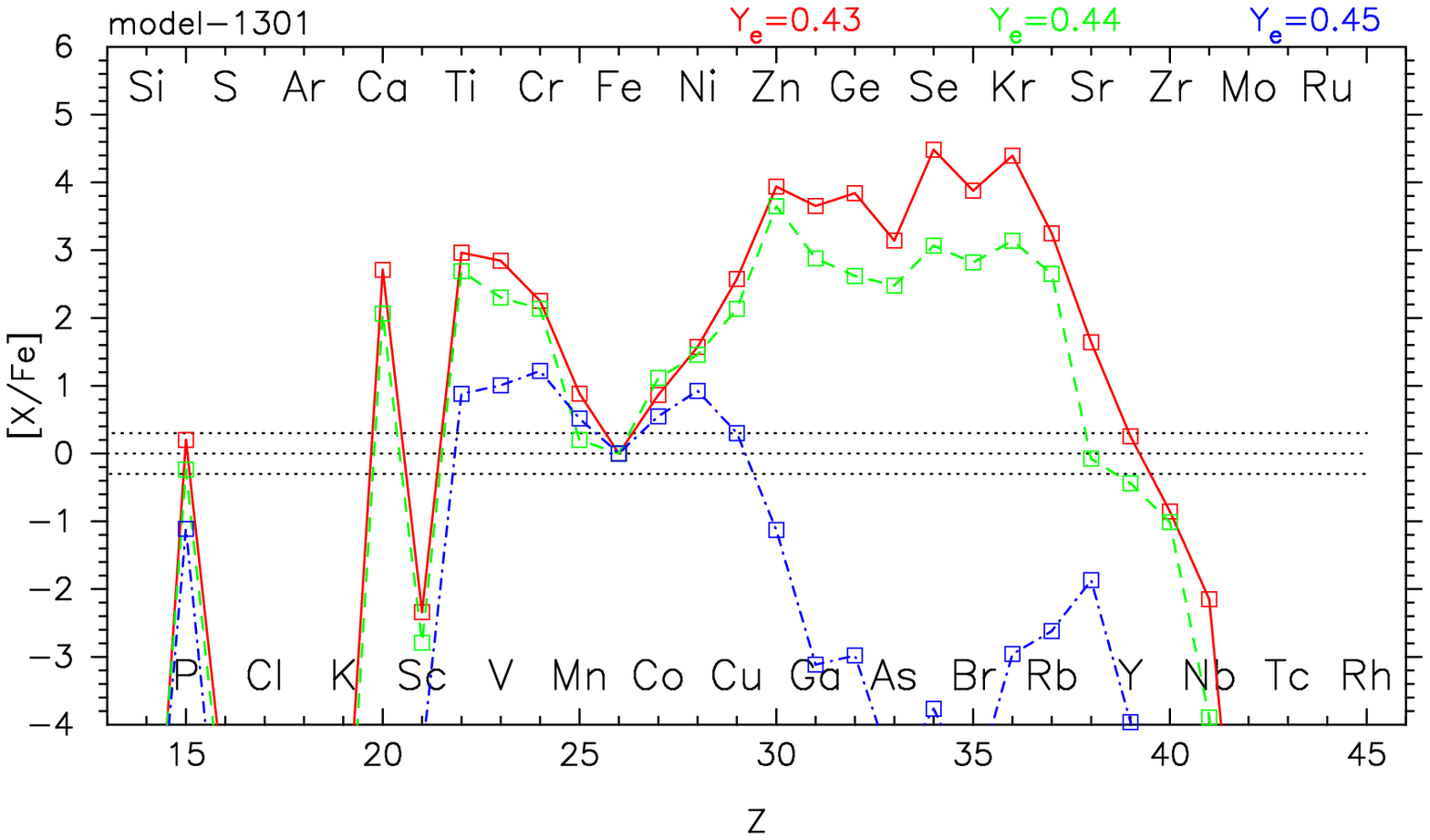}
\includegraphics[width=80mm]{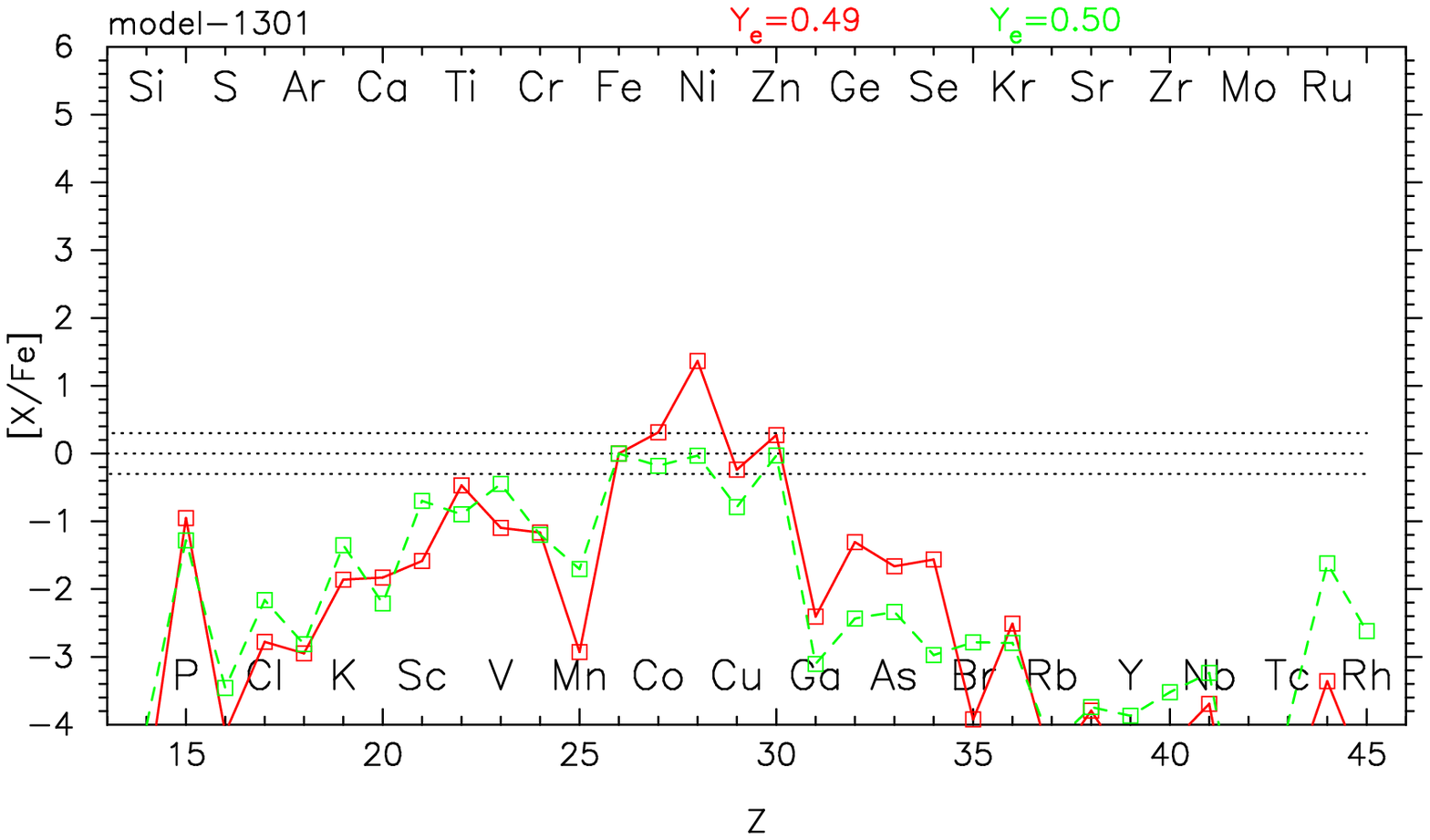}

\caption{Abundance patterns in the Si-burning regions
of model-1301 with each $Y_e$.
The left panels: $Y_e$ = 0.40, 0.41, 0.42 and  $Y_e$ = 0.43, 0.44, 0.45 (from the top).
The right panels: $Y_e$ = 0.46, 0.47, 0.48 and  $Y_e$ = 0.49, 0.50 (from the top).
}
\end{figure}


\begin{figure}

\includegraphics[width=80mm]{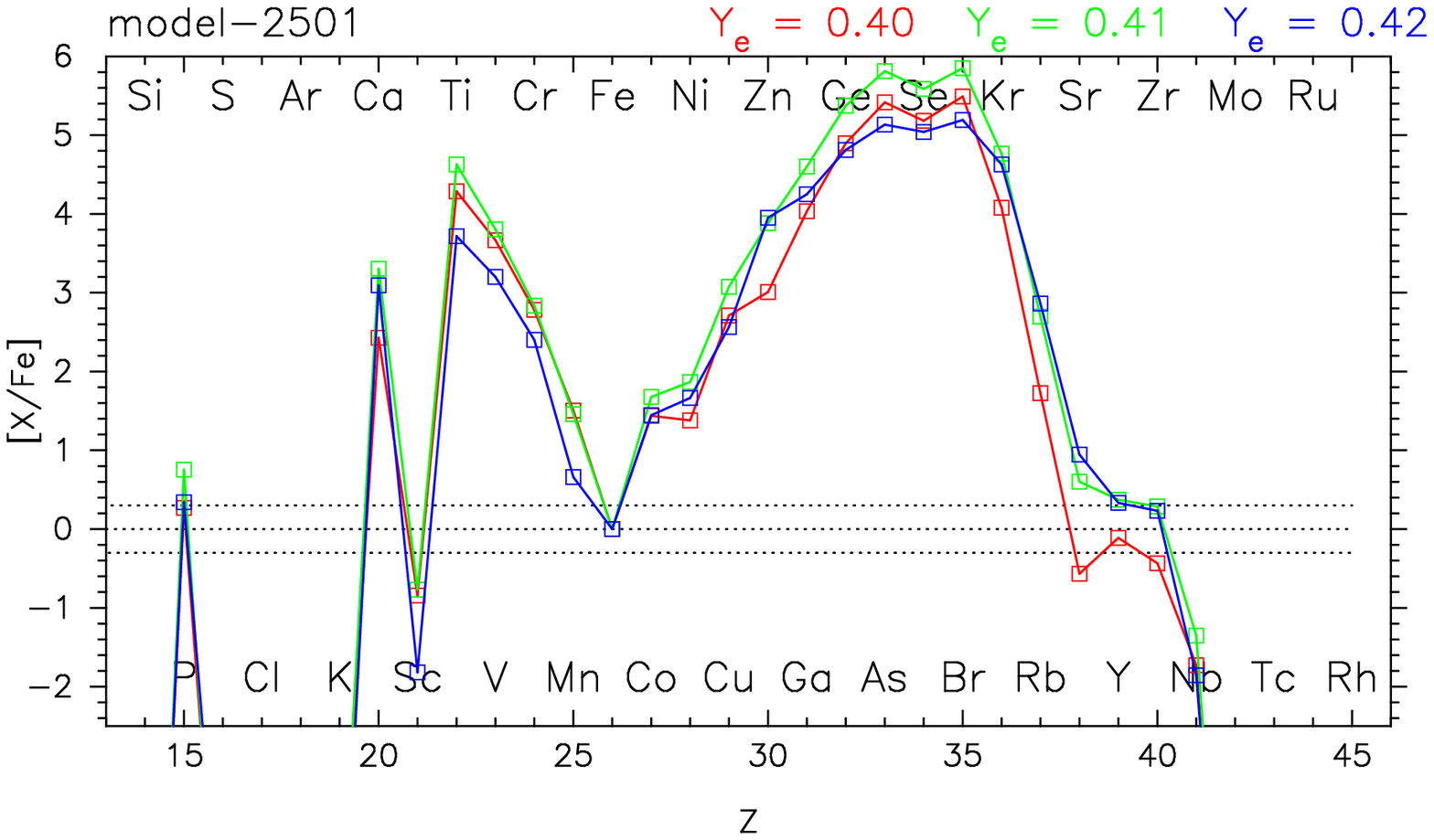}
\includegraphics[width=80mm]{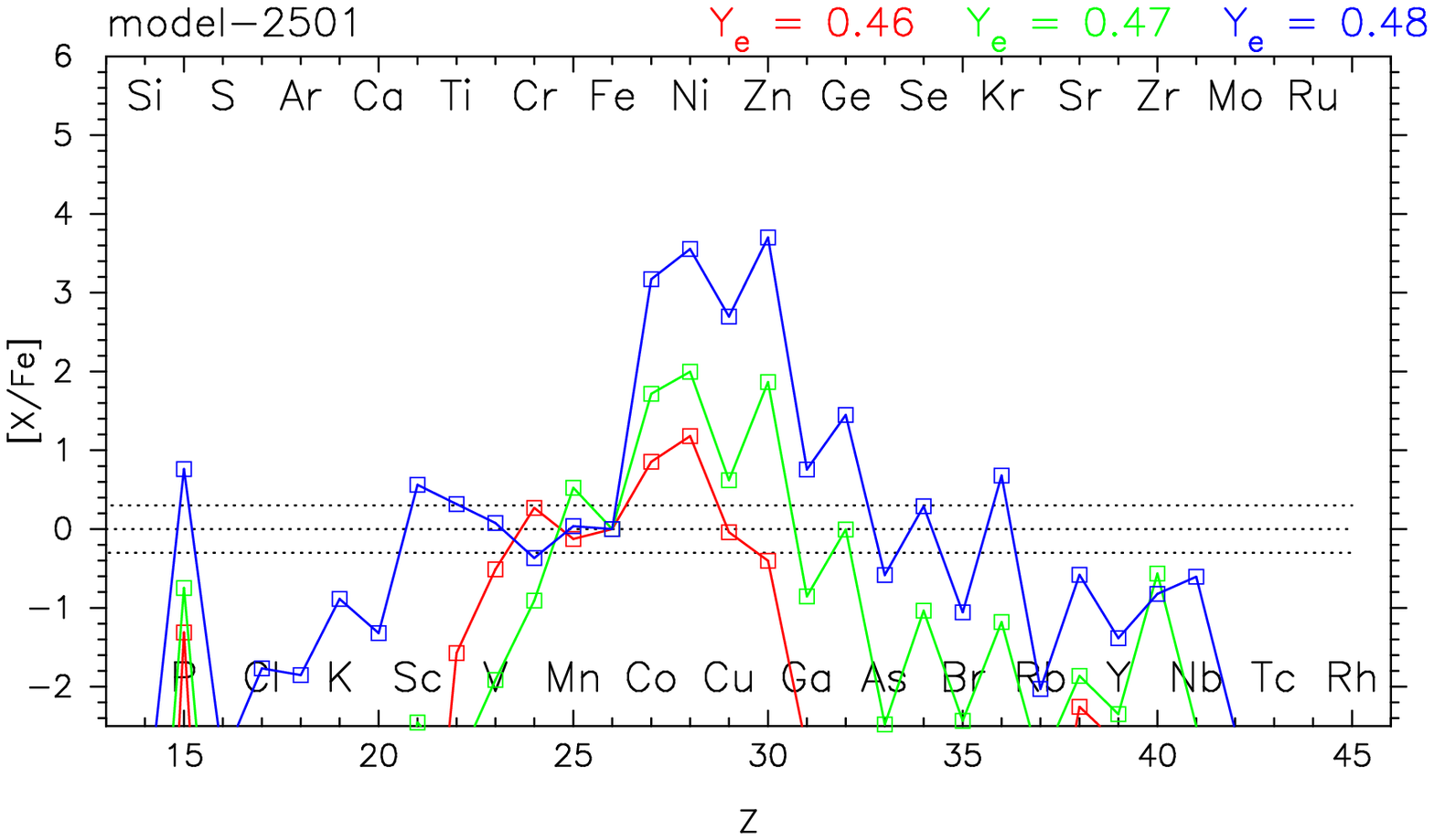}
\includegraphics[width=80mm]{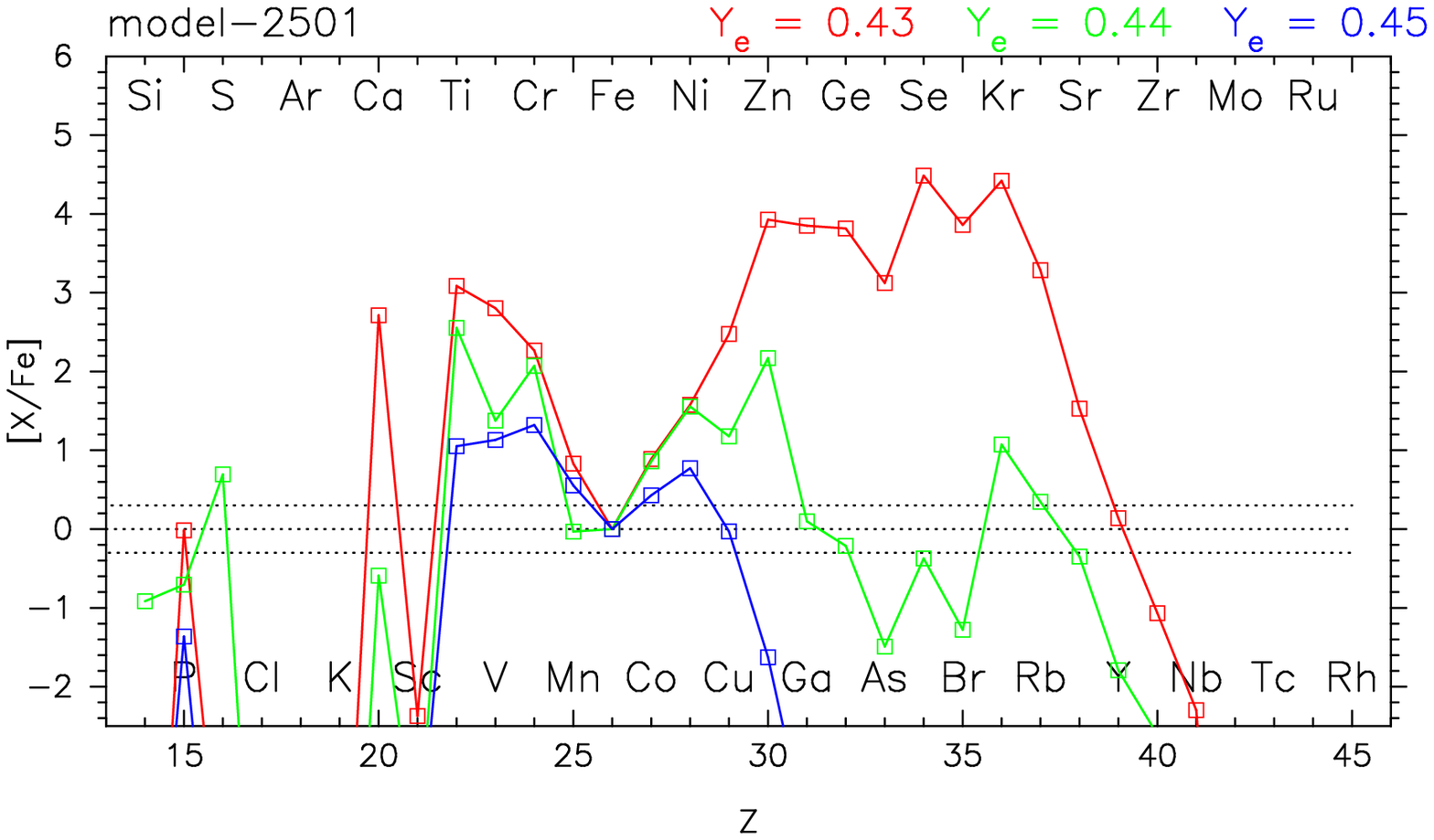}
\includegraphics[width=80mm]{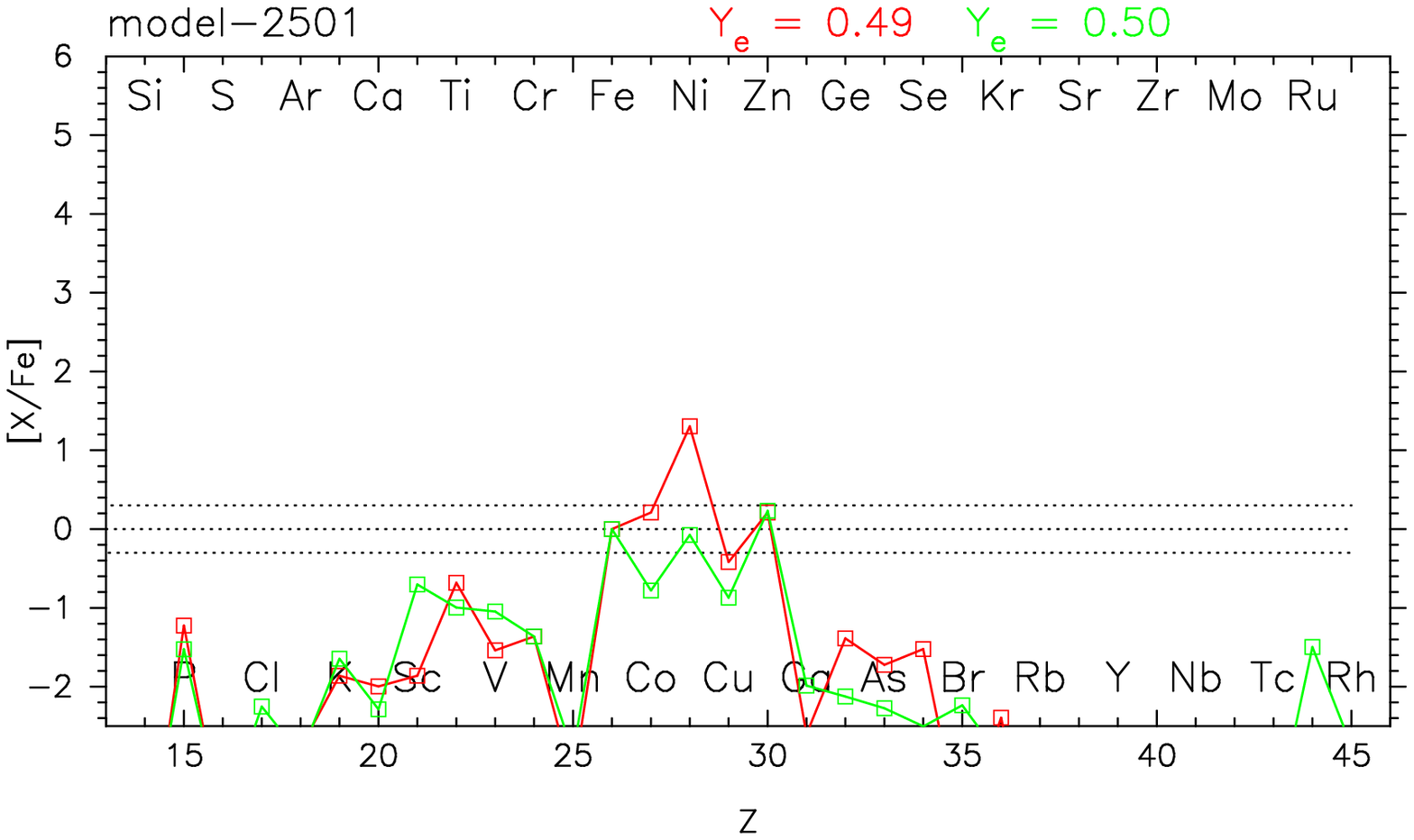}

\caption{Abundance patterns in the Si-burning regions of model-2501 with each $Y_e$.
The left panels: $Y_e$ = 0.40, 0.41, 0.42 and  $Y_e$ = 0.43, 0.44, 0.45 (from the top).
The right panels: $Y_e$ = 0.46, 0.47, 0.48 and  $Y_e$ = 0.49, 0.50 (from the top).
}
\end{figure}

\clearpage
\begin{figure}

\includegraphics[width=80mm]{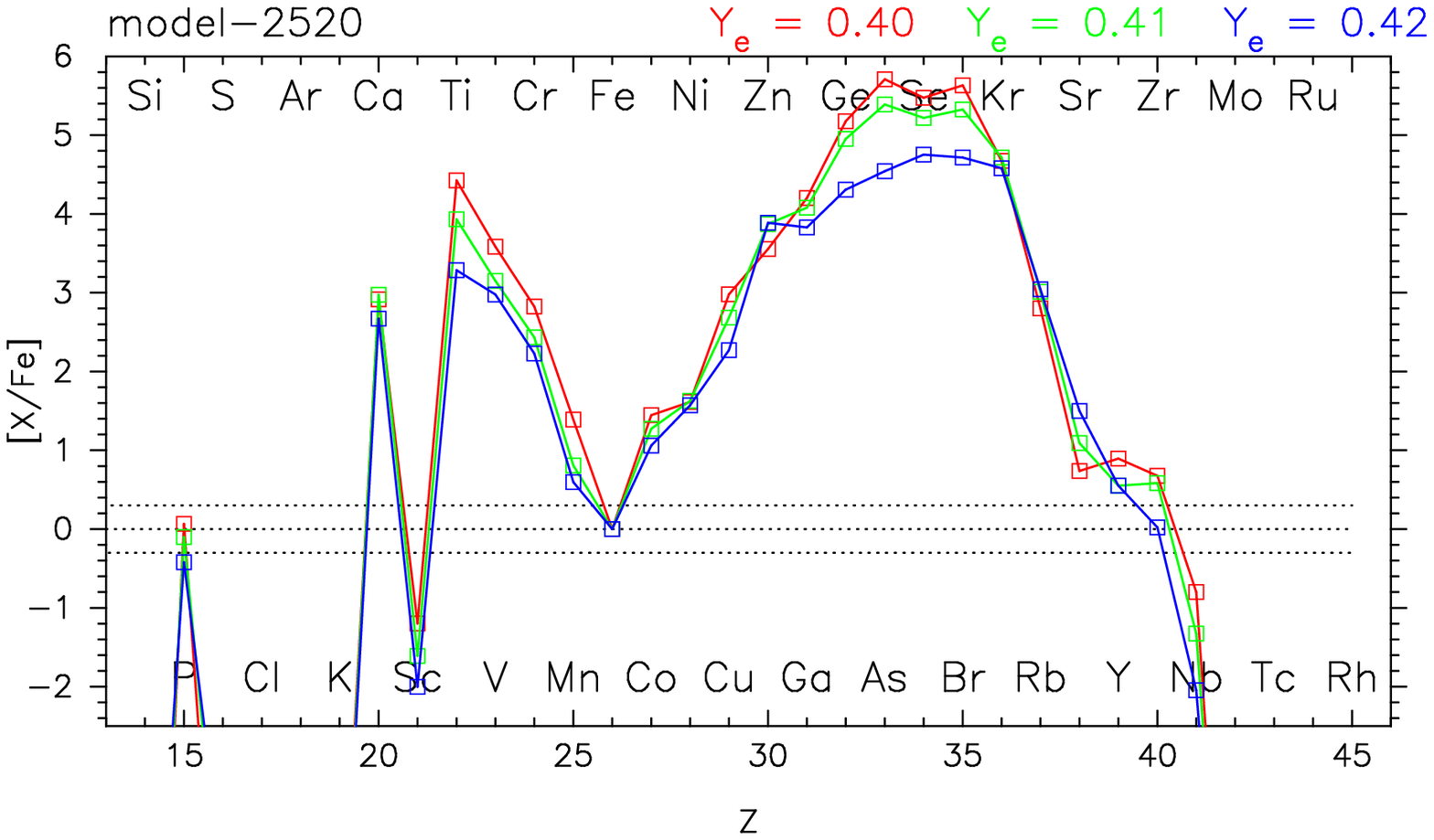}
\includegraphics[width=80mm]{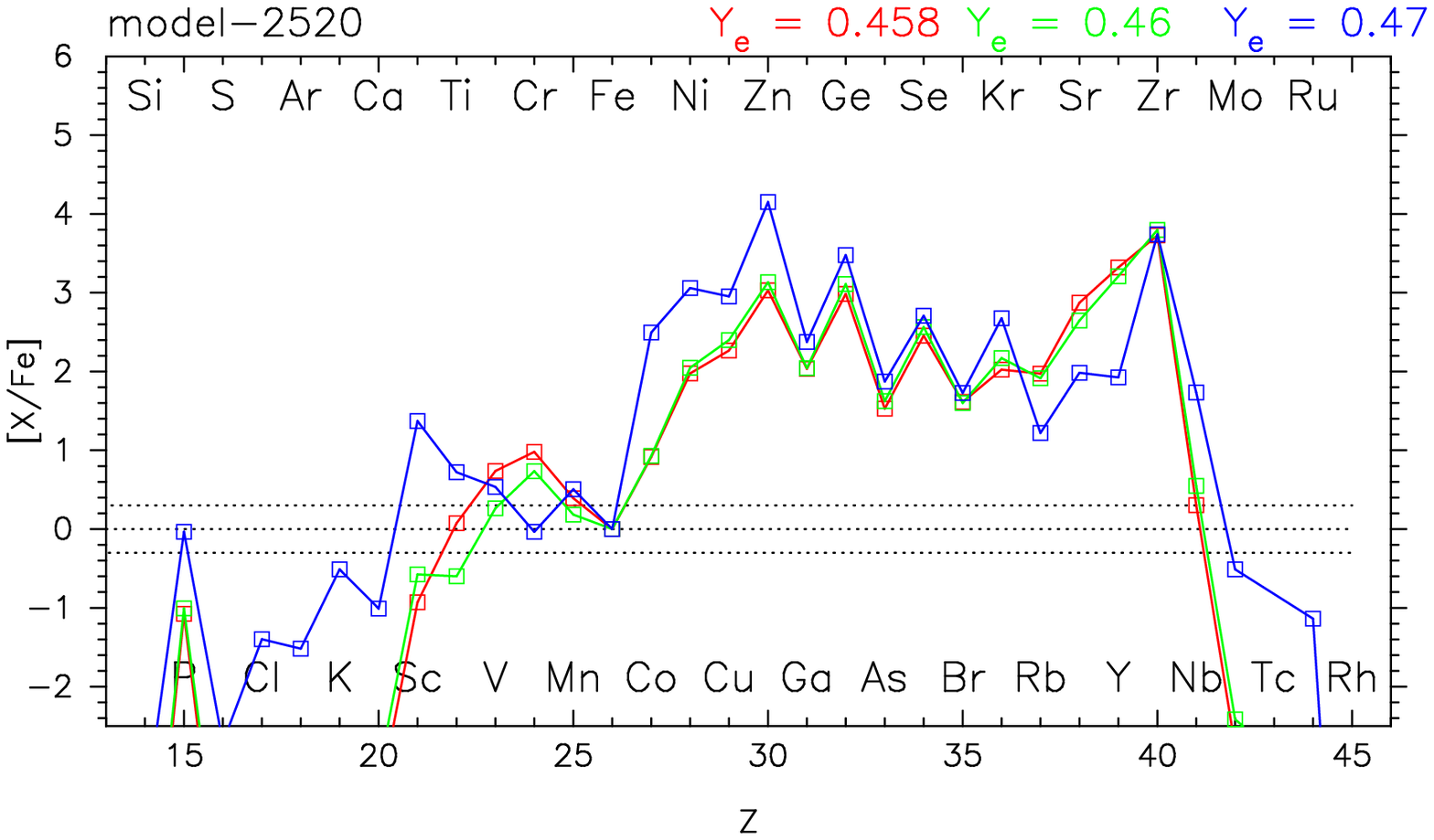}
\includegraphics[width=80mm]{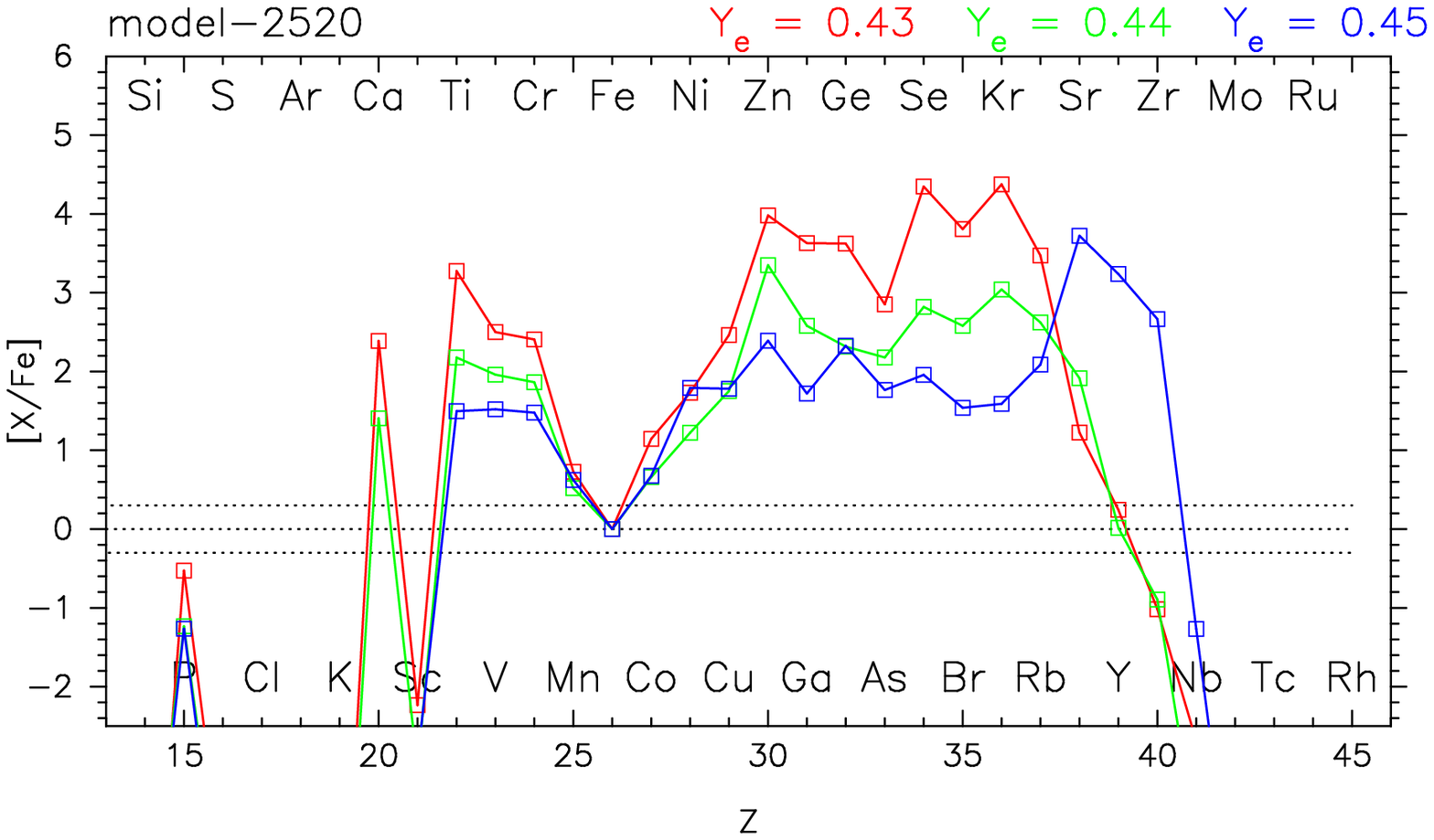}

\end{figure}

\begin{figure}

\includegraphics[width=80mm]{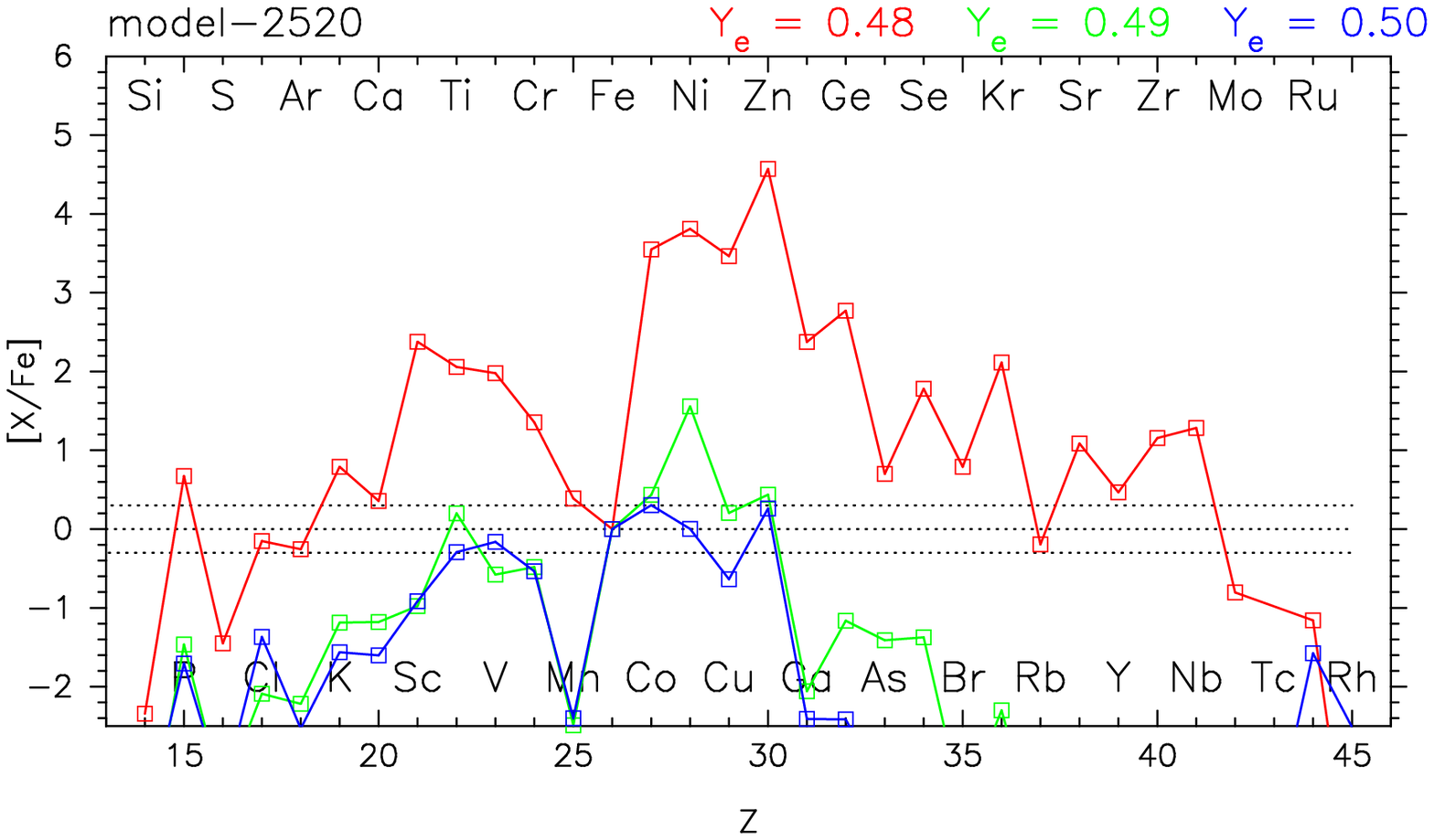}
\includegraphics[width=80mm]{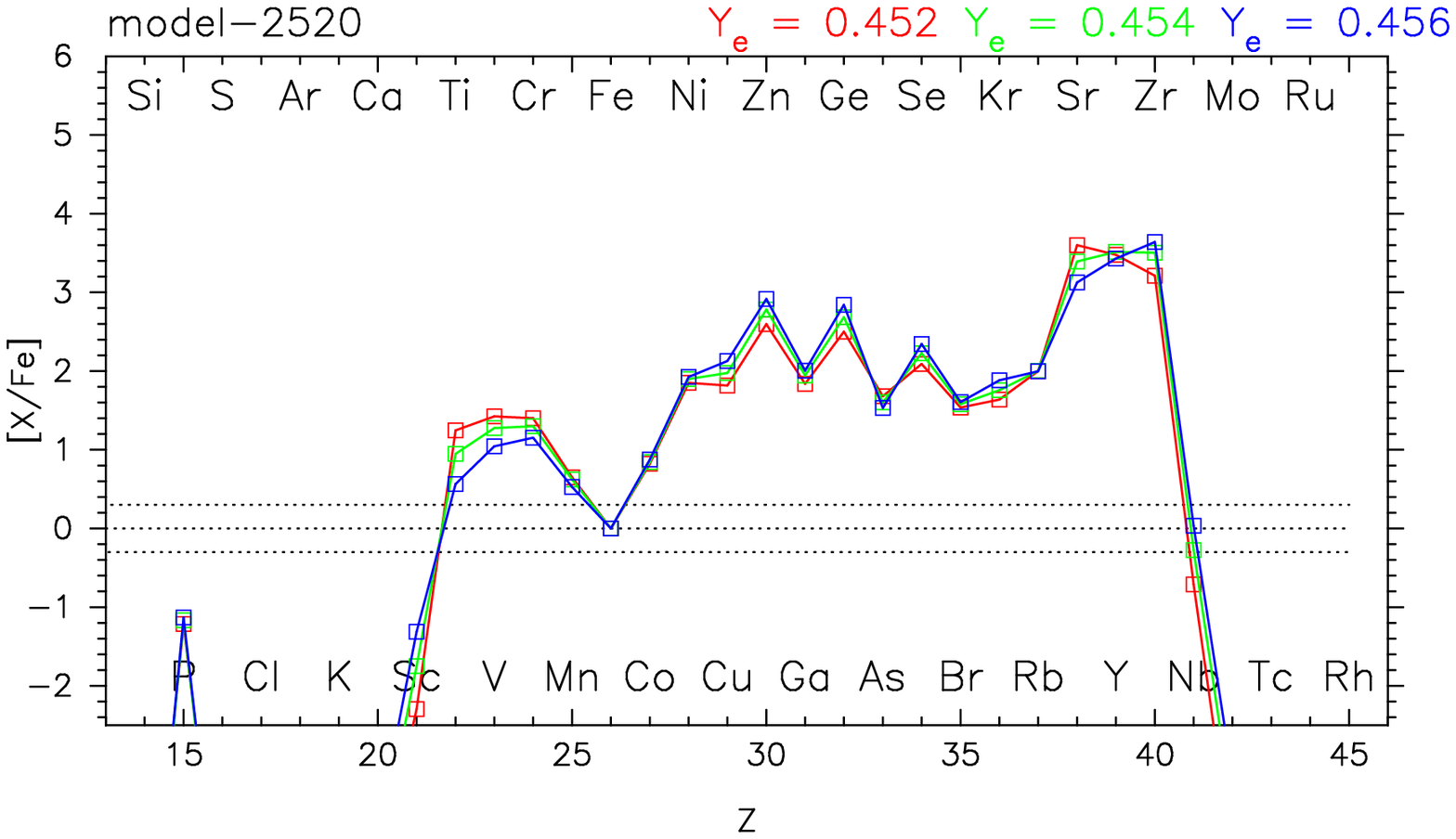}

\caption{Abundance patterns in the Si-burning regions of model-2520 with each $Y_e$.
The left panels: $Y_e$ = 0.40, 0.41, 0.42, $Y_e$ = 0.43, 0.44, 0.45 and $Y_e$ = 0.452, 0.454, 0.456 (from the top).
The right panels: $Y_e$ = 0.458, 0.46, 0.47 and $Y_e$ = 0.48, 0.49, 0.50 (from the top).}
\end{figure}

\clearpage
\begin{figure}

\includegraphics[width=80mm]{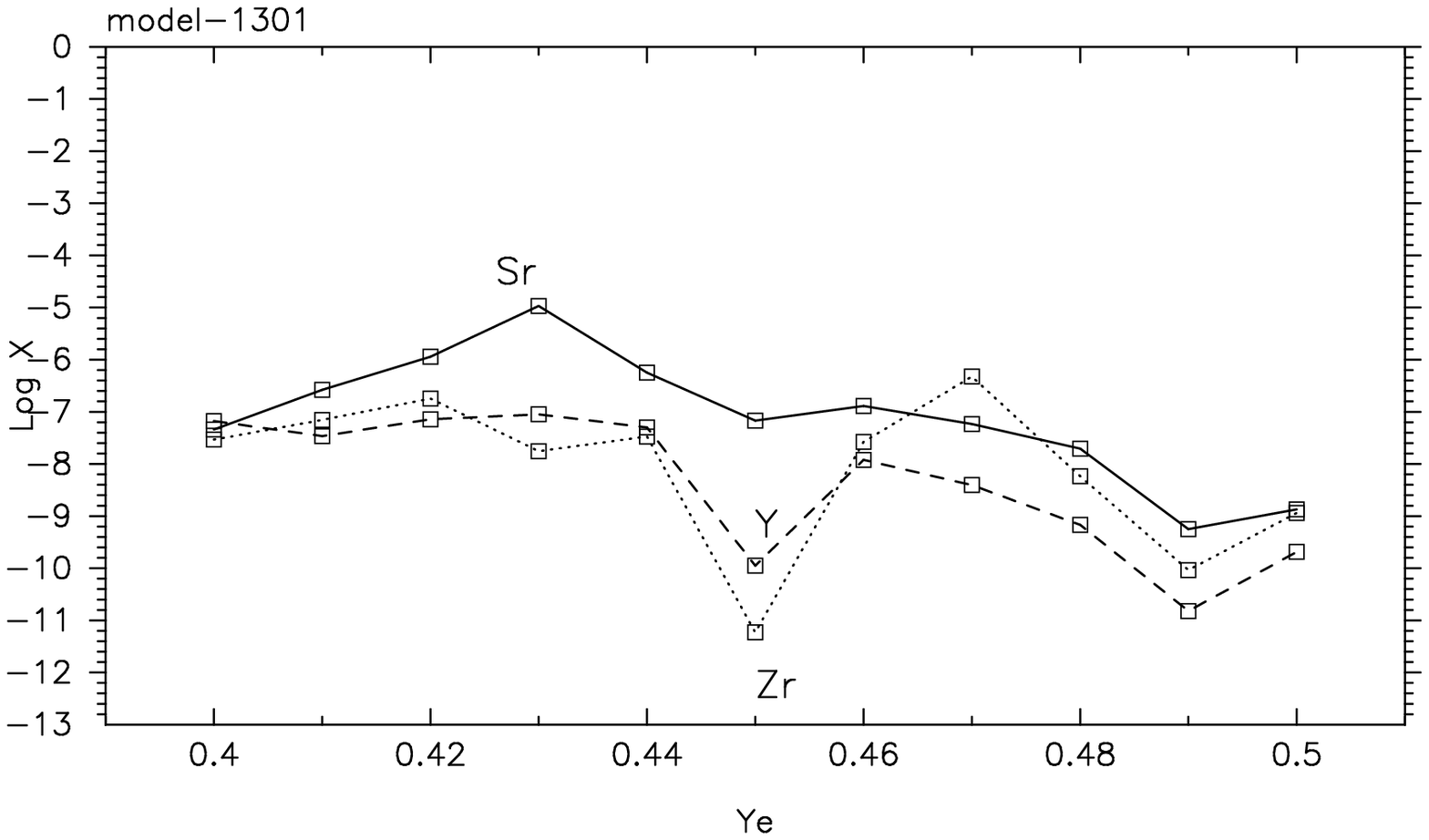}
\includegraphics[width=80mm]{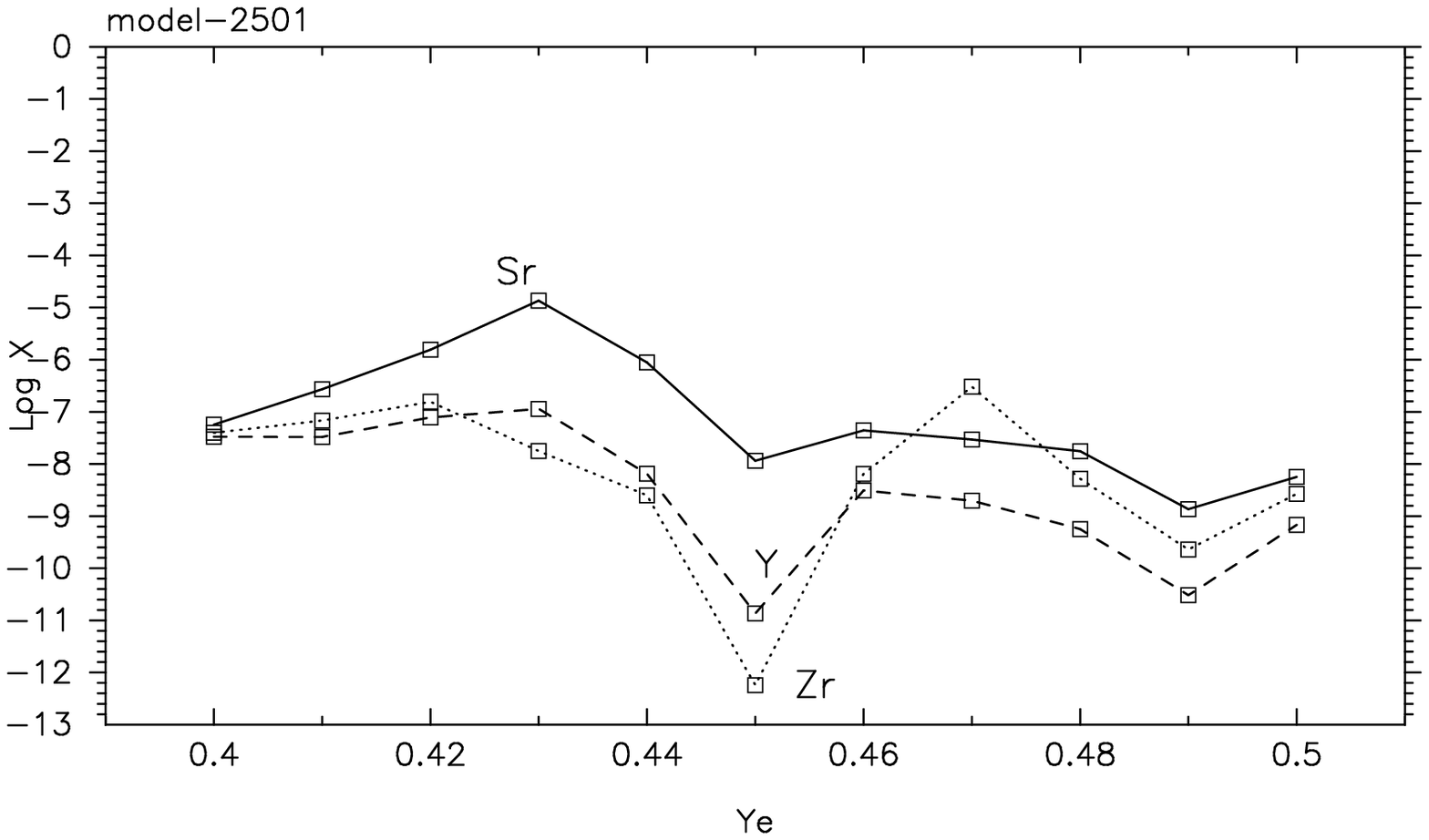}
\includegraphics[width=80mm]{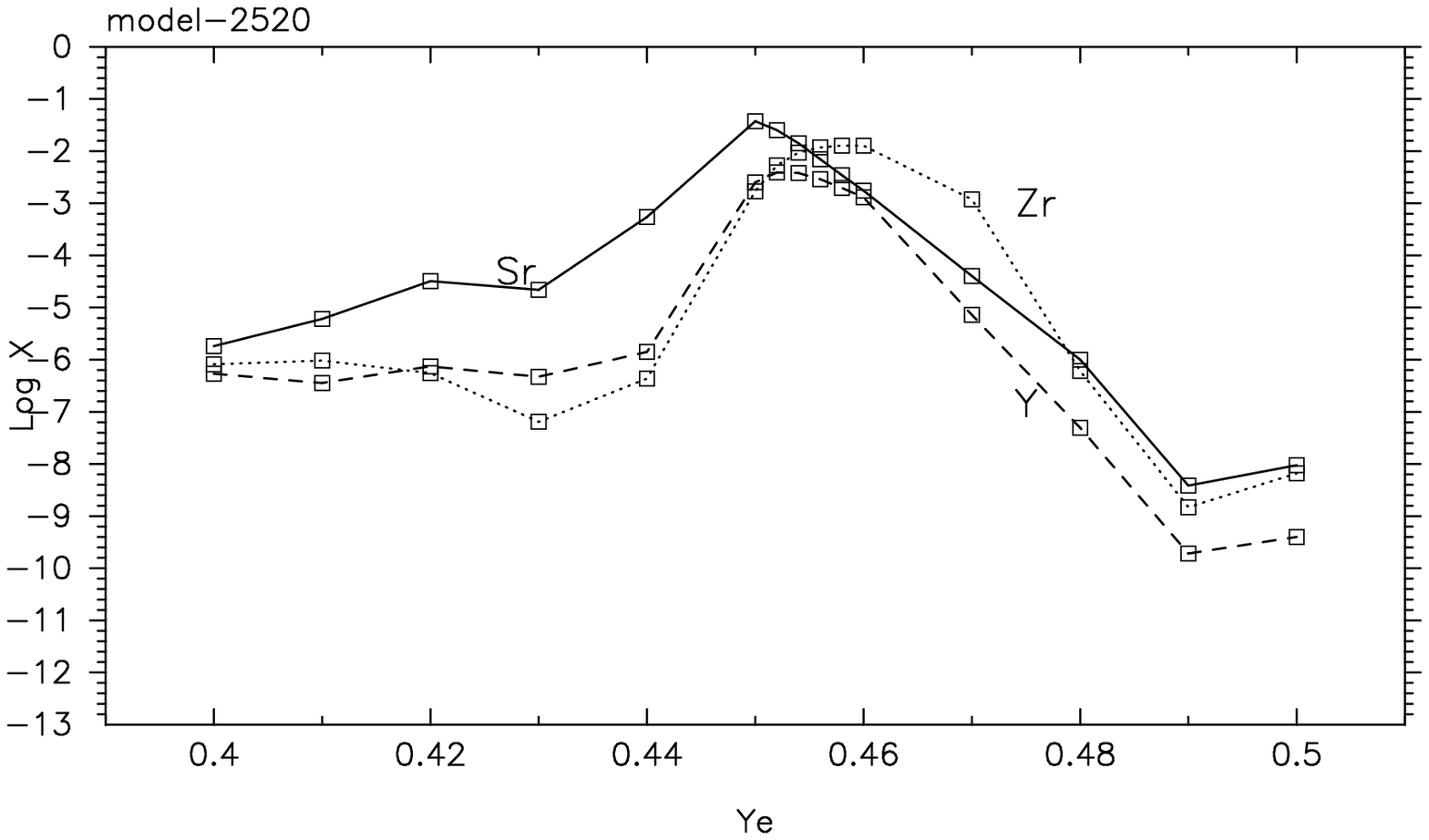}

\caption{The mass fraction in the Si-burning regions of Sr, Y and Zr.
The lines for
Sr, Y, and Zr are represented by 
the solid line, dashed line, and dotted lines, respectively.}
\end{figure}

\begin{figure}

\includegraphics[width=80mm]{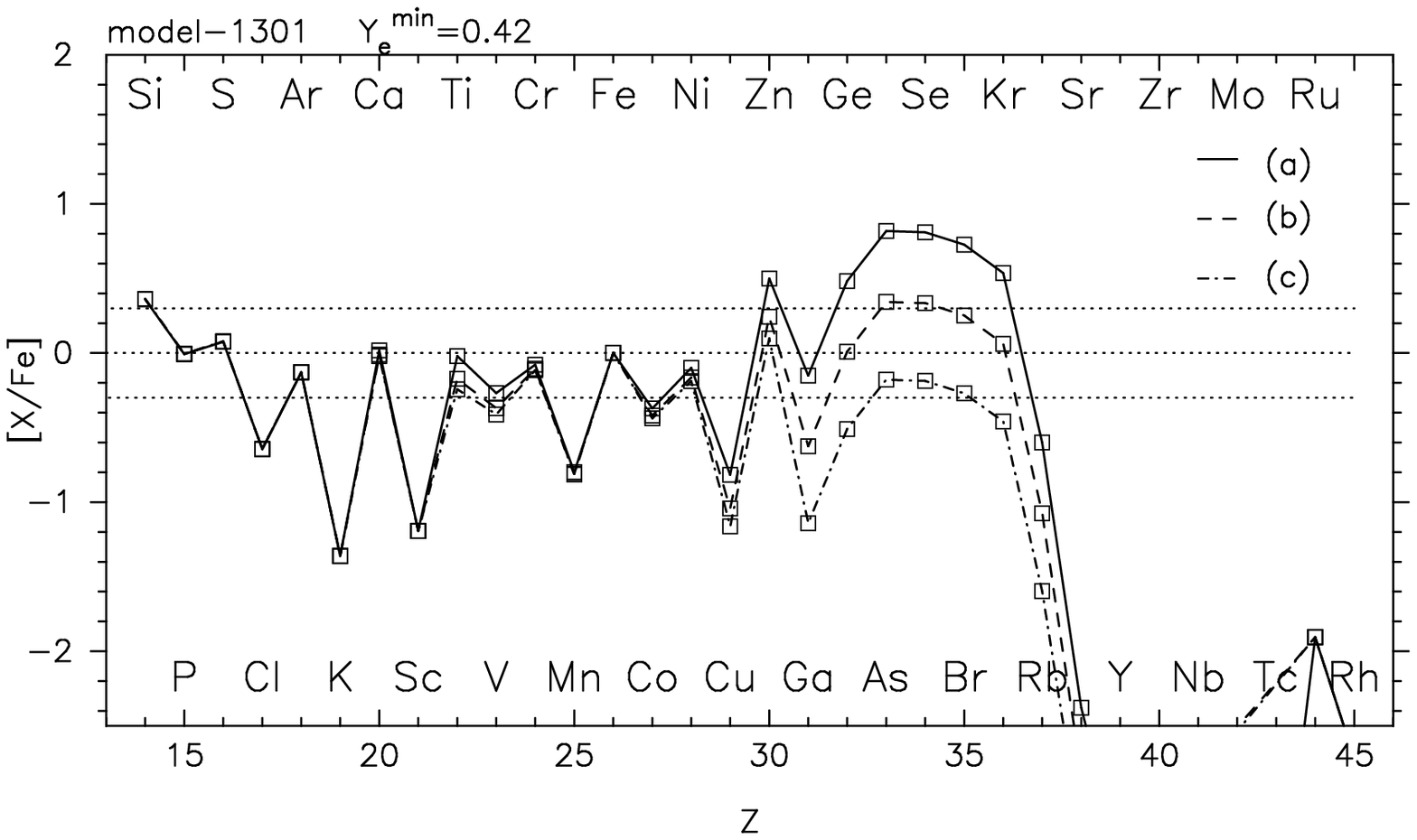}
\includegraphics[width=80mm]{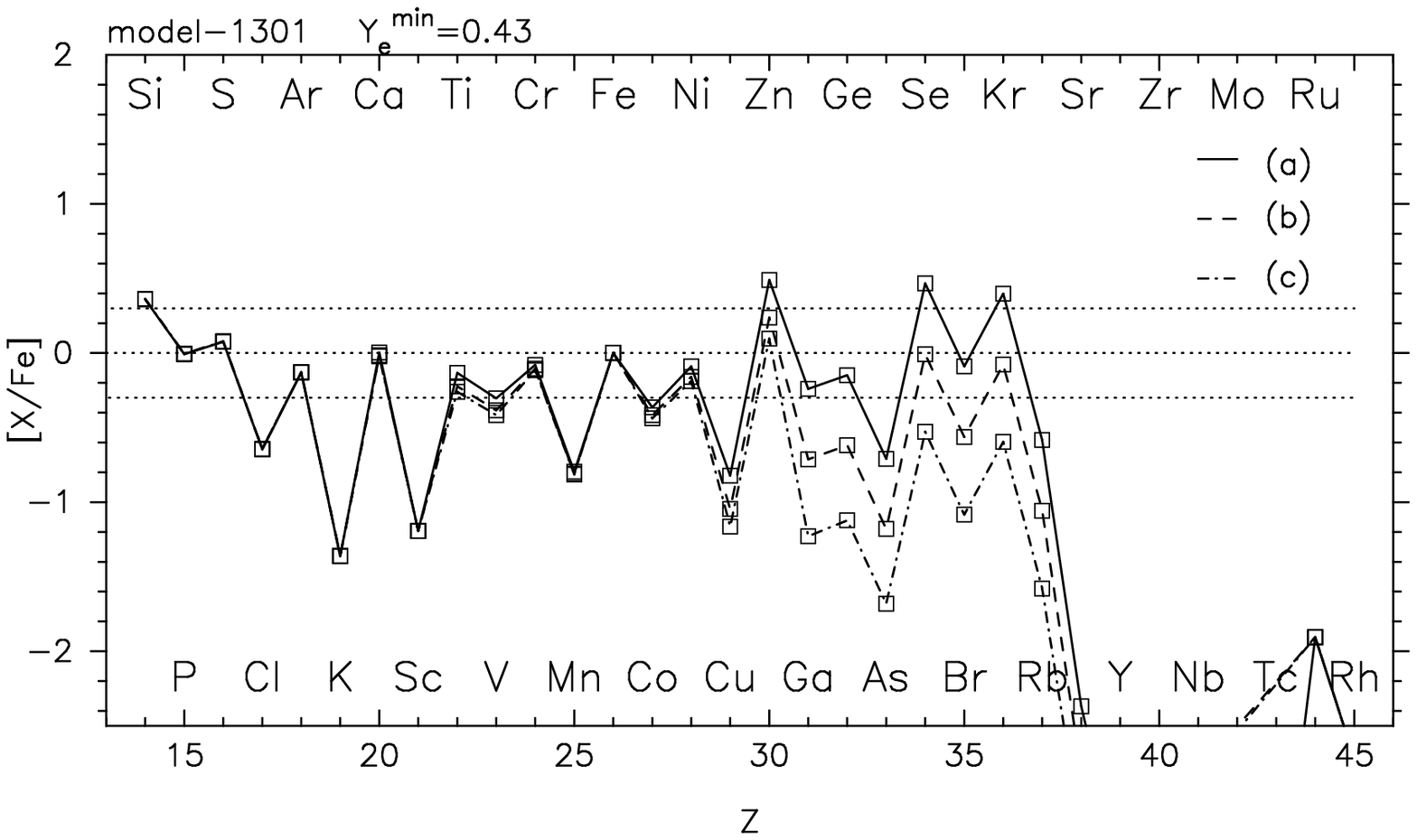}

\caption{Abundance patterns of the whole ejecta for model-1301 with each $Y_{\rm e}$ distribution
and $\Delta M$. Here, the abundance below $M_{\rm cut}$ is averaged for $Y_{\rm e}$
= $Y_{\rm e}^{\rm min} \sim$  0.50.
The parameters and some related numbers for models (a) are shown in Table 2.
For the models (b) and (c), $\Delta M$ of model (a) is divided by 3 and 10, respectively.
}
\end{figure}

\clearpage

\begin{figure}

\includegraphics[width=80mm]{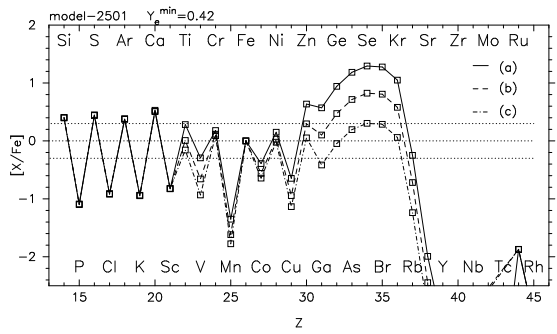}
\includegraphics[width=80mm]{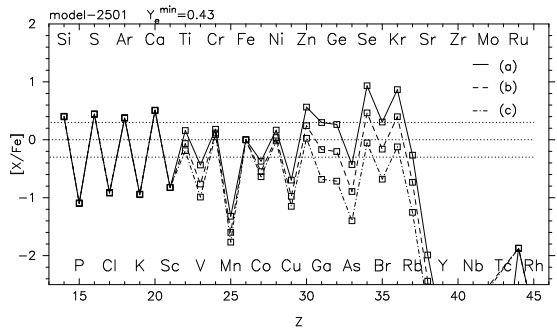}

\caption{Same as Figure 10, but for model-2501.
}
\end{figure}

\clearpage
\begin{figure}

\includegraphics[width=80mm]{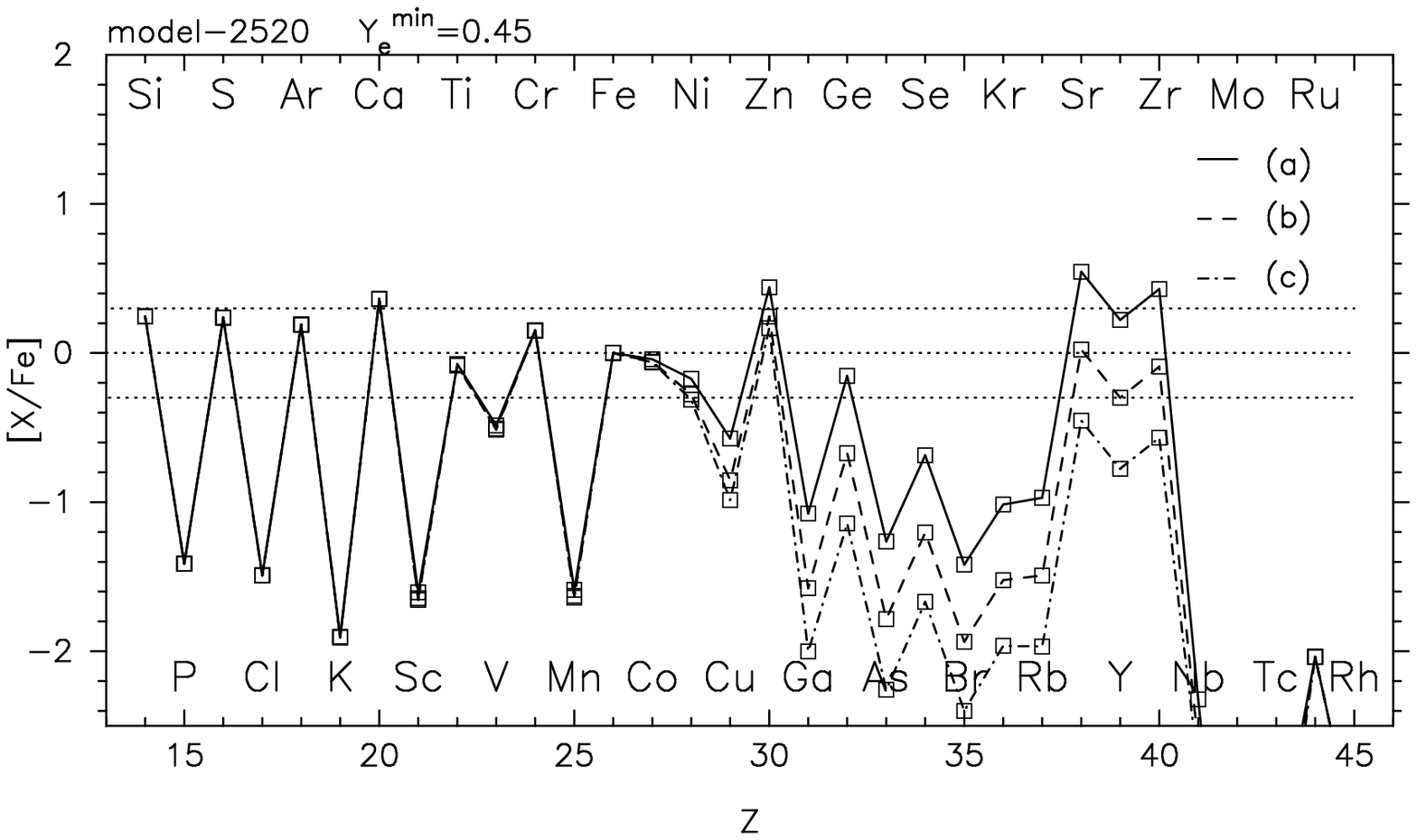}
\includegraphics[width=80mm]{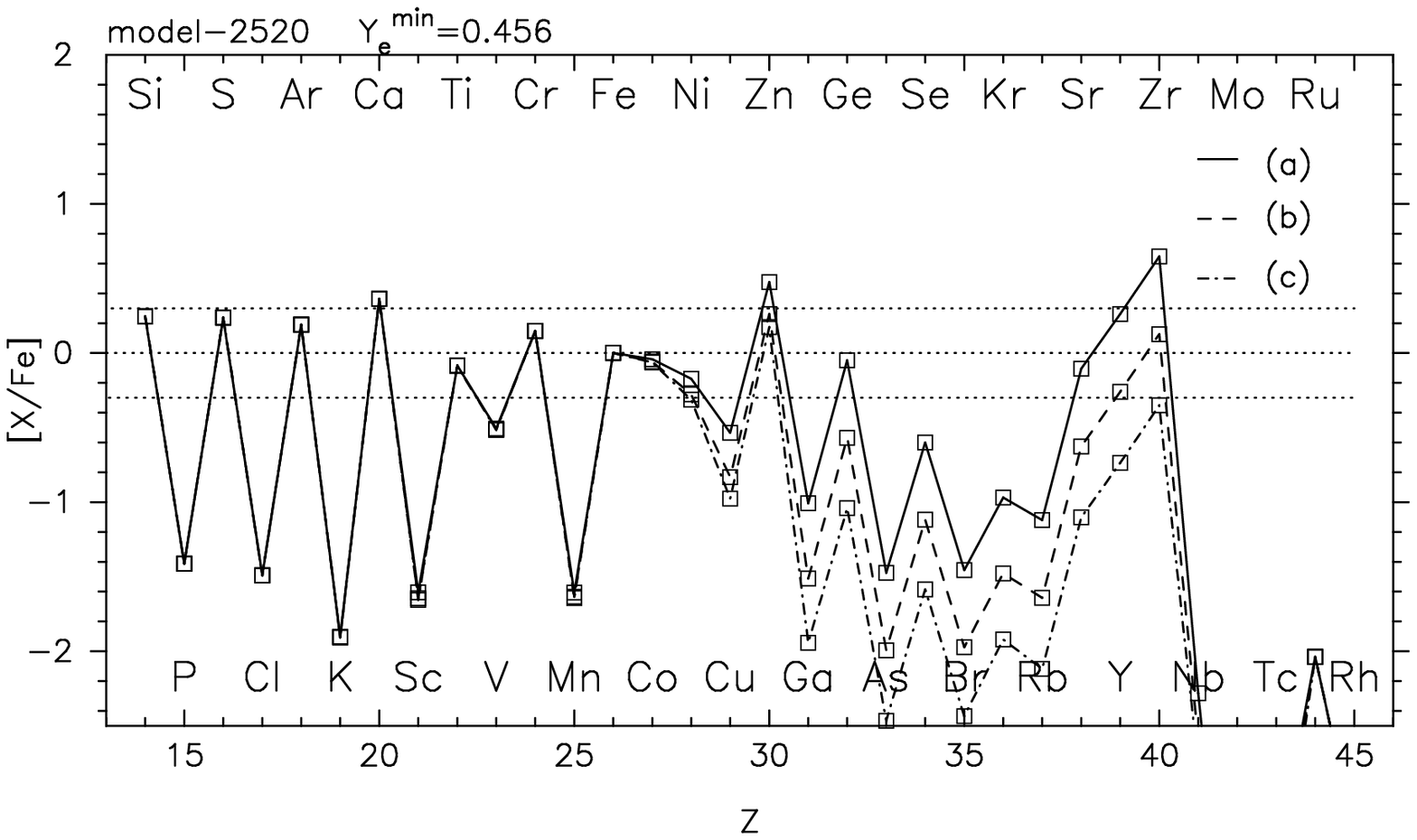}
\includegraphics[width=80mm]{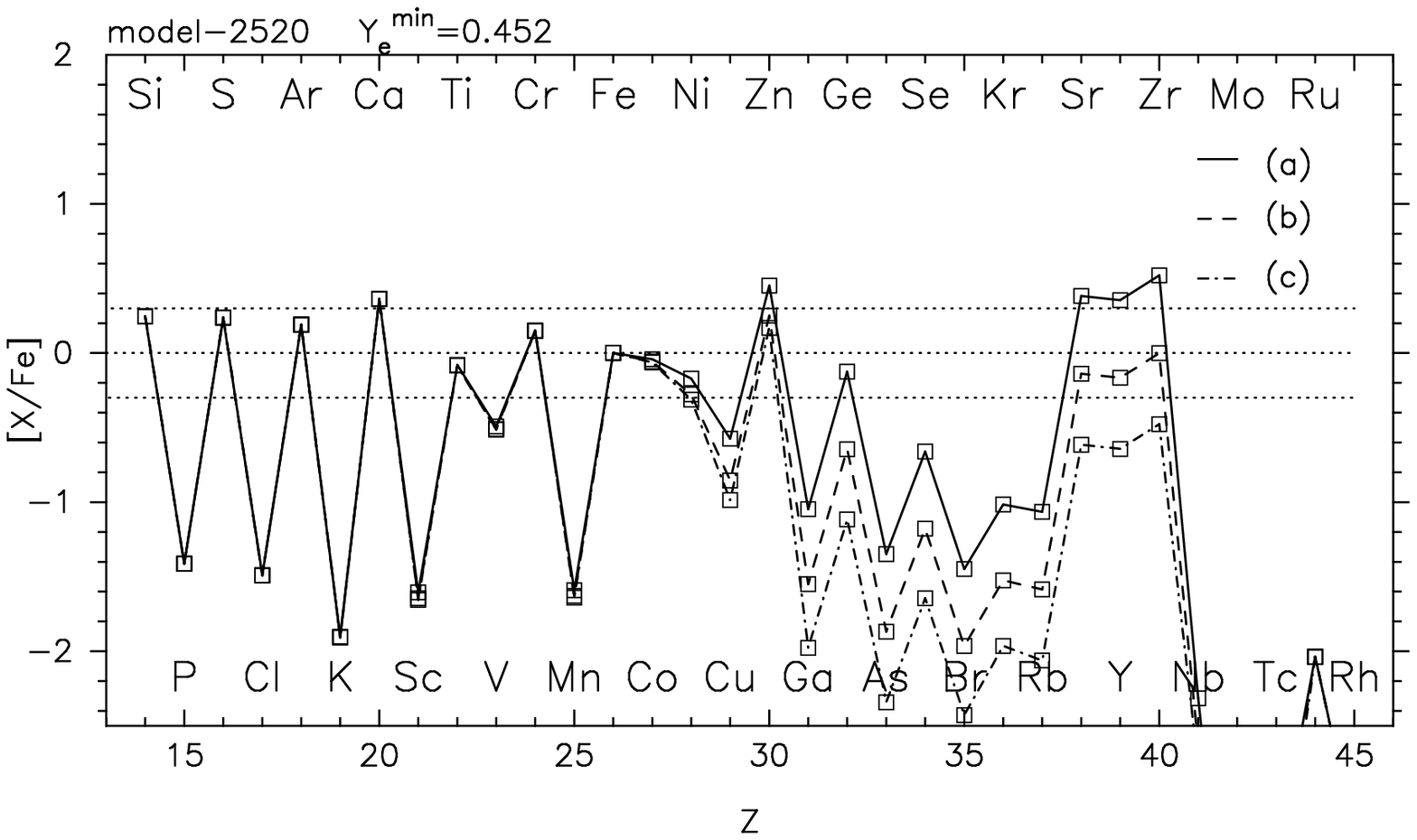}

\end{figure}

\begin{figure}

\includegraphics[width=80mm]{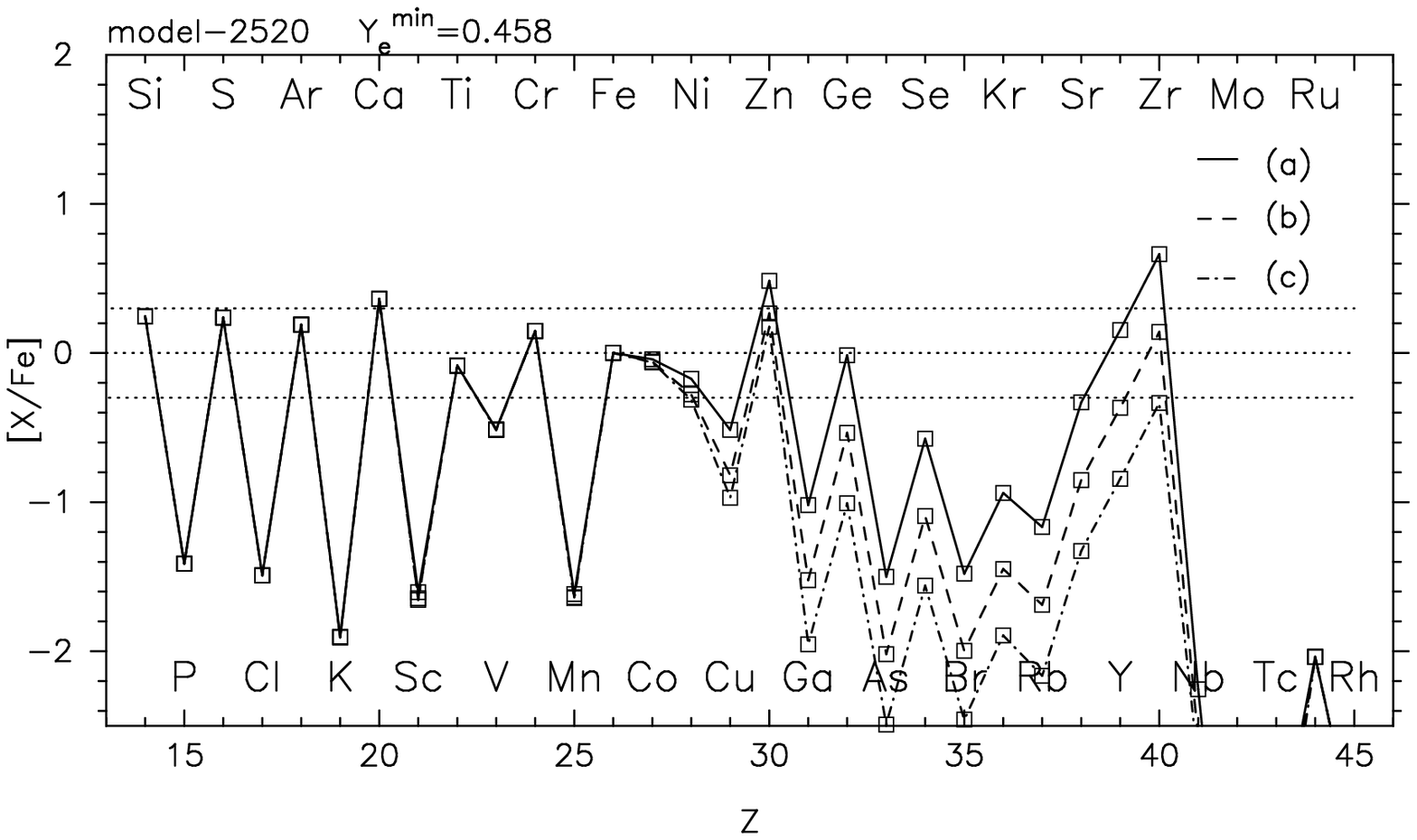}
\includegraphics[width=80mm]{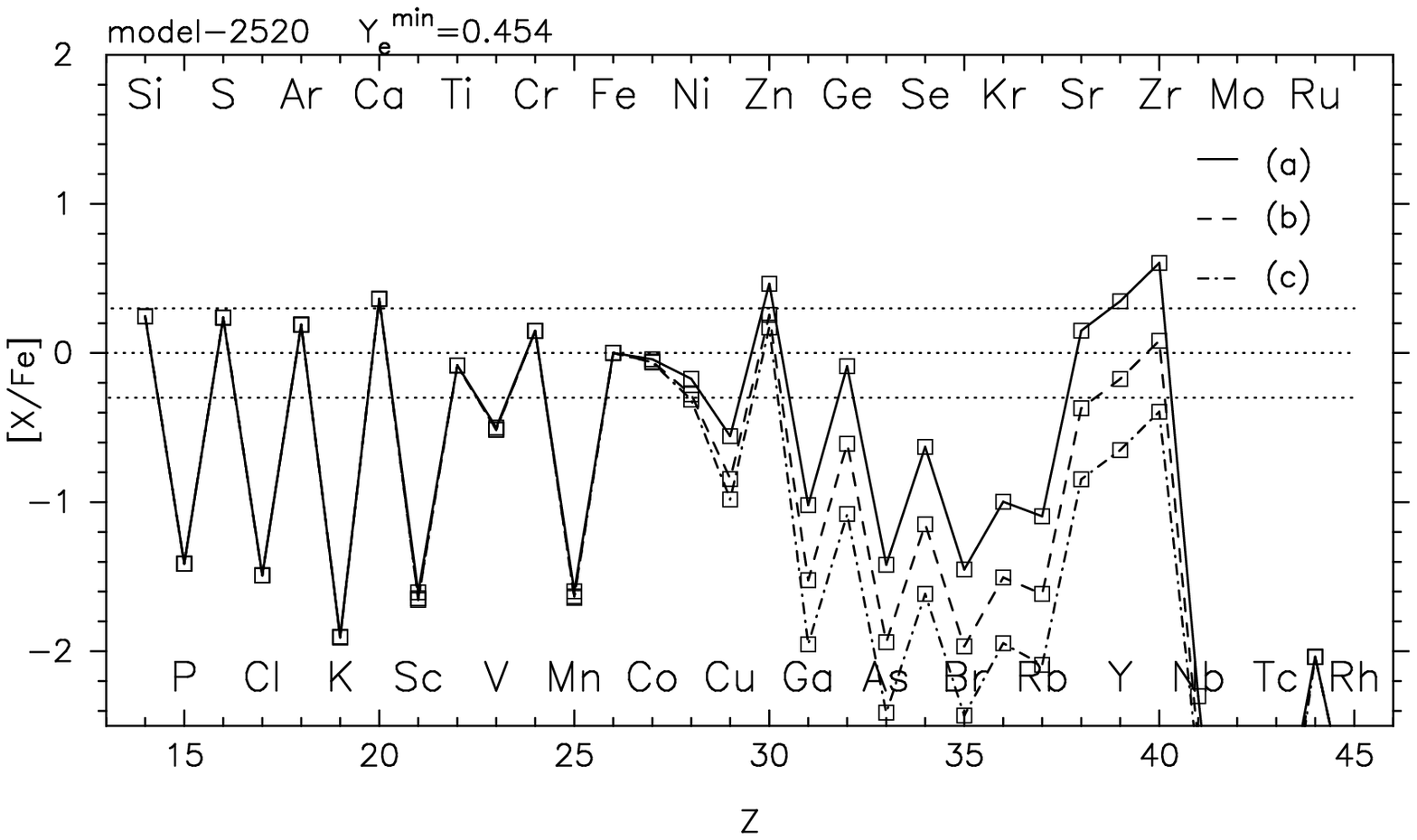}
\includegraphics[width=80mm]{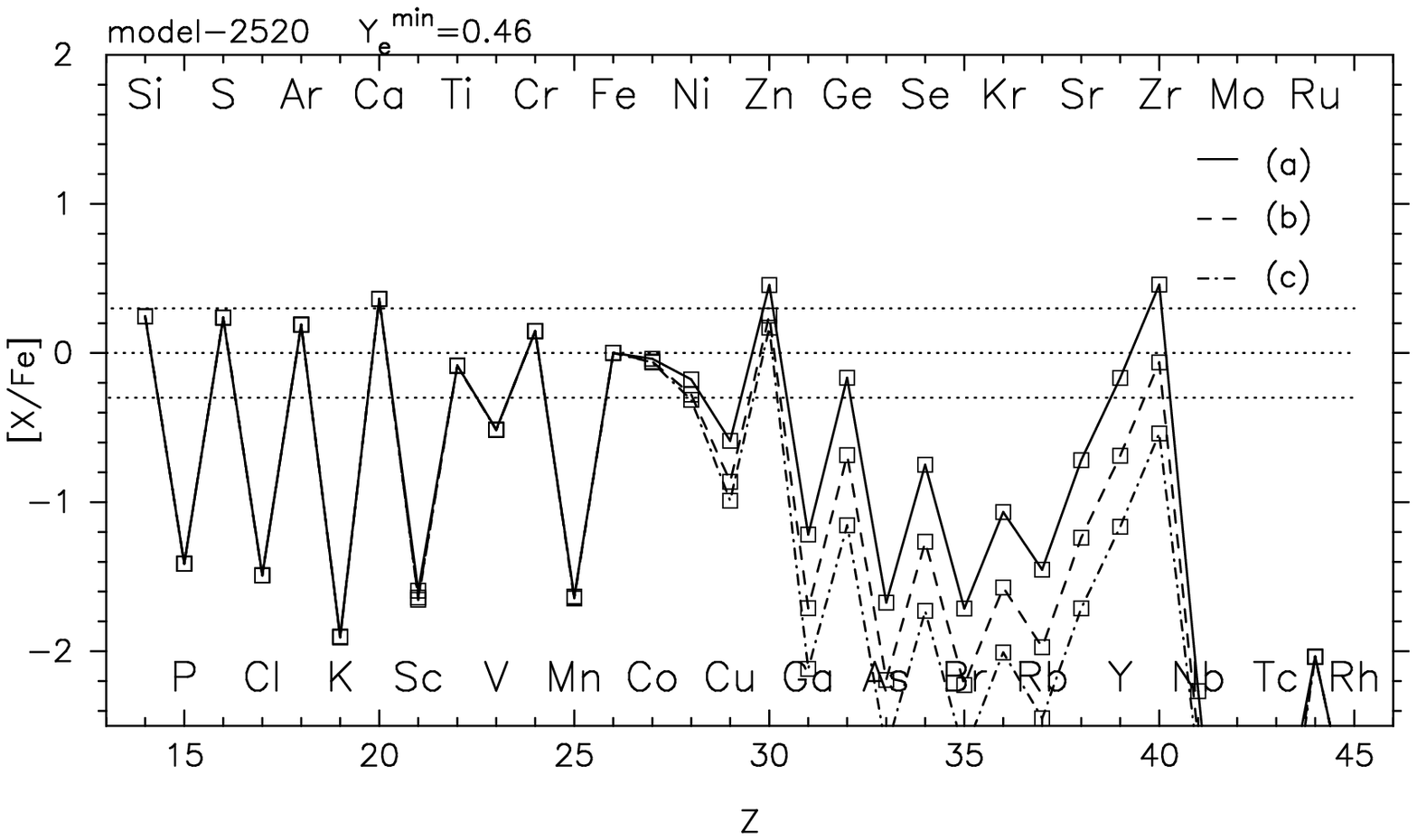}

\caption{Same as Figure 10, but for model-2520 with different
values of $Y_{\rm e}^{\rm min}$. 
}
\end{figure}

\clearpage
\begin{figure}

\includegraphics[width=80mm]{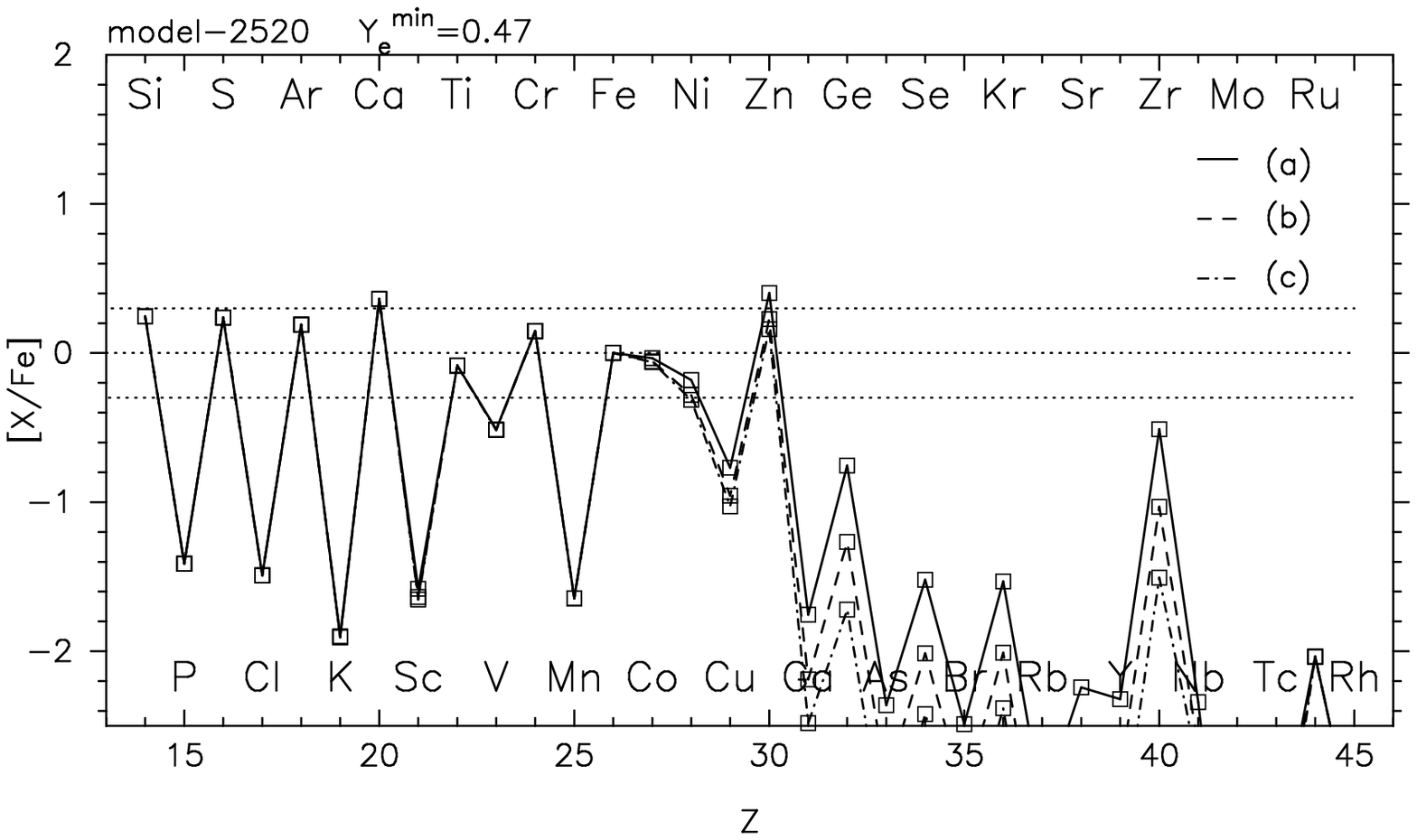}
\includegraphics[width=80mm]{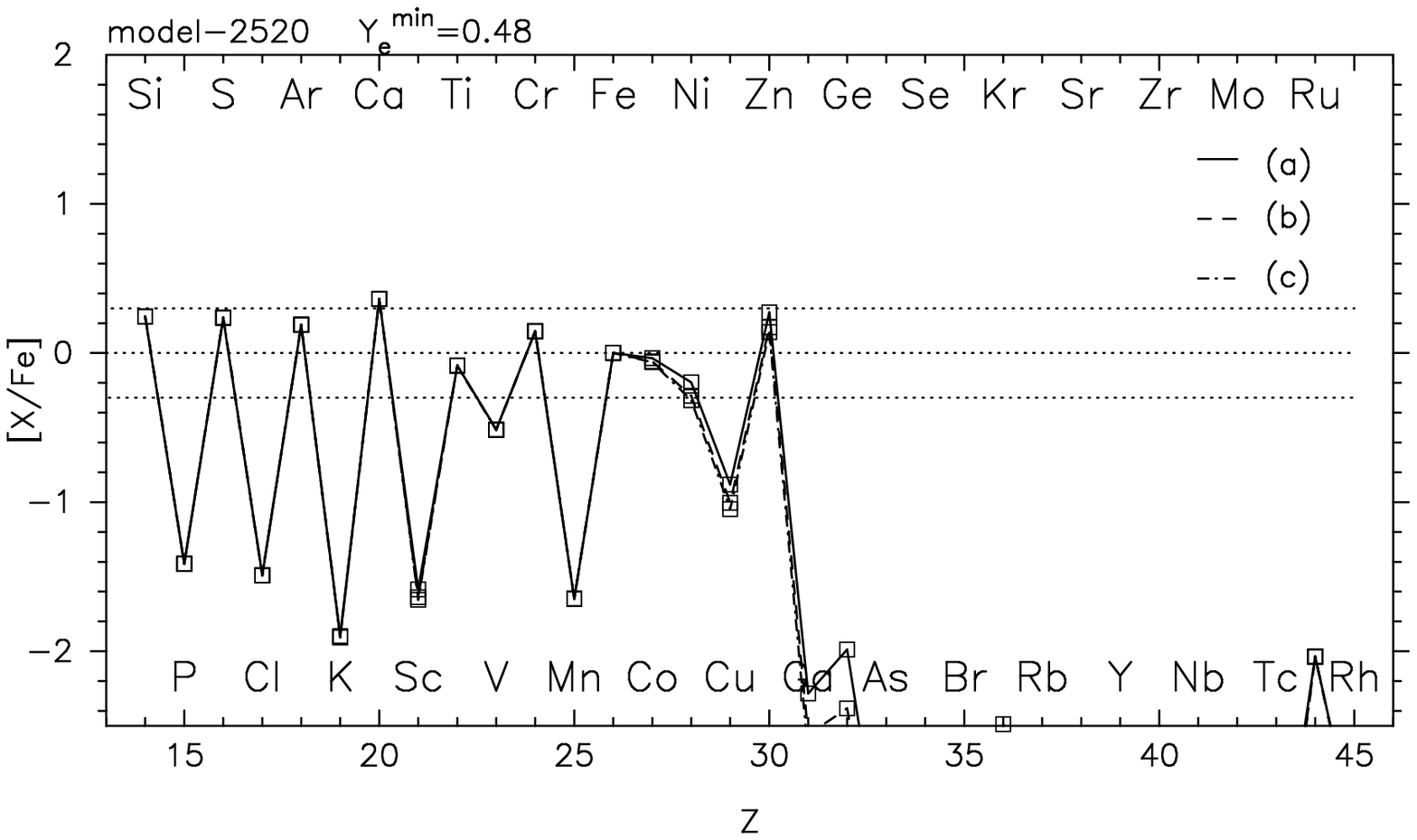}
\includegraphics[width=80mm]{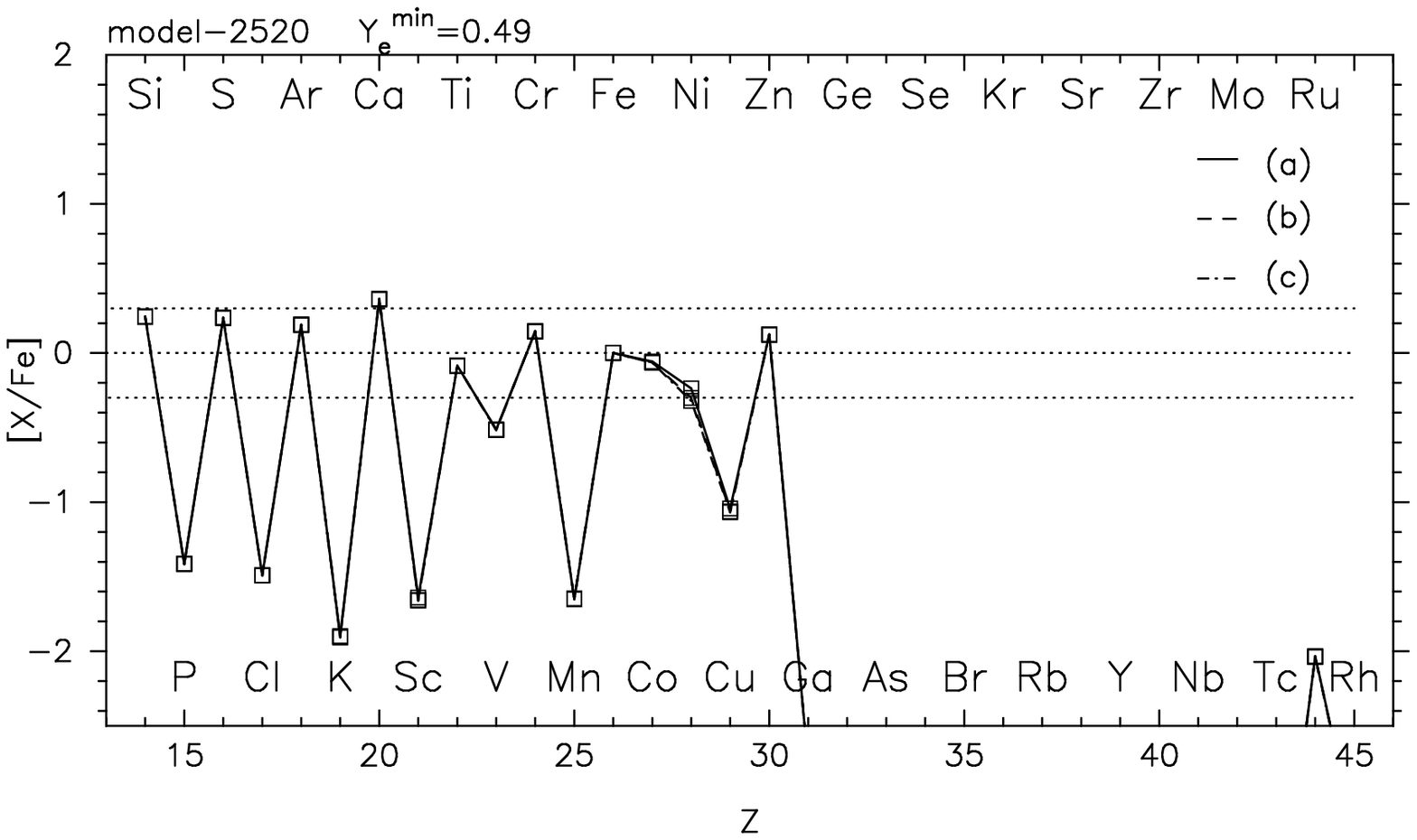}

\caption{Same as Figure 10, but for model-2520 with different
values of $Y_{\rm e}^{\rm min}$. 
}
\end{figure}

\begin{figure}

\includegraphics[width=80mm]{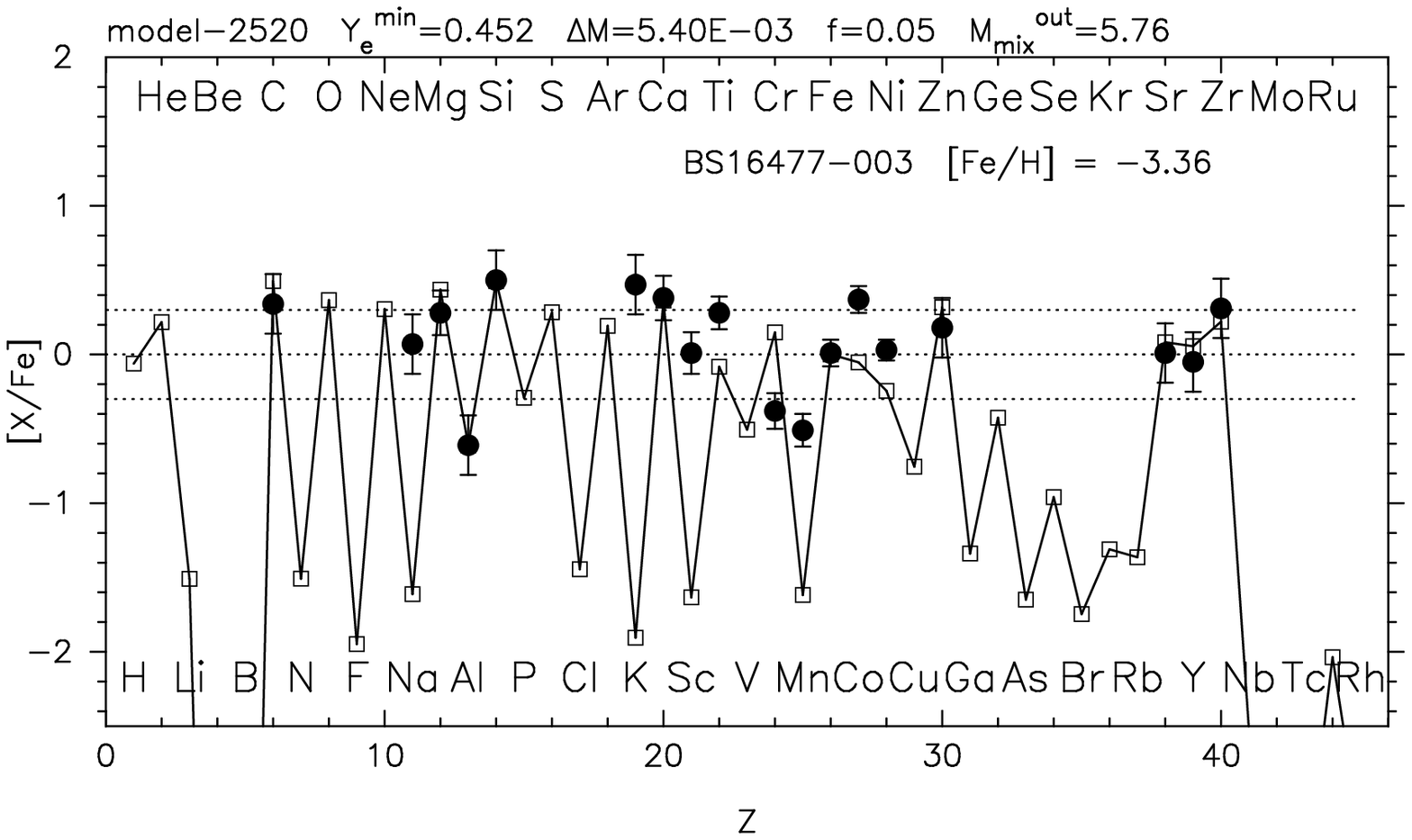}
\includegraphics[width=80mm]{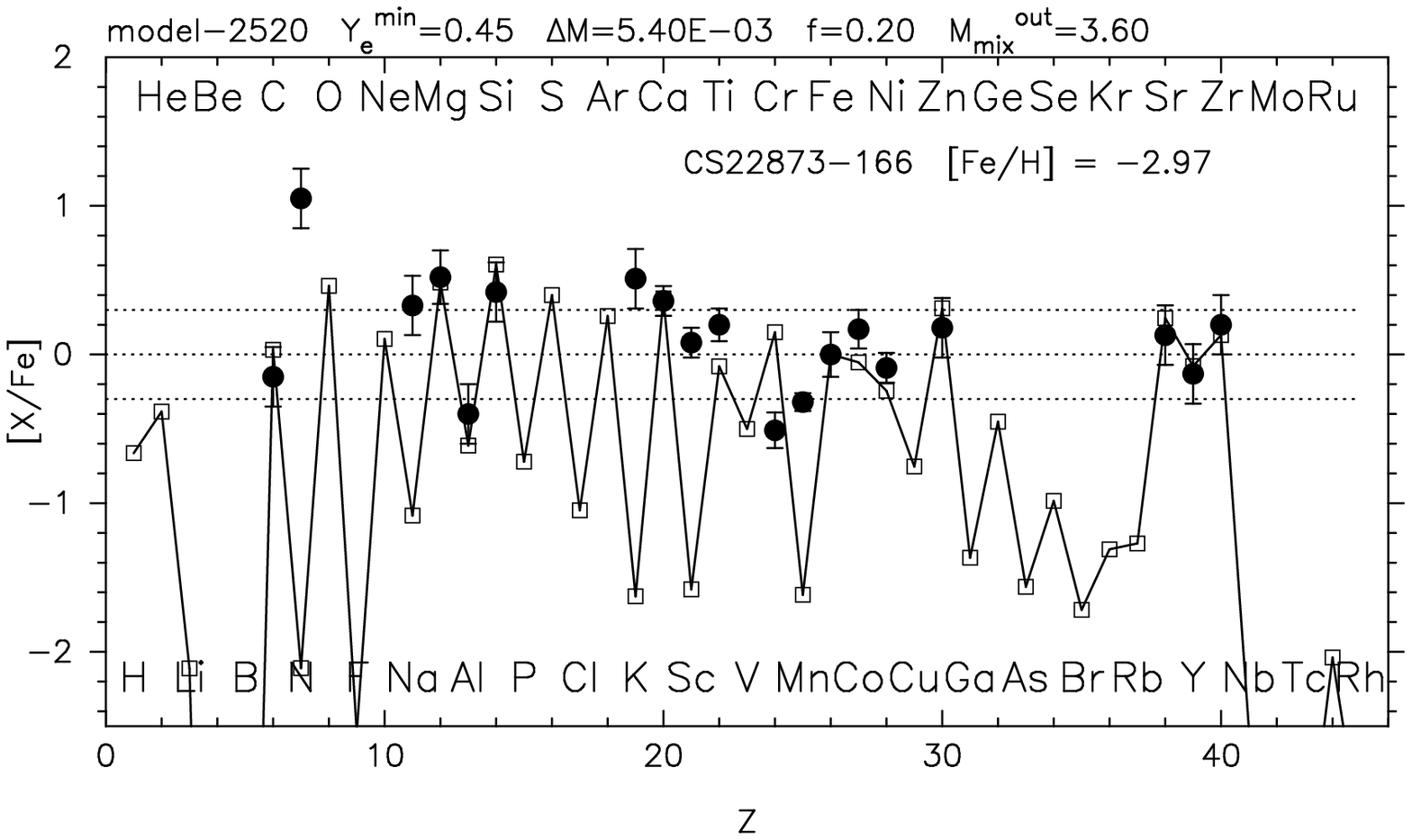}
\includegraphics[width=80mm]{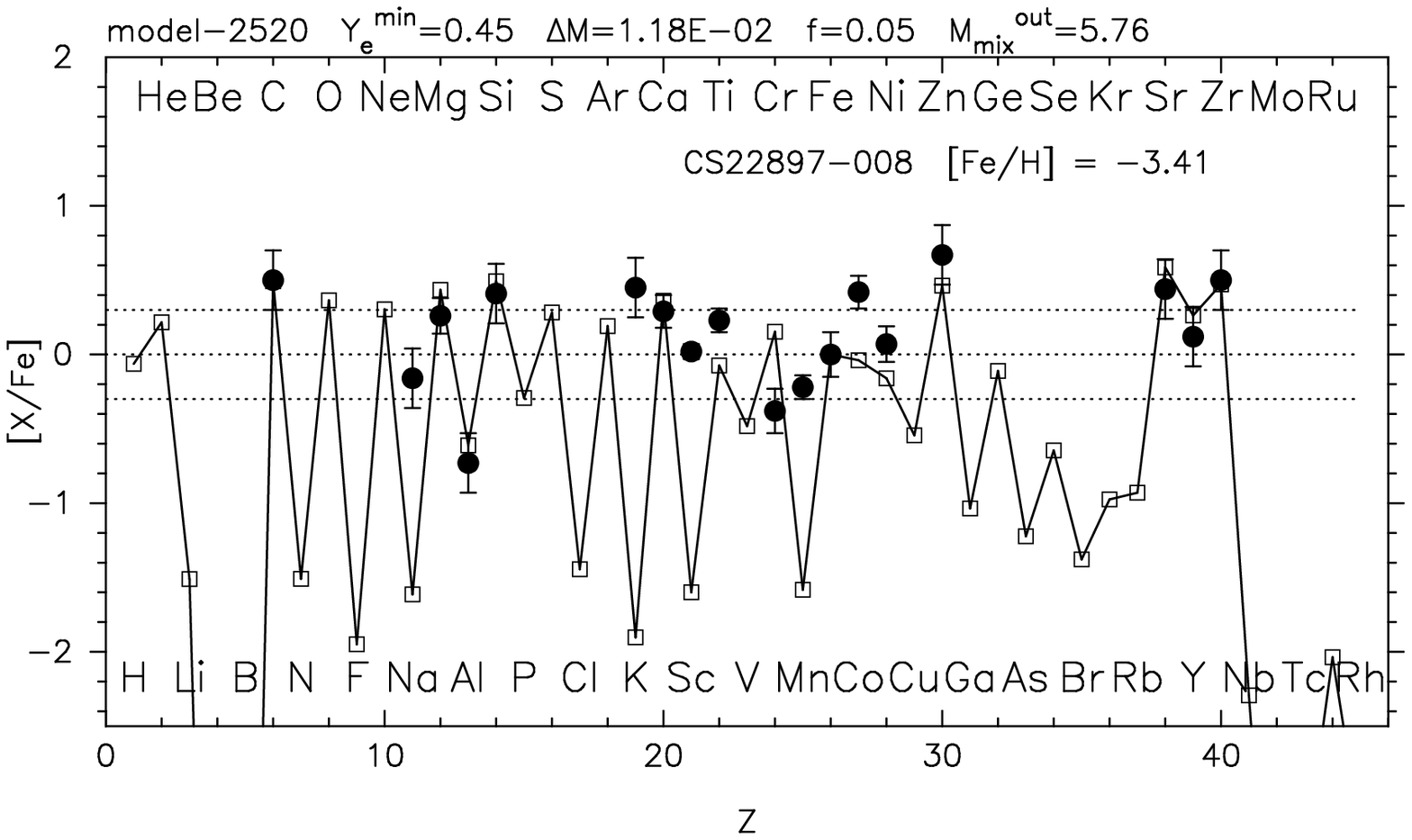}
\includegraphics[width=80mm]{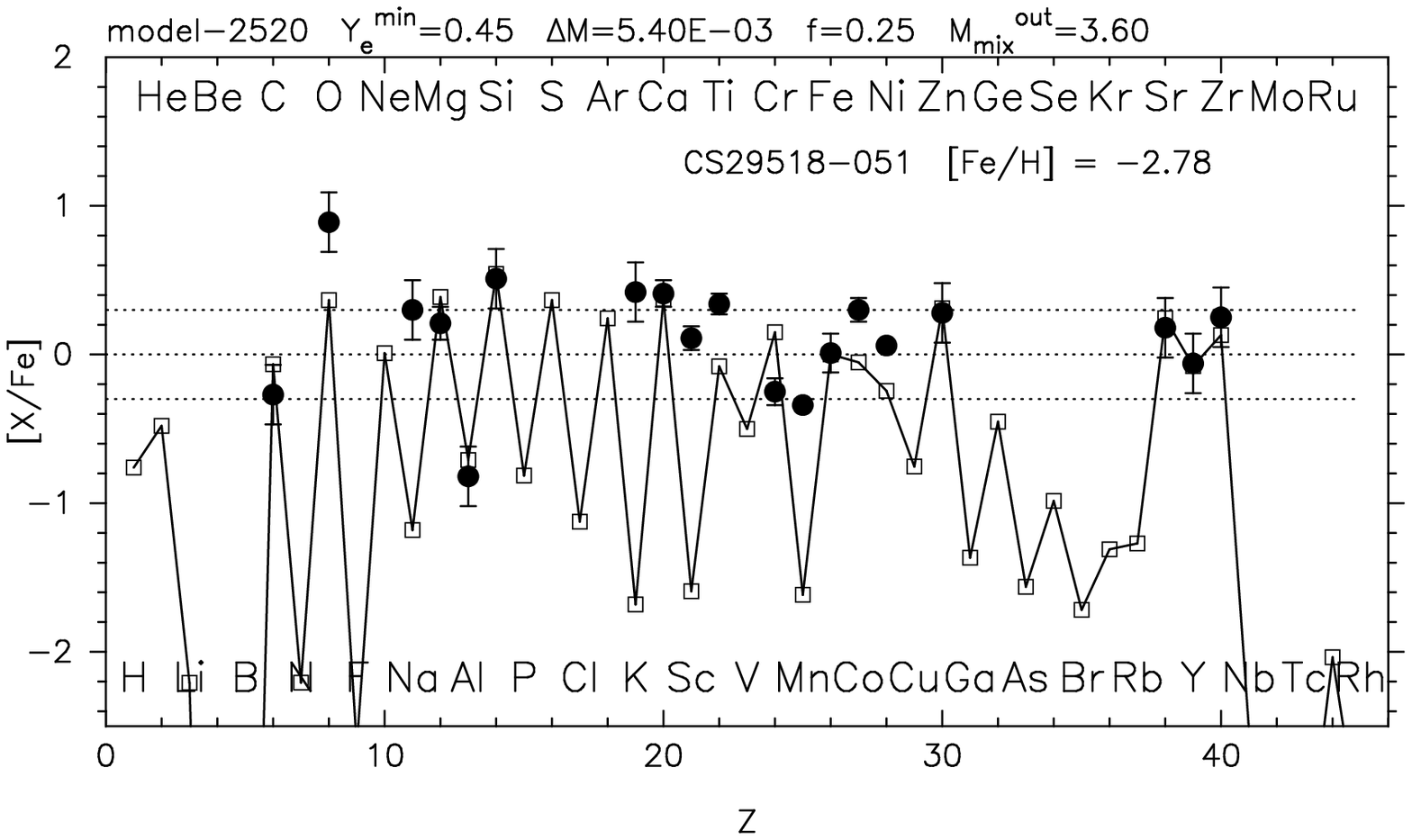}

\caption{Comparisons between the yields of our mixing-fallback models
and the abundance patterns of weak r-process stars.}
\end{figure}

\clearpage

\begin{figure}

\includegraphics[width=80mm]{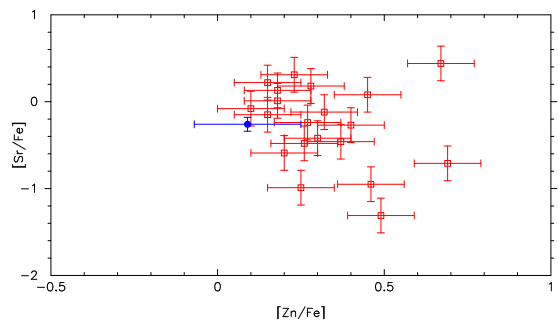}

\caption{Same as Figure 1, but for [Sr/Fe] vs. [Zn/Fe].
}
\end{figure}

\end{document}